\documentclass{article}
\usepackage{graphicx}

\begin{document}
\thispagestyle{empty}


{\LARGE Pasta structures in compact stars}
\bigskip

{{Toshiki Maruyama}\\
Advanced Science Research Center, Japan Atomic Energy Agency, Tokai, Ibaraki, 319-1195 Japan}

{{Toshitaka Tatsumi}\\
Department of Physics, Kyoto University, Kyoto 606-8502, Japan}

{{Tomoki Endo\footnote{Present address: Research Center for Nuclear Physics (RCNP), Osaka University, Ibaraki, Osaka 567-0047, Japan}}\\
Department of Physics, Kyoto University, Kyoto 606-8502, Japan
}

{{Satoshi Chiba}\\
Advanced Science Research Center, Japan Atomic Energy Agency, Tokai, Ibaraki, 319-1195 Japan}

\bigskip


We review our recent works about ``pasta'' structures following the first-order 
phase transition in dense matter, 
which correspond to the structured mixed phases with geometrical symmetries.
Three kinds of phase transitions at different density ranges 
are examined as the stages of pasta structures: liquid-gas phase
transition at subnuclear density, kaon condensation and hadron-quark
phase transition at high density.
Charge density as well as particle density is non-uniform there. 
A consistent treatment of the Coulomb potential and
the particle densities is presented
and a peculiar role of the Coulomb potential is elucidated:
the physical picture of the Maxwell construction will be effectively recovered.
It largely influences the density regime of pasta structures by the charge screening effect.

\newpage



\section{Introduction}

Recently many efforts have been made to reveal ``new form of matter'' under
extreme conditions such as high-temperature and/or high-density.
We may find such form through high-energy heavy-ion experiments or 
observation of compact stars. 
In this review article we discuss non-uniform structures of cold matter at high density, 
which is relevant for compact stars. 
Actually compact stars are good laboratories for nuclear physics and
elementary particle physics to study matter at extreme conditions \cite{sha,tam,glebook,web}.

Nuclear matter in the ground state,
which is approximately realized in atomic nuclei,
consists of the same number of protons and neutrons
and has the density of $\rho_0\simeq 0.16$ fm$^{-3}$ 
``the normal nuclear density'' or ``the saturation density'', and the
binding energy of around $-16$ MeV.
This is called ``saturation property'' of nuclear matter.
However, matter in the stellar objects has
variety of density and chemical component due to the presence of gravity.
In fact, at the surface of neutron stars,
there exists a region where the density is lower than $\rho_0$
over a couple of hundreds meters.
The pressure of such matter is retained by degenerate electrons,
while baryons are clusterized and have little contribution to the pressure.
Due to the gravity the pressure and the density increase
in the inner region
(in fact, the density at the center amounts to several times 
the normal nuclear density).
Charge neutral matter consists of neutrons and the same number of protons
and electrons under chemical equilibrium.
Since the kinetic energy of degenerate electrons is much higher than
that of baryons, the electron fraction (or the proton one) decreases
with increase of density and thus neutrons become the main component
and seep out from the clusters.
In this way baryons come to contribute to the pressure as well as electrons.
At a certain density, other components such as hyperons and strange
mesons may emerge.
For example, negative kaon condensation, expected to be of a first-order
phase transition (FOPT), remarkably softens the equation of state of matter.
At even higher density, hadron-quark deconfinement transition may occur
and quarks in hadrons are liberated.
This phase transition is also considered to be of first-order.

If we assume uniform matter,
a FOPT brings about a thermodynamical instability.
In other words, matter should have the mixed phase around
the critical density.
Consider isotherm of water at room temperature for example.
At low density (or pressure) water exists in the vapor phase.
With increase of density (or pressure), droplets of water appear
in the vapor phase.
Finally  all the volume is filled with liquid water at a certain density 
(or pressure).
The equation of state (EOS), i.e.\ the relation of pressure and density, 
at low density and high density is given by those of the vapor and liquid phases, respectively.
In between, the EOS of the mixed phase is obtained by the Maxwell construction.
Required conditions are equalities of chemical potential, temperature and pressure
between the vapor and liquid phases.

Phase transitions in nuclear matter are different from that in water.
First, nuclear matter consists of two chemically independent components, 
i.e.\ baryons and electrons. 
Then the  equalities of both baryon and electron chemical potentials 
between two phases are required by the Gibbs conditions in the mixed phase.
Therefore the EOS of the mixed phase cannot be obtained simply by the
Maxwell construction, which is relevant only for single component.
Secondly, those components are electrically charged and the mixed phase is
no more uniform.
This point is important to the geometrical structure of the mixed phase.
In the case of the mixed phase of electrically neutral particles, each phase 
has arbitrary geometry 
when surface tension is negligible
or two phases are separated when the surface tension is strong.
On the other hand the mixed phase in nuclear matter should possess some
specific size and shape due to the balance between the surface tension 
and the Coulomb interaction.
To minimize the surface energy plus the Coulomb energy,
matter is expected to form a structured mixed phase, i.e.\ a lattice of 
lumps of a phase with a geometrical symmetry embedded in the other phase.

\begin{figure}
\includegraphics[width=0.6\textwidth]{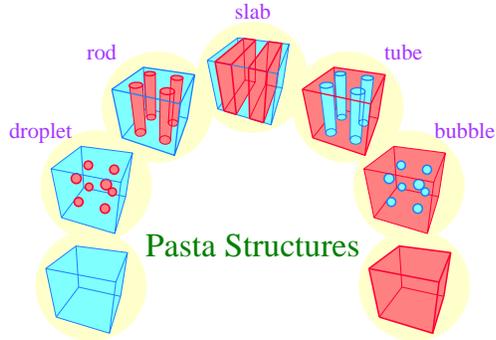}
 \caption{Schematic picture of pasta structures.
 Phase transition from blue phase (left-bottom) to red phase (right-bottom) 
 is considered.
 }
\label{pasta}
\end{figure}

At very low densities, nuclei in matter are expected to form the
Coulomb lattice embedded in the electron sea, that
minimizes the Coulomb interaction energy. 
With increase of density, ``nuclear pasta'' structures (see Fig.\ \ref{pasta}) 
emerge as a structured mixed phase \cite{Rav83} in the liquid-gas phase transition, 
where stable nuclear shape may change from droplet to rod, slab, tube, and to bubble. 
Pasta nuclei are eventually dissolved into uniform matter at a certain nucleon 
density below the saturation density $\rho_0\simeq 0.16$ fm$^{-3}$. 
The name ``pasta'' comes from rod and slab structures 
figuratively spoken as ``spaghetti'' and ``lasagna''.
Such low-density nuclear matter exists in the collapsing stage of
supernovae and in the crust of neutron stars.
Supernova matter is relevant to liquid-gas transition
of non-beta-equilibrium nuclear matter with a fixed proton mixing ratio
and the low-density neutron star matter is relevant to 
neutron-drip transition of beta-equilibrium nuclear matter.

The appearance of the pasta structures could have important consequences
for the various neutron star phenomena or supernova.
If they appear at an early stage of the neutron star evolution nuclear
pasta structures could
cause a drastic change of the neutrino opacity \cite{horowitz},
and consequently influence the subsequent neutron star cooling \cite{tsuru,yak,pag}. 
It may also affect the matter resistance to the stress and consequently 
the glitch phenomena \cite{MI95}, etc.

A number of authors have investigated the low-density nuclear
matter using various models
\cite{Rav83,Has84,Wil85,Oya93,Lor93,Cheng97,maru98,kido00,Gen00,Gen02}.
Roughly speaking, the favorable nuclear shape is determined by a
balance between the surface and Coulomb energies. 
In most of the previous studies the rearrangement effect of the density
profile of the charged particles due to the Coulomb interaction has been discarded. 
In Ref.\ \cite{Gen03} the electron screening effect has been studied 
and it has been found that this effect is of minor importance. 
However, the rearrangement of the proton profiles as the consequence 
of the Coulomb repulsion was not shown up in their model.
The structured mixed phase is also expected in the phase transitions at
higher density, 
like kaon condensation and 
hadron-quark phase transition.
In these cases, the charge screening effect may be important
because the local charge density can be significantly high.

As it was argued in a number of works, 
the FOPT to a $K^-$ condensate state might occur
in neutron stars at densities several times larger than the
nuclear saturation density $\rho_0$
(see \cite{kn86,BLRT20,MTT93,T95,KVK95,PREPL00,RBP00,GS99,CG00,CGS00,NR00,YT00,M01}
and refs therein).
Refs \cite{kn86,BLRT20,MTT93,T95,PREPL00,RBP00,GS99,CG00,CGS00,NR00,YT00}
and many other studied possibility of the
$s$-wave kaon condensation, whereas refs \cite{KVK95,M01}
studied different possibilities for the $p$-wave condensation.
It was concluded that the kaon condensation in neutron stars
may indeed occur and it is the FOPT.
Such phenomena 
would lead to interesting consequences in the physics of neutron stars:  
softening of EOS may give a possibility of the delayed 
collapse of protoneutron stars to the low-mass black holes, 
and the nucleon Urca process under background kaons 
may give a fast cooling mechanism of neutron 
stars \cite{T95,YT00,L96P97,bb94,bkpp88,t88,fmtt94,pb90}.
Kaon condensation as a FOPT may also lead to an expected existence of
a wide region of the mixed phase, as was suggested in
earlier works, cf.\ \cite{GS99,CG00,CGS00}, which consists of kaon
condensed phase (dense phase) and usual nuclear matter (dilute phase)
with some geometrical structures ({\it kaonic pasta}).

Next we consider the hadron-quark mixed phase.
Within the currently accepted theory,
the deconfinement transition is believed to occur in hot and/or
high-density matter, although its mechanism is not yet well understood.
Many
people have studied this transition with model
calculations and first-principle calculations, like lattice QCD \cite{rev}.
Above the critical temperature and/or density, the quarks inside each hadron
are liberated, and we can understand the resulting situation as 
quark matter interacting through gluon exchange.
Theoretically, static and dynamic properties of quark matter have been
 extensively studied for quark-gluon
 plasmas (QGP), color superconductivity \cite{alf1,alf3} and magnetism
 \cite{tat1,tat2,tat3}.
Experimentally, evidence of quark matter has been searched for in
relativistic heavy-ion collisions (RHIC) \cite{rhic1,mul}
and in relics of the early universe and in compact stars \cite{glebook,web,mad3,chen}.

Because many theoretical calculations have suggested that the deconfinement
transition should be of first order
in the low temperature, high density regime \cite{pisa,latt},
 we assume it to be a first-order phase transition.
If the
deconfinement transition is of first order,
it is natural to believe that in the phase transition the system is
in a mixed phase.
Actually, the hadron-quark mixed phase has been considered during
hadronization in RHIC \cite{rhic21,rhic22,rhic23}
and in the transition region from hadron matter to quark matter in neutron
stars \cite{gle2,gle4,pet,alf2}.

Our purpose here is 
to investigate low-density nuclear pasta structure, kaonic pasta structure,
and hadron-quark pasta structure self-consistently within  
the mean-field approximation. 
In particular, 
we figure out how the charge screening effects modify
the results obtained disregarding these effects in a self-consistent manner.

This review article is organized as follows.
In Sec.\ 2, investigations on the mixed phase is reviewed.
In Sec.\ 3 and 4, the nuclear pasta structure in low-density nucleon matter
and the kaonic pasta structure at higher densities are studied.
Then the finite-size effects on the pasta structures are
discussed in Sec.\ 5.
In Sec.\ 6 and 7, pasta structures in hadron-quark phase 
transition and the charge-screening effects are discussed.
Finally, summary and concluding remarks are given in Sec.\ 8.

\section{General remarks about the treatment of the mixed phases}

\subsection{Bulk calculation and the finite-size effects}

Consider the mixed phase composed of two different phases denoted by I and II.
Then the Gibbs conditions require the pressure balance and the equality of
the chemical potentials between two phases  
for phase equilibrium \cite{gug}.%
\footnote{We consider here matter at zero temperature}
 For a multi-component system with more than
one chemical potential, as is common in neutron-star matter, we must
impose the equality condition for each chemical potential in order  
to fulfill  the condition of the 
physico-chemical equilibrium. 
More definitely, we
hereafter consider the charge chemical potential ($\mu_Q$) and 
the baryon-number
chemical potential ($\mu_B$) respecting two conservation laws in neutron-star
matter: $\mu_Q^{\rm I}=\mu_Q^{\rm II}$ and $\mu_B^{\rm I}=\mu_B^{\rm II}$. 
On the other hand, the first condition is not fulfilled in the Maxwell construction,
since the $\it local$ charge neutrality is implicitly imposed, while
only the {\it global} charge neutrality must be satisfied. 

A naive application of the Gibbs conditions to the mixed phase composed
of two semi-infinite matter, when one ignores
the surface and Coulomb interaction, demonstrates a broad region of
the structured mixed phase (SMP),  cf.\ \cite{CG00,gle92}.
However the charge screening effect (caused by the non-uniform 
charged particle distributions) should be very important when 
the typical structure size is of the order of
the minimal Debye screening length in the problem.
It may largely  affect the stability condition of 
the geometrical structures in the mixed phases.

We shall see
that the Debye screening effects greatly modify the 
mechanical stability of SMP, and consequently largely limit the density
region of the mixed phase.
In the absence 
of SMP we effectively recover the picture of phase equilibrium 
given by the Maxwell
construction where two bulk phases are separated without 
spoiling the Gibbs conditions. 

\begin{figure}[h]
\begin{center}
\includegraphics[width=0.5\textwidth]{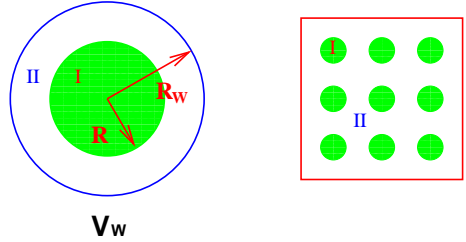}
\end{center}
 \caption{Equal droplets made of the phase I embedded in the phase II (right panel), 
 and the geometrical structure of the Wigner-Seitz cell (left panel)}
\label{ws}
\end{figure}

Consider SMP consisting of two phases I and II, 
where we assume spherical droplets of phase I with a radius $R$ to be
embedded in the matter of phase II and two phases are clearly separated
by sharp boundaries. 
We divide the whole space into the
equivalent Wigner-Seitz cells with a radius $R_{\rm W}$ (see Fig.\ \ref{ws}).
The volume of the cell is $V_{\rm W}=4\pi R_{\rm W}^3/3$ and that of the
droplet is $V=4\pi R^3/3$. 

\begin{figure}[h]
\begin{center}
\includegraphics[width=0.45\textwidth]{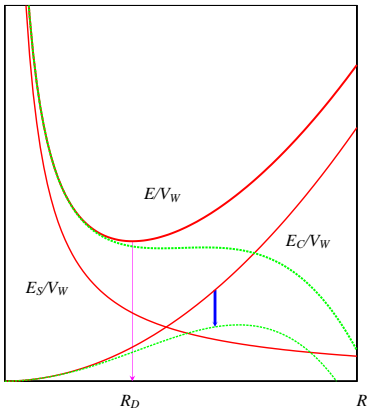}
\end{center}
\caption{Schematic view of the total energy and each contribution 
in the bulk calculations (solid curves). 
 Screening effect reduces the Coulomb energy, shown by
 the thick arrow.}
\label{bulk}
\end{figure}

A bulk calculation proceeds as follows \cite{pet}.
First, consider two semi-infinite matter with a volume fraction $f_V$,
separated by a sharp boundary. 
By applying the global
charge-neutrality condition and the pressure balance condition under the
chemical equilibrium, we can get $f_V$ for each density.
Then we use the volume fraction thus determined to describe the
geometrical structure in each Wigner-Seitz cell.
For a given volume fraction 
$f_V=(R/R_{\rm W})^3$, the total energy $E$ may be written as the sum of the
volume energy $E_V$, the Coulomb energy $E_C$ and the surface energy $E_S$,
\begin{equation}
E=E_V+E_C+E_S.
\label{energy}
\end{equation}
We further assume, for simplicity, that baryon number ($\rho_B^\alpha$) and
charge ($\rho_{\rm ch}^\alpha$) densities are uniform in each phase
$\alpha,~\alpha={\rm I},{\rm II}$ as in semi-infinite matter. 
Then, $E_V$ can be written as
$E_V/V_{\rm W}=f_V\epsilon^{\rm I}(\rho_B^{\rm I})+(1-f_V)\epsilon^{\rm II}(\rho_B^{\rm II})$ in terms
of the energy densities $\epsilon^\alpha,~\alpha={\rm I},{\rm II}$. 
The surface energy $E_S$ may 
be represented as $E_S/V_{\rm W}=f_V\times 3\tau/R$ in terms of the surface
tension $\tau$. 
The Coulomb energy $E_C$ is given by 
\begin{equation}
E_C/V_{\rm W}=f_V\times \frac{16\pi^2}{15}\left(\rho_{\rm ch}^{\rm I}-\rho_{\rm ch}^{\rm II}\right)^2R^2.
\end{equation}
The optimal value of $R$, which we call $R_{\rm D}$, 
is determined by the minimum condition,
\begin{equation}
\left.\frac{\partial (E/V_{\rm W})}{\partial R}\right|_{R=R_{\rm D}}=0,
\end{equation}
for a given $f_V$ (see Fig.\ \ref{bulk}). 
Since $E_V$ does not depend on $R$, 
we can {\it always} find a minimum as a result of the competition
between the Coulomb and the surface energies, satisfying the 
well-known relation, $E_S=2E_C$.
However, such bulk calculations have been proved to be too crude for the 
discussions of SMP. 
Instead, a careful consideration of 
the interface of two phases is required. 
As a defect of the bulk calculations they ignore the {\it finite-size effects}. 
In particular, they have the inconsistent treatment
of the Coulomb potential; they do not use the Poisson equation, so that
the charge density profiles are assumed ab initio to be uniform 
and the Coulomb potential is assumed to be $1/r$. 
If one properly solves the Poisson
equation, one should have the screening effect as a result of the rearrangement
of the charge-density distribution. 
Hence, the radius $R_{\rm D}$ should be not too large compared with 
the Debye screening length 
$\lambda_{\rm D}$, 
\begin{eqnarray}
\lambda_{\rm D}^{-2}&=&\sum_i(\lambda_{\rm D}^i)^{-2}, \\
{1/\lambda_{\rm D}^i}^2&=&4\pi \frac{\partial\rho_{Q}}{\partial\mu_i},
\end{eqnarray} 
in order the above treatment to be justified,
the suffix $i$ runs over the particle species. 
Otherwise, the Coulomb
energy is reduced by the charge screening effect, 
which should lead to a {\it mechanical
instability} of SMP in some cases (Fig.~\ref{bulk}).

We have been recently exploring the effect of the charge screening 
in the context of the various structured mixed phases
\cite{voskre,voskre1,emaru1,maruKaon,dubna,maru1}.
In fact, we have examined the mixed phase at the
hadron-quark transition, kaon condensation and of nuclear pasta, 
and found that in cases of the hadron-quark transition and kaon condensation
the mixed phase might be largely limited by the charge screening
and surface effects.
In the case of the hadron-quark deconfinement transition, for example, 
$ \lambda_{\rm D}^q\simeq 5$ fm
and $\lambda_{\rm D}^p, \lambda_{\rm D}^e$ are of the same order as $\lambda_{\rm D}^q$, for
a typical density with $\mu_B\simeq 1$ GeV. 
We shall see in the following that $R_{\rm D}$ is typically of the same order as
$\lambda_{\rm D}\sim \lambda_{\rm D}^q$, and the mechanical stability of the droplet
is much affected by the screening effect. 
 
Besides the screening effect, note that the surface tension and the pressure
balance condition are inconsistently incorporated in the bulk
calculations; actually the pressure balance condition is imposed before
introducing the surface tension.

\subsection{Gauge invariance and the meaning of chemical potentials}

When we use the idea of the density functional theory \cite{drez,parr}, 
the thermodynamic potential is given as a functional of the particle
densities $\rho_i({\bf r})$, 
\begin{equation}
\Omega(\rho_i({\bf r}))=\int d^3r\epsilon_{\rm kin+str}(\rho_i({\bf r}))
+E_V-\sum_i\mu_i\int d^3r\rho_i({\bf r}),
\end{equation}
where $\mu_i$ are the chemical potentials and $\epsilon_{\rm kin+str}$
stands for the contributions of the kinetic energy and the strong
interaction energy.
The Coulomb interaction energy 
$E_V$ is also expressed in terms of particle densities,
\begin{equation}\label{enden2}
E_V=\frac{1}{2}\sum_{i,j}\int d^3 r\,d^3 r'
\frac{Q_i \rho_i ({\bf r}) Q_j \rho_j ({\bf r}') }
{\mid {\bf r}-{\bf r}'\mid},
\end{equation}
with $Q_i$ being the particle charge ($Q =-e <0$ for the electron). 
Then the chemical potentials are given as 
\begin{equation}\label{eom}
\mu_i=\frac{\partial\epsilon_{\rm kin+str}}
{\partial\rho_i}- N^{{\rm{ch}}}_i V
({\bf r})  ,~~~ N^{{\rm{ch}}}_i =Q_i/e,
\end{equation}
with the electric potential $V({\bf r})$: 
\begin{equation}\label{other1}
V({\bf r} ) =- \sum_i\int d^3r'\, \frac{e Q_i \rho_i
({\bf r}') }{\mid {\bf r}-{\bf r}'\mid }
\end{equation}
generated by the particle distributions.

We must keep the gauge invariance through the calculation: $V$
can be arbitrarily shifted by a constant $V_0$, $V({\bf r})\rightarrow V({\bf r}) -V_0$. 
Formally varying Eq.~(\ref{eom}) with respect to $V({\bf r})$ or
$\mu_i$ we have the matrix form relation,
\begin{eqnarray}\label{matrix}
A_{ij}\frac{\partial\rho_{j}}{\partial V}=
N^{{\rm ch}}_i ,\,\,\,\, A_{ij}B_{jk}
=\delta_{ik},
\end{eqnarray}
where  matrices $A$ and $B$ are defined as
\begin{eqnarray}\label{a}
A_{ij} \equiv\frac{\delta^2 E_{\rm kin+str}}
{\delta\rho_i\delta\rho_j},\,\,\,\,
B_{ij}\equiv\frac{\partial\rho_i}{\partial\mu_j}.
\label{matrix1}
\end{eqnarray}
Eqs.~(\ref{matrix}) and (\ref{matrix1})  reproduce the gauge-invariance
relation,
\begin{equation}
\frac{\partial\rho_i}{\partial V}=N^{\rm ch}_j \frac{\partial\rho_j}{\partial\mu_i},
\label{gauge}
\end{equation}
clearly showing that constant-shift of the chemical potential is
compensated by the gauge transformation of $V({\bf r})$:
$\mu_i\rightarrow \mu_i+N_i^{{\rm ch}}V_0$, as
$V({\bf r})\rightarrow V({\bf r}) -V_0$. 
Hence chemical potential
$\mu_i$ acquires physical meaning only after fixing the
gauge of $V({\bf r})$. 

\begin{figure}[h]
\begin{center}
\includegraphics[width=0.4\textwidth]{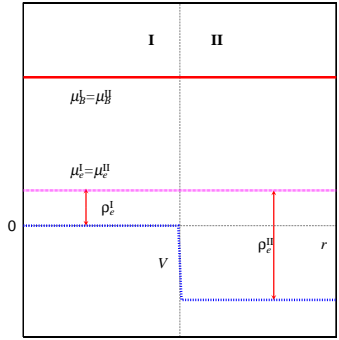}
\end{center}
\caption{Relation between the charge chemical potential $\mu_Q(=\mu_e)$
 and the electron number density $\rho_e$ in the presence of the Coulomb
 potential $V$. 
 Fulfilling the Gibbs conditions, $\mu_B^{\rm I}=\mu_B^{\rm II},~\mu_e^{\rm I}=\mu_e^{\rm II}$, 
 we can change $\rho_e$ in
 two phases as in the Maxwell construction, if $V$ suitably changes from one
 phase to another.}
\label{gainv}
\end{figure}

We reconsider the relation between the Gibbs conditions and the Maxwell
construction from this view point. 
Consider a schematic situation for simplicity, where two
semi-infinite matter denoted by I and II are separated by a sharp boundary.
As has been  mentioned, at first glance the Maxwell construction
apparently  violates the Gibbs conditions, 
especially the equilibrium condition for the charge
chemical potential $\mu_Q(=\mu_e)$ in our context. 
However, correctly speaking,
when we say $\mu_e^{\rm I}\neq \mu_e^{\rm II}$ within the Maxwell construction, it means nothing
but the difference in the electron number density $\rho_e$ in  two
phases, $\rho_e^{\rm I}\neq \rho_e^{\rm II}$; this is simply because $\rho_e=\mu_e^3/(3\pi^2)$, 
if the Coulomb
potential is {\it absent}. 
Once the Coulomb potential is taken into account, 
using Eq.~(\ref{eom}), $\rho_e$ can be written as 
\begin{equation}
\rho_e=\frac{(\mu_e-V)^3}{3\pi^2}. \label{enumdens}
\end{equation}
Thus we may have $\mu_e^{\rm I}=\mu^{\rm II}_e$ and $\rho_e^{\rm I}\neq \rho_e^{\rm II}$
simultaneously, with the {\it different values of $V$}, $V^{\rm I}\neq V^{\rm II}$
(see Fig.~\ref{gainv}).
We shall see that if the Coulomb interaction is properly taken into
account, the resultant EOS looks similar to that given by the Maxwell construction.

\section{Low-density nuclear pasta structures}

\subsection{Relativistic mean-field (RMF) treatment of nucleon matter}

Here we explain how to investigate the property of nuclear matter
at low density.
Exploiting the idea of the density functional theory within the RMF model, we can formulate
equations of motion to  study non-uniform nuclear matter numerically, cf.\ \cite{drez,parr}.
The RMF model with fields of mesons and baryons introduced
in a Lorentz-invariant way is not only relatively simple for numerical calculations,
but also sufficiently
realistic  to reproduce bulk properties of finite nuclei
as well as the saturation properties of nuclear matter \cite{maru05}.
In our framework, the Coulomb interaction is properly included in the 
equations of motion for nucleons and electrons and for meson mean fields,
and we solve the Poisson equation for the Coulomb potential $V$
self-consistently with those equations.
Thus the baryon and electron density profiles, as well as the meson
mean fields, are determined in a fully
consistent way with the Coulomb interaction.

\subsubsection{Thermodynamic potential and the equations of motion}

We start with the thermodynamic potential for
the system of neutrons, protons, electrons and mesons including kaons
\begin{equation}\label{Omega-tot}
\Omega = \Omega_N+\Omega_M +\Omega_e.
\end{equation}
The first term
\begin{equation}
\Omega_N  =
  \sum_{a=p,n}
  \int  d^3r
  \left[
  \int_0^{k_{{\rm F},a}}
  { d^3k \over 4\pi^3}
  \sqrt{{m_N^*}^2+k^2}-\rho_a\nu_a
  \right]
\label{OmegaN}
\end{equation}
is the contribution of nucleons with the local Fermi momenta
$k_{{\rm F},a}({\bf r}); a=n,p$, $m_N^*({\bf r})=m_N-g_{\sigma N}\sigma({\bf r})$
is the effective nucleon mass and $m_N$ is the nucleon mass in the vacuum.
Nucleons couple with $\sigma$, $\omega$ and $\rho$ mesons and thereby,
\begin{eqnarray}
\nu_n({\bf r})&=&\mu_n-g_{\omega N}\omega_0({\bf r})+g_{\rho N}R_0({\bf r}),\\
\nu_p({\bf r})&=&\mu_p+
{V({\bf r})}-g_{\omega N}\omega_0({\bf r})-g_{\rho N}R_0({\bf r}),
\nonumber
\end{eqnarray}
where $\mu_n$ and $\mu_p$ are neutron and proton chemical potentials
and  $g_{\sigma N}$, $g_{\omega N}$ and  $g_{\rho N}$
are coupling constants between corresponding fields.

The second term  in  (\ref{Omega-tot}) incorporates
the scalar ($\sigma$) and vector ($\omega_0, R_0$) mean fields,
\begin{eqnarray}
\Omega_M &=& \int   d^3r \Biggl[
  {(\nabla\sigma)^2 + m_\sigma^2\sigma^2 \over2} + U(\sigma) \nonumber\\
 &&{}-{(\nabla\omega_0)^2 + m_\omega^2\omega_0^2 \over2}
  -{(\nabla R_0)^2 + m_\rho^2R_0^2\over2}  \Biggr]  ,
\label{OmegaM}
\end{eqnarray}
where $m_\sigma$, $m_\omega$ and $m_\rho$ are the field masses, and
$U(\sigma)={1\over3}bm_N(g_{\sigma N}\sigma)^3+{1\over4} c(g_{\sigma N}\sigma)^4$
is the nonlinear potential for the scalar field.

The third term in (\ref{Omega-tot}) contains the contribution of
the Coulomb field  (described by the potential $V({\bf r})$)
and the contribution of relativistic electrons,
\begin{equation}
\Omega_e = \int d^3r \left[
-{1\over8\pi e^2}(\nabla {V})^2-{(\mu_e-{V})^4\over12\pi^2}
\right],
\end{equation}
where $\mu_e$ is the electron chemical potential.

Temperature $T$ is kept to be zero in the present study.

For nucleons and electrons we used the local-density approximation, i.e.,
nucleons and electrons are described by their local densities.
This approximation  has its sense only if the typical length of the change 
of the nucleon density is larger than the inter-nucleon distance.
Derivative terms of the  particle densities can be  incorporated in
the quasi-classical manner by  the derivative expansion within
the density functional theory \cite{drez,parr}.
Their contribution to the energy can be reduced to a  surface tension term.
Here we simply discard those derivative terms, as a first-step approximation.
Thus we discard the contribution of the nucleon fields to the surface tension assuming
that it is smaller than the corresponding contribution of the meson fields that we retain.
In the case when we suppress derivative terms of the
nucleon densities they follow changes of the meson $\sigma$, $\omega$, $\rho$
mean fields and the Coulomb fields that have derivative terms.
Note that we have fitted our model to properly describe
finite nuclei (see below) without including nucleon derivative terms.
If we introduced them it would need to re-adjust the model parameters.
We should  bear in mind that for small structure
sizes quantum effects become prominent.
For simplicity we disregard these effects.
Thus we may properly describe only rather large-size structures  within this scheme.

\begin{table*}
\caption{
Parameter set used in RMF in our calculation.
}
\begin{center}
\begin{tabular}{ccccccccccccc}
\hline
$g_{\sigma N}$ &
$g_{\omega N}$ &
$g_{\rho N}$ &
$b$ &
$c$ &
$m_\sigma$ &
$m_\omega$ &
$m_\rho$ &
\\
\hline\\
6.3935 &
8.7207 &
4.2696 &
0.008659 &
0.002421 &
 400 MeV &
 783 MeV &
 769 MeV &
\\
\hline
\end{tabular}
\end{center}
\end{table*}

\subsubsection{Equations of motion and numerical procedure}

Equations of motion for the mean fields 
and for the Coulomb potential are obtained
from the variational principle: ${\delta\Omega\over\delta\phi_i({\bf r})}=0$
($\phi_i=\sigma,\omega_0,R_0 ,V$), 
\begin{eqnarray}
\nabla^2\sigma({\bf r}) &=& m_\sigma^2\sigma({\bf r}) +{dU\over d\sigma}
-g_{\sigma N}(\rho_n^{s}({\bf r}) +\rho_p^{s}({\bf r}))
,\label{EOMsigma}\\
\nabla^2\omega_0({\bf r}) &=& m_\omega^2\omega_0({\bf r}) 
-g_{\omega N} (\rho_p({\bf r})+\rho_n({\bf r}))
,\label{EOMomega}\\
\nabla^2R_0({\bf r}) &=& m_\rho^2R_0({\bf r}) -g_{\rho N}
(\rho_p({\bf r})-\rho_n({\bf r}))
,\label{EOMrho}\\
\nabla^2V({\bf r}) &=& 4\pi e^2\rho_{\rm ch}({\bf r}),
\label{EOMpoisson}
\end{eqnarray}
with 
the proton and neutron scalar densities 
\begin{eqnarray}
\rho_a^{s}({\bf r})&=& 
\int_0^{k_{{\rm F},a}({\bf r})} \frac{d^3k}{4\pi^3} 
\frac{m_N^*({\bf r)}}{\sqrt{m_N^*({\bf r})^2+k^2}}, \quad (a=p,n),
\label{rhos}
\end{eqnarray}
and the charge density 
\begin{eqnarray}
\rho_{\rm ch}({\bf r})&=& \rho_p({\bf r})-\rho_e({\bf r}).
\label{eq:poisson}
\end{eqnarray}
For nucleons and electrons, ${\delta\Omega\over\delta\rho_a({\bf r})}=0$
($a=n,p,e$) give the expressions of the chemical potentials; 
\begin{eqnarray}
\mu_e&=&\left({3\pi^2}\rho_e({\bf r})\right)^{1/3}+V({\bf r}),\label{EOMmuE}\\
\mu_n&=&\mu_B\;=\;\sqrt{k_{{\rm F},n}({\bf r})^2+{m_N^*({\bf r})}^2}
\nonumber\\&&\quad
{}+g_{\omega N}\omega_0({\bf r})
-g_{\rho N}R_0({\bf r}) \label{EOMmuN},\\
\mu_p&=&\mu_B-\mu_e = \sqrt{k_{{\rm F},p}({\bf r})^2+{m_N^*({\bf r})}^2}
\nonumber\\&&\quad
{}+g_{\omega N}\omega_0({\bf r})
+g_{\rho N}R_0({\bf r})-V({\bf r}).
\label{EOMmuNp}
\end{eqnarray}
The last two equations are the standard relations between the local
nucleon densities and chemical potentials.
We have assumed that the system is in the chemical equilibrium
with respect to the weak, electromagnetic and strong interactions
and we introduced the baryon chemical potential $\mu_B =\mu_n$
and the charge chemical potential, i.e.\ the electron chemical potential, $\mu_e$,
according to the corresponding conserved charges.

To solve the above coupled equations numerically,
the whole space is divided into equivalent Wigner-Seitz cells with a radius $R_{\rm W}$.
The geometrical shape of the cell changes as follows:
sphere in three-dimensional (3D) calculation, cylinder in 2D and slab in 1D, respectively.
Each cell is globally charge-neutral and all physical quantities
in the cell are smoothly connected to those of the next cell
with zero gradients at the boundary.
Every point inside the cell is represented by the grid
points (number of grid points $N_{\rm grid}\approx 100$) and
differential equations for fields are solved by the relaxation method
for a given baryon-number density under constraints of the global charge neutrality.
Details of the numerical procedure are explained in Ref.\ \cite{maru05}.
To fix the gauge of $V({\bf r})$, we choose the following condition as
\begin{equation}
V(R_{\rm W})=0.
\end{equation}

\subsubsection{Parameter set and finite nuclei}

\begin{figure}
  \includegraphics[width=.6\textwidth]{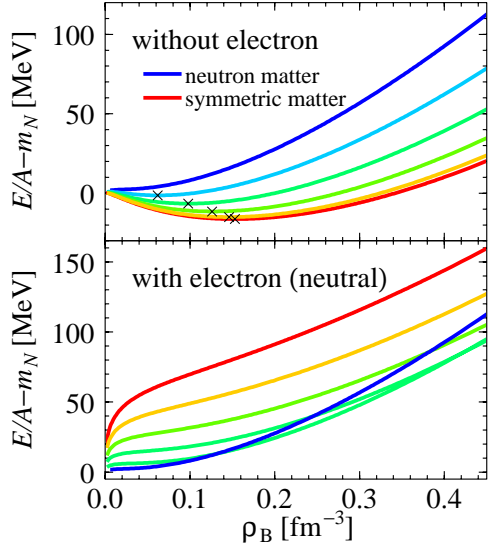}
  \caption{EOS of uniform nucleon matter with different 
  proton mixing ratios.
  Red line indicates symmetric matter (proton mixing ratio $Y_p=0.5$)
  and blue line indicates neutron matter ($Y_p=0$).
  Lines with intermediate colors show the case of $Y_p=0.4$, 0.3, 0.2 and 0.1.
  Crosses on the lines show the minimum points.
  Upper panel is the case without electron and the Coulomb energy,
  while the lower panel is the case of charge-neutral matter
  which includes energy of electrons.
}
\label{EOSunif}
\end{figure}

Parameters of the RMF model are chosen to reproduce saturation properties
of symmetric nuclear matter:
the minimum energy per nucleon $-16.3$ MeV at $\rho =\rho_0 \equiv 0.153$ fm$^{-3}$,
the incompressibility $K(\rho_0) =240$ MeV, the effective nucleon  mass
$m_N^{*}(\rho_0)=0.78 m_N$; $m_N =938$ MeV, and the isospin-asymmetry
coefficient $a_{\rm sym}=32.5$ MeV.
Coupling constants and meson masses used in our calculation are listed in Table 1.
Figure \ref{EOSunif} shows the binding energy per nucleon of uniform
nucleon matter. 
The proton mixing ratio $Y_p$ is the number ratio of
protons ($Z$) to total baryons ($A$), $Y_p=Z/A$.
The saturation property mentioned above is seen for symmetric nuclear matter 
($Y_p=0.5$) in the upper panel of Fig.\ \ref{EOSunif}.
Note that if one imposes the charge neutrality by inclusion of electrons,
the saturation property cannot be seen (lower panel of Fig.\ \ref{EOSunif}).

For the study of non-uniform nuclear matter, the ability to reproduce
the bulk properties of finite nuclei should be essential.
We check how it works to describe finite nuclei.
In this calculation, for simplicity, we assume  the spherical shape of nuclei.
The electron density is set to be zero. 
Therefore, neither the global charge neutrality
condition nor the local charge-neutrality condition is imposed.

\begin{figure}
  \includegraphics[width=.35\textwidth]{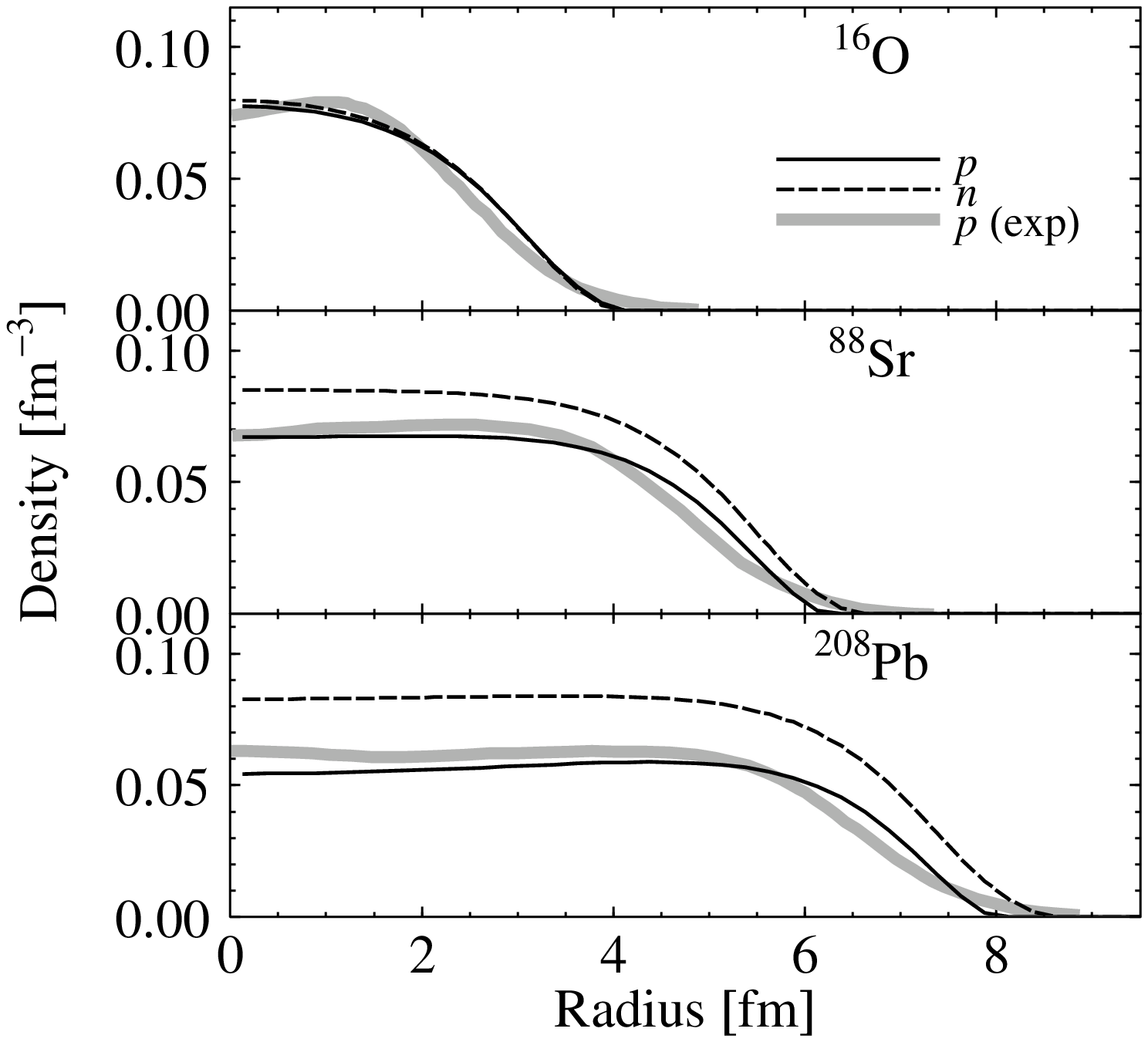}
  \includegraphics[width=.35\textwidth]{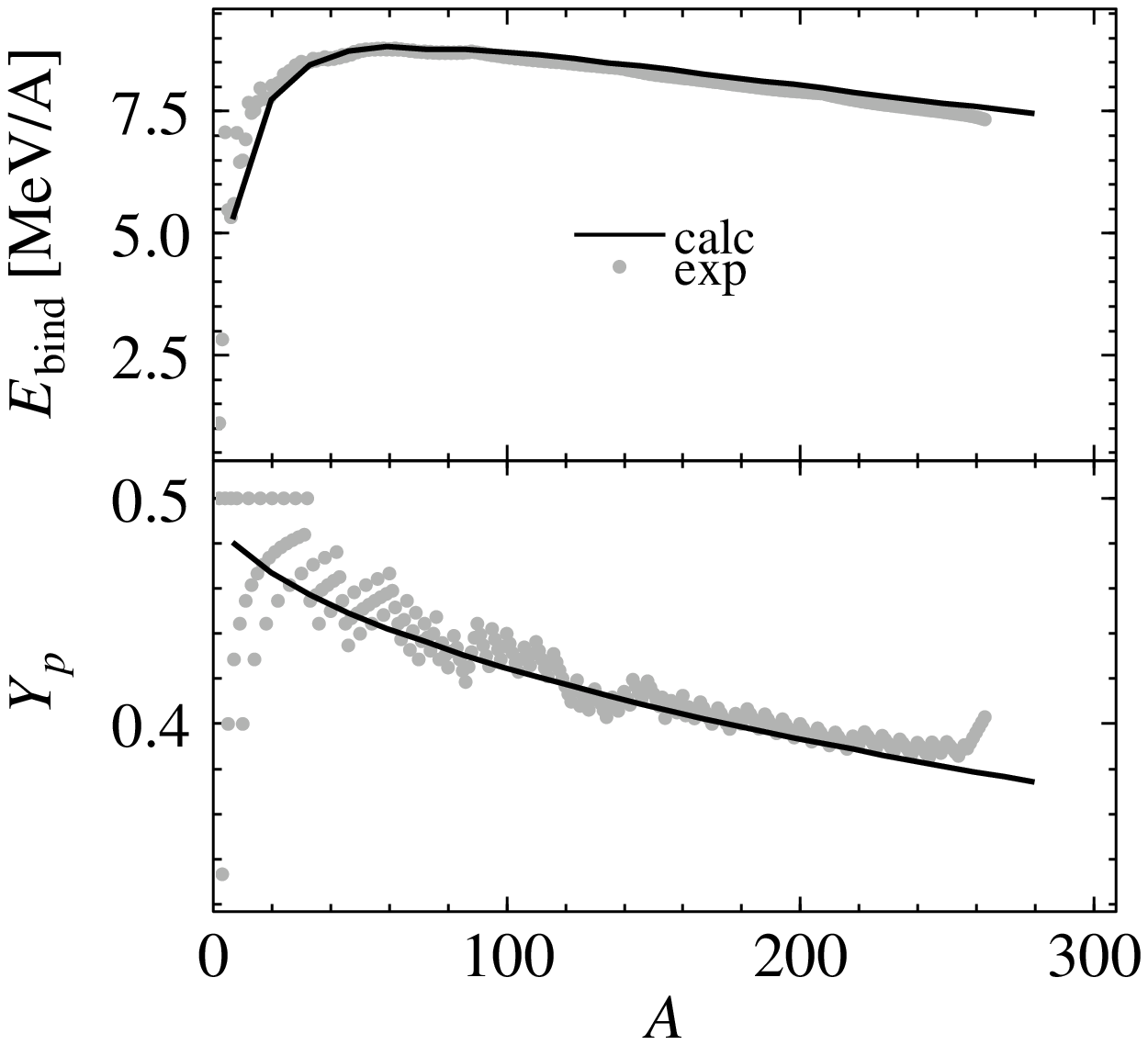}
  \caption{Properties of finite nuclei obtained with the present RMF model.
  Left: the density profiles of typical nuclei.
The proton number densities (solid curves) are compared with the experimental data.
Right: binding energy per nucleon $-(E/A-m_N)$ and the proton mixing ratio $Y_p$
of finite nuclei.
}
\label{finite}
\end{figure}

In Fig.~\ref{finite} (left panel) we show the density profiles of some typical nuclei.
One can see how well our framework may reproduce the density profiles.
To get a still better fit, especially around the surface region,
we might need to include the derivative terms of the nucleon densities,
as we have already remarked.
Fine structures seen in the empirical density profiles,
which may come from the shell effects (see, e.g., a proton density dip
at the center of a light $^{16}$O nucleus),
cannot be reproduced by the mean-field theory.
The effect of the rearrangement of the proton density distribution
is seen in heavy nuclei. 
Protons repel each other, 
which enhances their density near the surface of heavy nuclei. 
This effect is analogous to the charge screening effect for the Coulomb
potential in a sense that the proton
distribution is now changed not on the scale of  the nuclear radius, 
but on another length scale, that we  will call the proton Debye screening
length (see Eq.\ (\ref{Deb}) below).
It gives rise to important consequences for the pasta structures 
since the proton Debye screening
length is typically less than the lump size.
The optimal value of the proton mixing ratio $Y_p$ is obtained by imposing 
the beta-equilibrium condition for a given baryon number.
Figure \ref{finite}
(right panel) shows the baryon-number dependence of the binding
energy per baryon $E_{\rm bind}$ and the proton mixing ratio $Y_p$. 
We can see that the bulk properties of finite nuclei 
(density, binding energy, and proton to baryon number ratio) 
are satisfactorily reproduced for our present purpose.

Note that in our framework we must use a sigma mass 
$m_{\sigma}=400$ MeV \cite{centelles93}, a slightly smaller value 
than that one usually  uses, to get an appropriate fit. 
If we used a popular value $m_\sigma\approx 500$ MeV, finite nuclei would be
over-bound by about 3 MeV/$A$. 
The actual value of the sigma
mass (as well as the omega and rho masses) has little relevance
for the case of infinite nucleon matter, since it enters the
thermodynamic potential only in the combination
$\widetilde{C}_{\sigma}=g_{\sigma N}/m_{\sigma}$. 
However meson masses  are important characteristics of finite nuclei and of
other non-uniform nucleon systems, like those in pasta. 
The effective meson mass characterizes the typical scale for the
spatial change of the meson field and consequently it affects the
value of the effective surface tension.

\subsection{Nucleon matter at fixed proton mixing ratios}\label{Ypfix}

\begin{figure*}
\includegraphics[width=.31\textwidth]{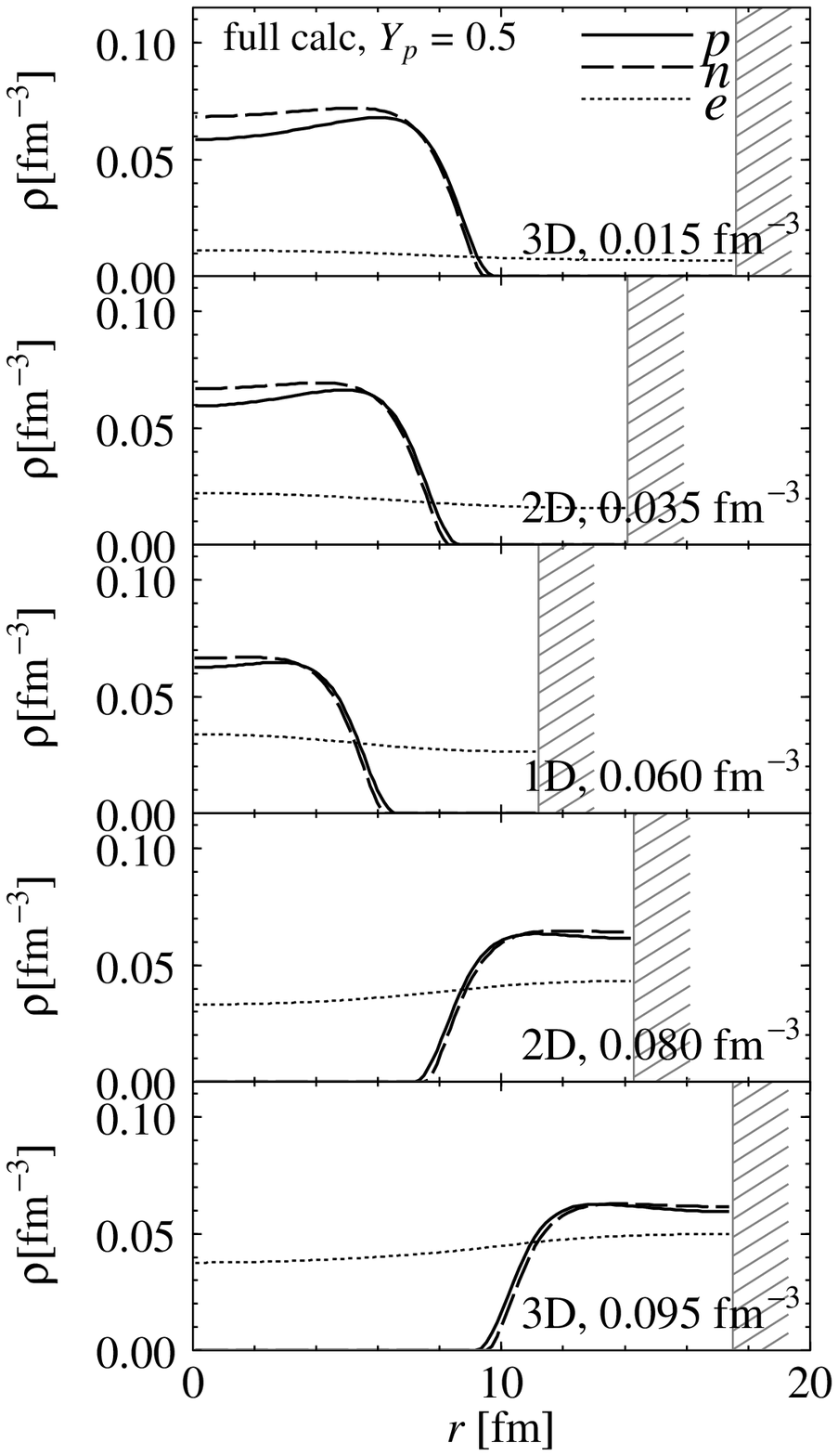}
\includegraphics[width=.31\textwidth]{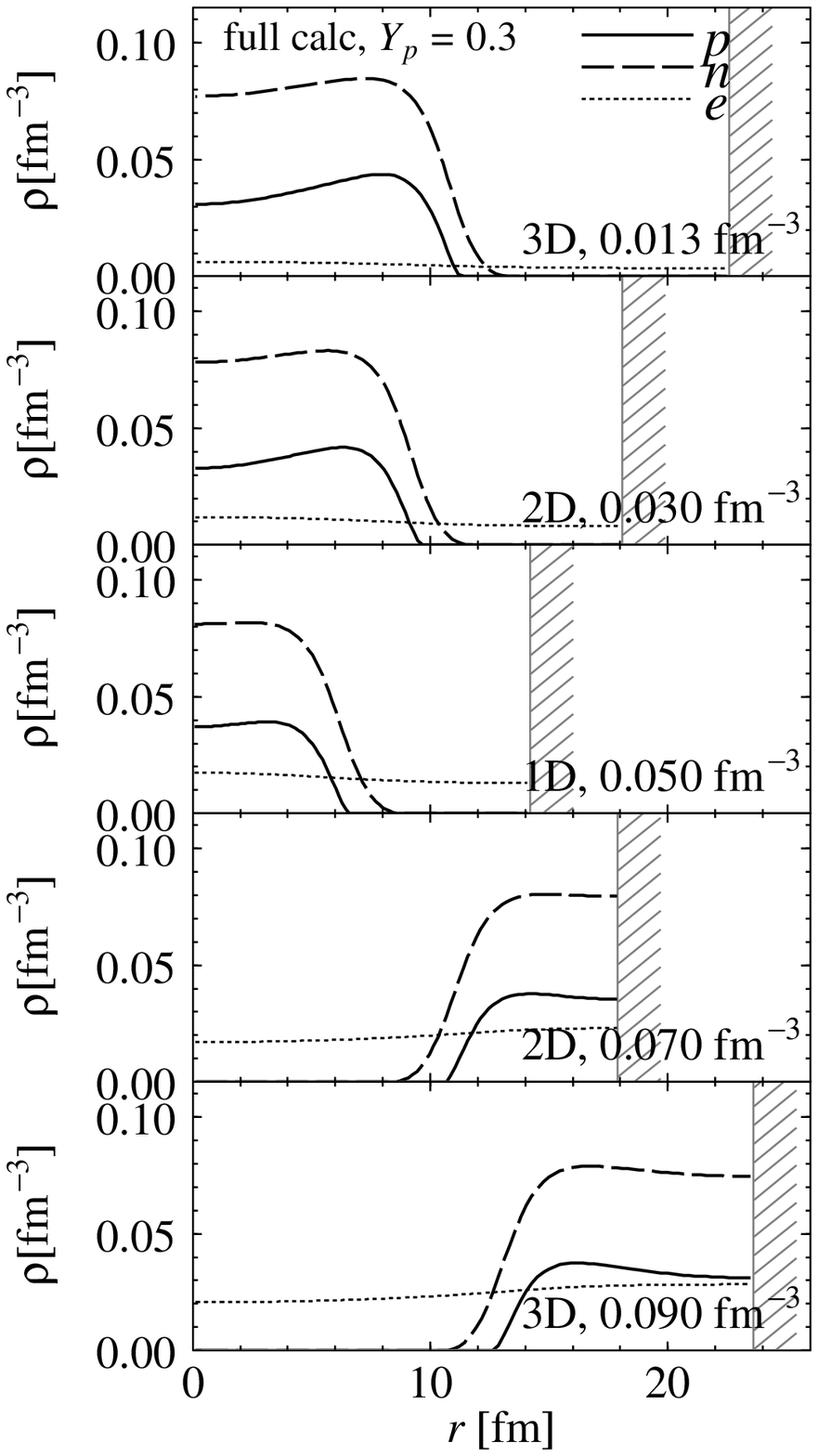}
\includegraphics[width=.31\textwidth]{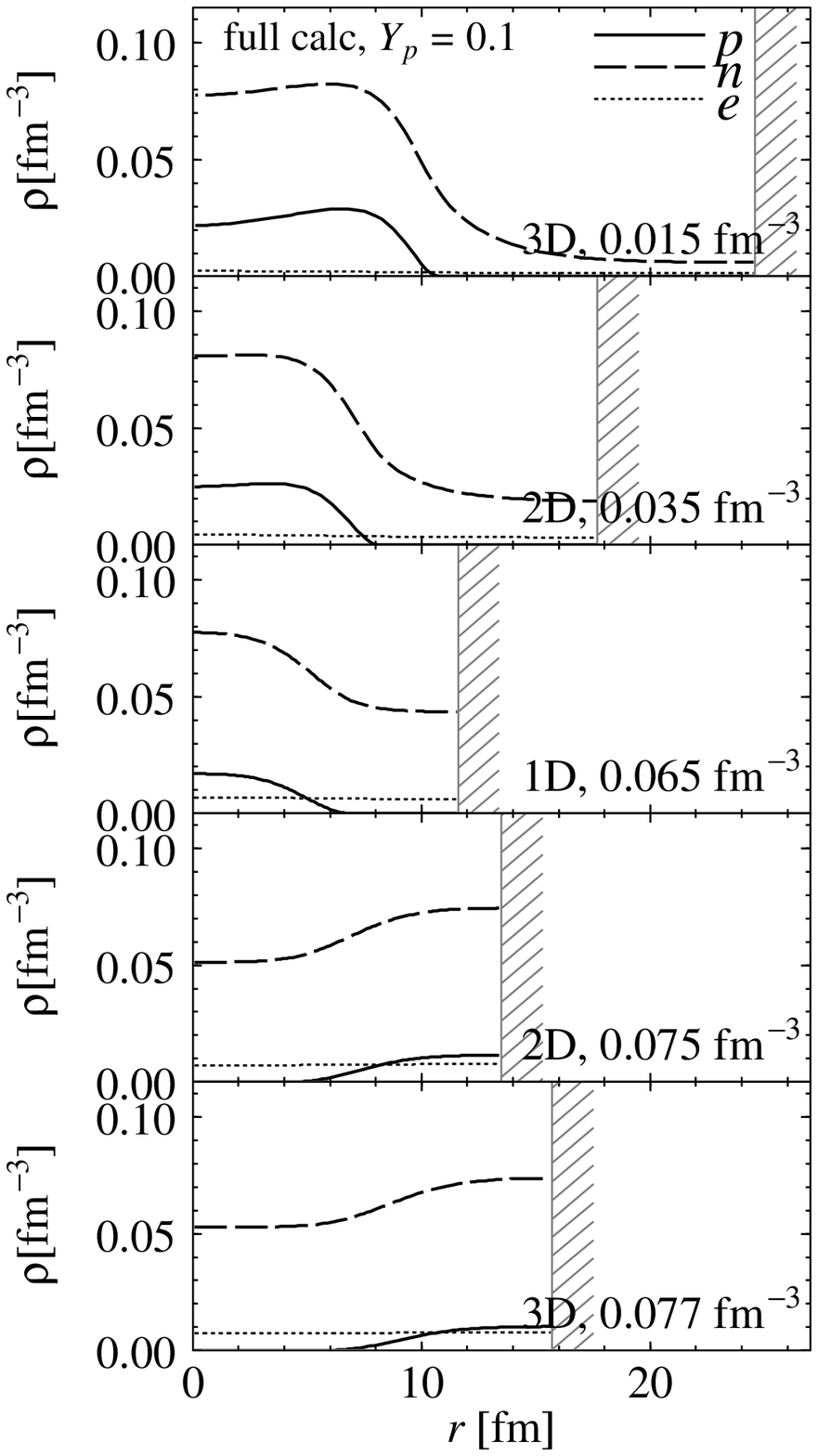}
\caption{
Examples of the density profiles in the cell for symmetric nuclear
 matter with $Y_p$=0.5 (left panel) and for asymmetric matter
 with $Y_p=0.3$ (center panel) and 0.1 (right panel).
}
\label{proffixfull}
\end{figure*}
\begin{figure*}[t]
\includegraphics[width=.32\textwidth]{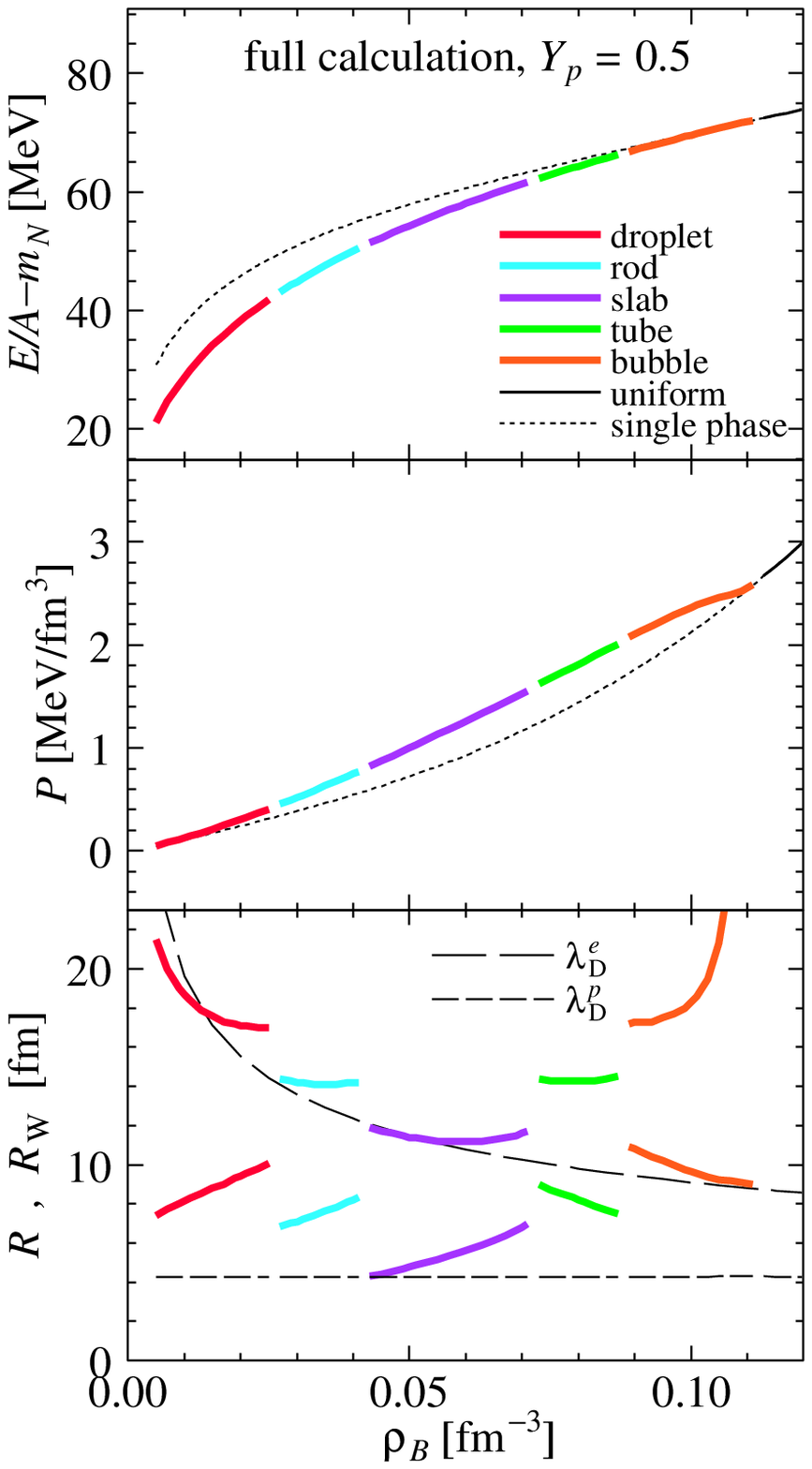}
\includegraphics[width=.32\textwidth]{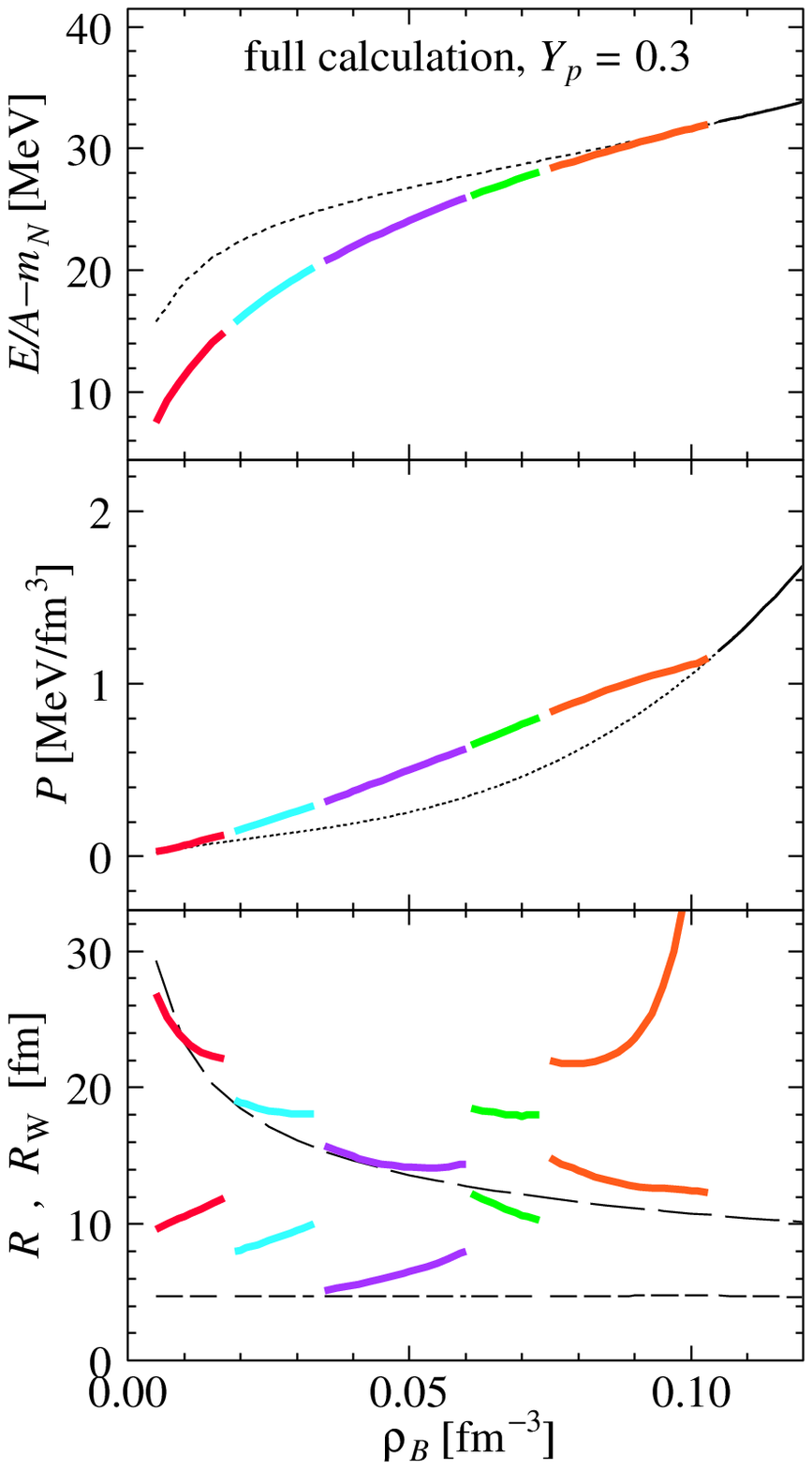}
\includegraphics[width=.32\textwidth]{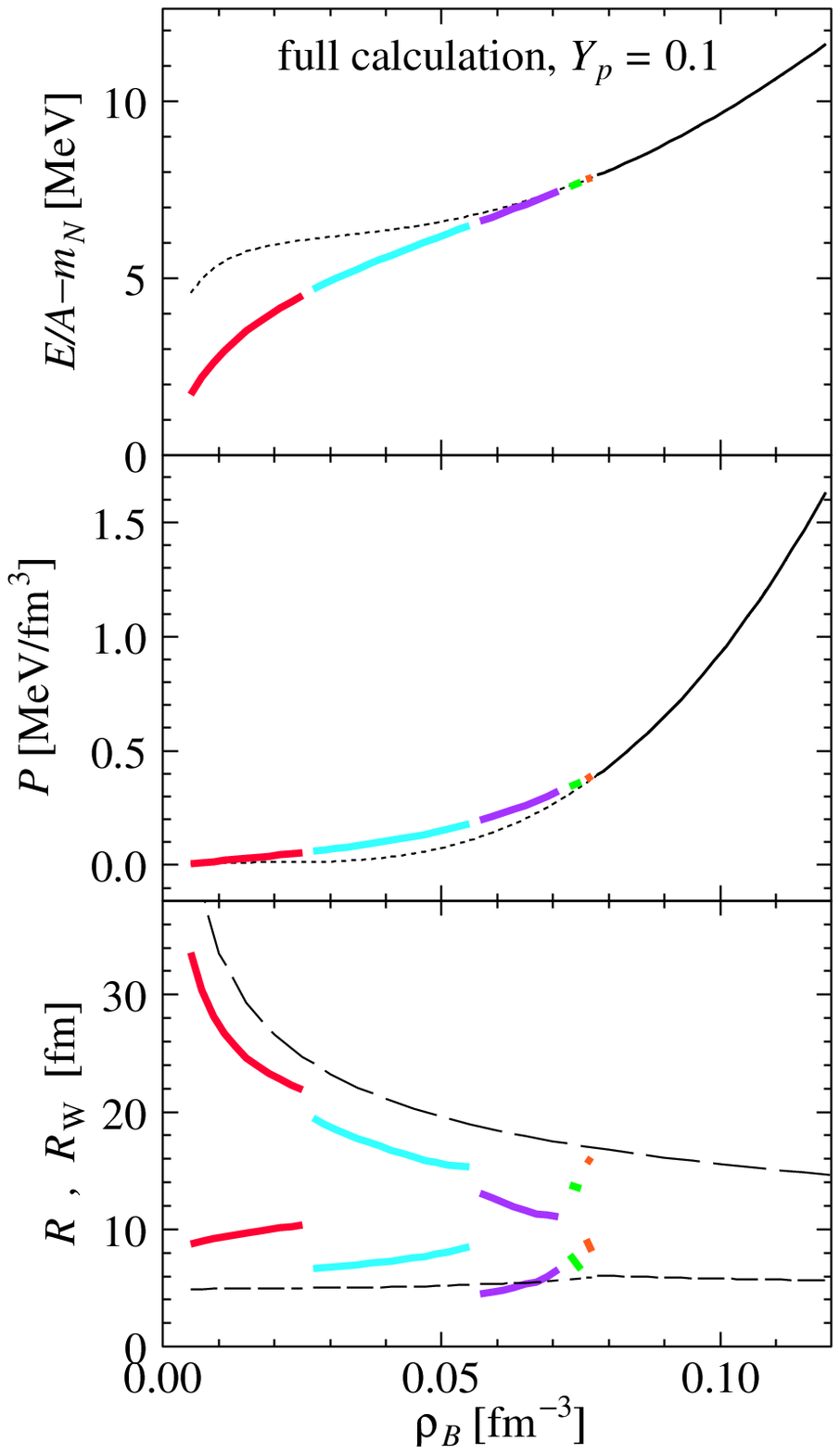}
\caption{
Binding energy per nucleon (top), pressure (middle), and the cell and lump sizes (bottom) 
for symmetric nuclear matter with $Y_p$=0.5 (left panel), and for asymmetric matter with
$Y_p=0.3$ (center panel) and 0.1 (right panel).
}
\label{eosfixfull}
\end{figure*}

First, we concentrate on the discussion of the behavior of 
nucleon matter at fixed values of the proton mixing ratio $Y_p$.
Particularly, we explore the proton mixing ratios $Y_p=0.1$, 0.3, and 0.5.
The cases $Y_p=0.3$ -- 0.5 should be relevant for supernova matter
and for newly born neutron stars. 
Figure \ref{proffixfull} shows
some typical density profiles inside the Wigner-Seitz cells. 
The geometrical dimension of the cell  is denoted as ``3D''
(three-dimensional sphere), etc. 
The horizontal axis in each panel
denotes the radial distance from the center of the cell. 
The cell boundary is indicated by the hatch.
 From the top to the bottom the configuration corresponds to
droplet (3D), rod (2D), slab (1D), tube (2D), and bubble (3D).
The nuclear ``pasta'' structures are clearly manifested.
For the lowest $Y_p$ case ($Y_p =0.1$), the neutron density is finite at any point:
the space is filled by dripped neutrons.
The value of $Y_p$ above which neutrons drip
is around 0.26 in our 3D calculation, for example.
For a higher  $Y_p$, the neutron density
drops to zero outside the nucleus. 
The proton number density always drops to zero outside the nucleus.
We can see that the charge screening effects are pronounced. 
Due to the spatial rearrangement of electrons the electron
density profile becomes no more uniform. 
This non-uniformity of the electron distribution is
more pronounced for a higher $Y_p$ and a higher density. 
Protons repel each other.
Thereby the proton density profile substantially deviates from the step-function. 
The proton number is enhanced 
near the surface of the lump.

The equation of state (EOS) for the sequence of geometric structures is
shown in Fig.~\ref{eosfixfull} (top panels)
as a function of the averaged baryon-number density.
Note that the energy $E/A-m_N$ also includes the
kinetic energy of electrons, which makes the total pressure
positive (middle panels). 
The lowest-energy configurations are selected among
various geometrical structures. 
The most favorable configuration
changes from the droplet to rod, slab, tube, bubble, and to the
uniform one (the dotted thin curve) with an increase of density.
The appearance of non-uniform structures in matter results in a
softening of EOS: the energy per baryon
gets lower up to about 15 MeV$/A$ compared to uniform matter.

The bottom panels in Fig.~\ref{eosfixfull} are the cell radii
$R_{\rm W}$ and the lump radii $R$ versus averaged baryon number density. 
The radius $R$ is defined by way of a density fluctuation as
\begin{eqnarray}
R&=&
\cases{
\displaystyle
R_{\rm W} { \langle\rho_p\rangle^2 \over \langle\rho_p^2\rangle }, 
&\hbox{(for droplet, rod, and slab)}\cr
\displaystyle
R_{\rm W} \left(1-{\langle\rho_p\rangle^2 \over \langle\rho_p^2\rangle}\right), 
&\hbox{(for tube and bubble)} 
}
\end{eqnarray}
where the bracket ``$\langle \rangle$'' indicates the average along
the radial (for 3D and 2D cases) or perpendicular (1D) direction in the cell.
Dashed curves show the Debye screening lengths of
electrons and protons calculated as
\begin{eqnarray}\label{Deb}
\lambda^{a}_{\rm D}&=&\left(4\pi e^2{d\rho_a^{\rm av}\over d\mu_a}\right)^{-1/2}\ \ \ (a=p,e),
\end{eqnarray}
where $\rho_p^{\rm av}$ is the proton number density
averaged inside the nucleus (the region with finite $\rho_p$)
and $\rho_e^{\rm av}$ is the electron density averaged inside the cell.
Actually doing more carefully we should introduce four Debye screening lengths
$\lambda^{a,<}_{\rm D}$ and $\lambda^{a,>}_{\rm D}$
with a separate averaging for the interior and the exterior of the nuclei.
However we observe that the proton number density is always zero
in the exterior region and $\lambda^{p,>}_{\rm D} =\infty$ thereby.
For electrons $\lambda^{e,<}_{\rm D}$
and $\lambda^{e,>}_{\rm D}$ are in general different
but both being large and of the same order of magnitude in
the pasta case under consideration. 
Therefore we actually do not need a more detailed analysis of these quantities.
Note that these values are obviously gauge invariant.
Numerically, the cell radii $R_{\rm W}$ for droplet, rod, and slab
configurations at $Y_p=0.5$ and 0.3 were proven to be close to
the electron screening length. 
For the tube, $R_{\rm W}$ is larger than ${\lambda}^{e}_{\rm D}$. 
For $Y_p=0.1$, in all cases  $R_{\rm W}$ is substantially
smaller than ${\lambda}^{e}_{\rm D}$ and  the electron screening should be much weaker, thereby.
In all cases, except for bubbles (at $Y_p=0.5$ and 0.3), the structure radii $R$ are
smaller than ${\lambda}_{\rm D}^{e}$.
This means that the Debye screening effect of electrons inside 
these structures should not be pronounced. 
For bubbles at $Y_p=0.5$ and 0.3, ${\lambda}_{\rm D}^{e}$
is substantially smaller than the cell size and the electron screening 
should be significant.
For $Y_p=0.5$, 0.3, 0.1 in all  cases (with the only exception $Y_p=0.1$ for slabs), 
the value ${\lambda}_{\rm D}^{p,<}$ is shorter than $R$.
Hence the density rearrangement of protons is essential
for the pasta structures, as it is indeed seen from the Fig.\ \ref{proffixfull}.

As we have mentioned one of the characteristics of our calculation
is that the obtained density profiles of particles in the cell is 
free from assumption and is consistent with the potential.
However, the usage of Wigner-Seitz cell is a significant limitation.
It is because we have to assume the geometrical symmetry.
At zero temperature, a specific pasta structure should be 
the distinct ground state for a given baryon density.
However, at a certain density near the boundary of two pasta structures,
some intermediate structures could be favorable, or at least
can be local-minimum state.
Furthermore, intermediate structure should not be negligible at finite temperatures.
Such intermediate states and incomplete pasta structures are reported 
in the studies using molecular dynamics \cite{maru98,Gen02,Gen03,Gen04,Gen05}.
In fact, molecular dynamics simulation is completely free from
assumption of nucleon distribution.
However, uniform electron background is assumed and beta-equilibrium condition
is not imposed in the molecular dynamics yet.
In this sense, the present mean-field calculation and the molecular dynamics 
for nuclear matter structure are complementary models to each other.

\subsection{Nucleon matter in beta equilibrium}

\begin{figure*}
\begin{minipage}{0.48\textwidth}
\includegraphics[width=.97\textwidth]{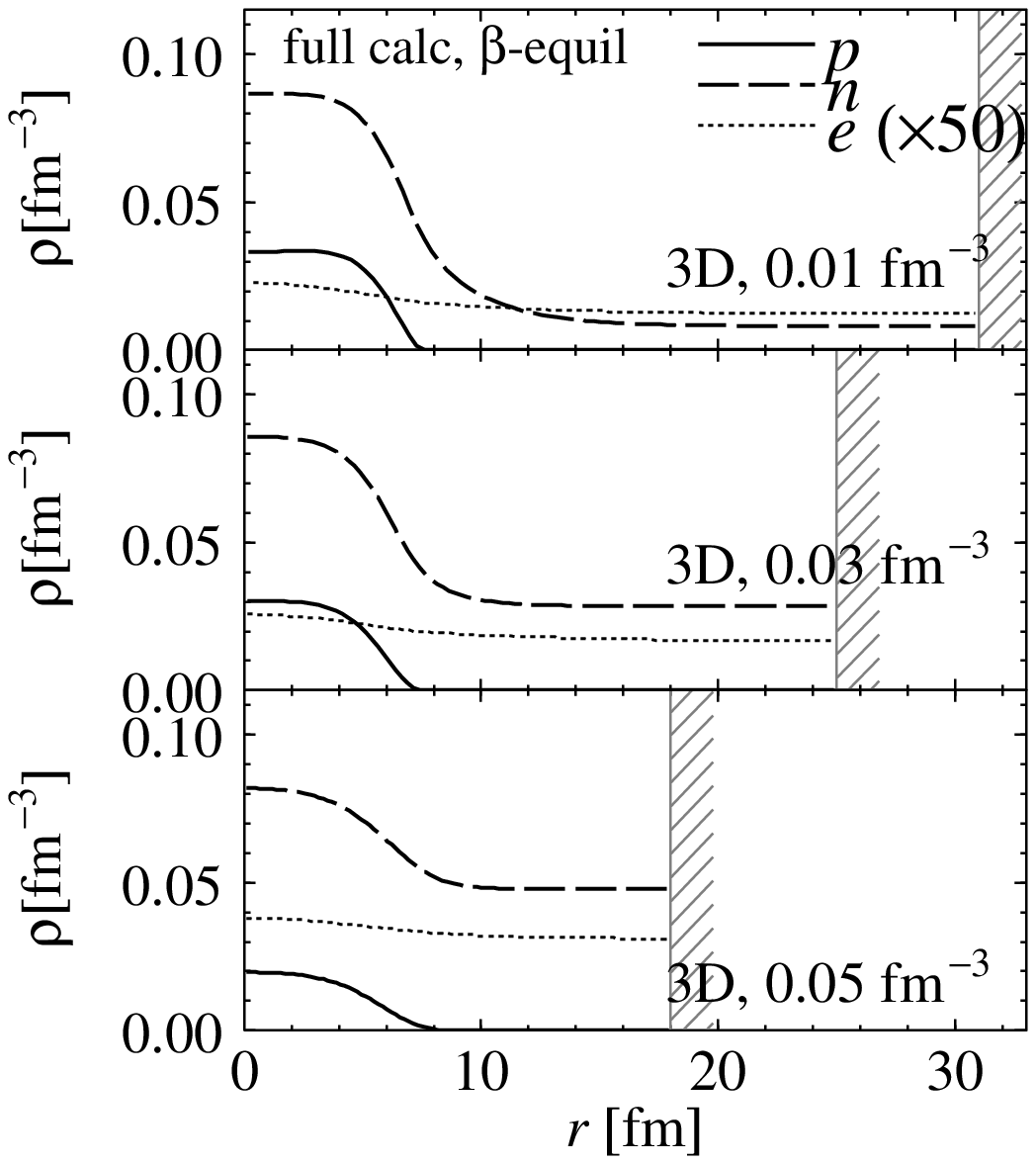}
\caption{
Density profiles in the cell for nuclear matter in beta equilibrium with
baryon-number densities, 0.01, 0.03 and 0.05 fm$^{-3}$ from the top to the bottom.
}
\label{profbeta}
\end{minipage}
\hspace{\fill}
\begin{minipage}{0.48\textwidth}
\includegraphics[width=.88\textwidth]{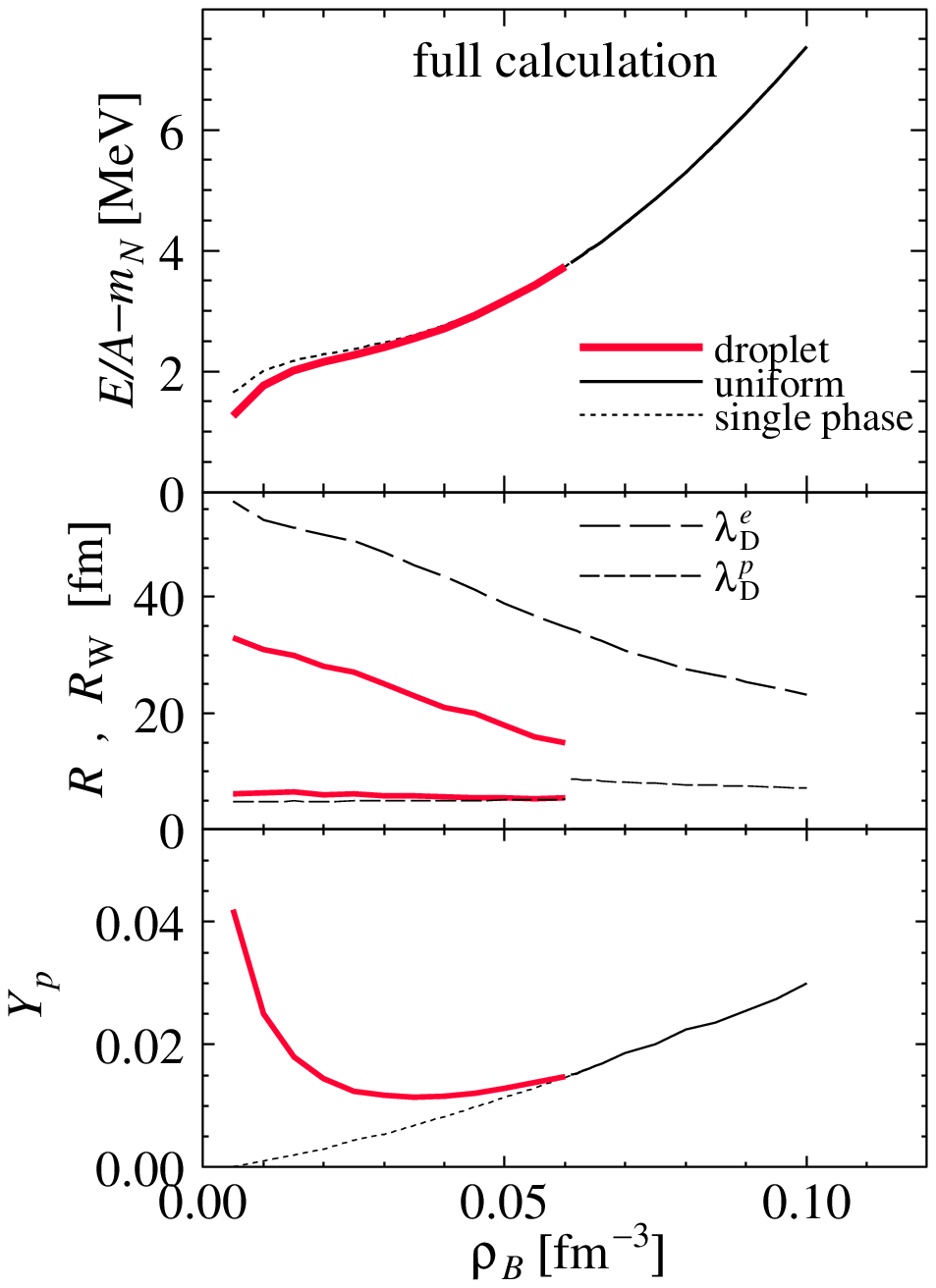}
\caption{
Binding energy (top), cell and lump sizes (middle), and proton mixing ratio (bottom)
in the cell for nuclear matter in beta equilibrium.
}
\label{eosbeta}
\end{minipage}
\end{figure*}

Next, we  consider neutron star matter at zero temperature, and
explore the non-uniform structures for nucleon matter in beta equilibrium. 
Figure \ref{profbeta} shows the density profiles for
different baryon number densities. 
The droplet structure itself is
quite similar to the case of the fixed proton mixing 
ratio $Y_p=0.1$ considered above. 
The apparently different feature in this case is
that only the droplet configuration appears as a non-uniform structure. 
It should be noticed, however, that the presence or
absence of the concrete pasta structure sensitively depends on the
choice of the effective interaction.

In Fig.~\ref{eosbeta} we plot
the energy per baryon (top), the cell and lump sizes (middle), and the proton
number ratio (bottom).
The effect of the non-uniform structure on EOS (the difference between
the energy of uniform matter and that of non-uniform one) is small.
However, the proton mixing ratio is significantly affected by the presence of the pasta 
at lower densities.
In the zero-density limit, the proton mixing ratio should converge to that of
the normal nuclei.
The droplet radius and the cell radius in the middle panel of Fig.~\ref{eosbeta}
are always smaller than the electron Debye screening length $\lambda_{\rm D}^{e}$.
Thereby the effect of the electron charge screening is small.
Unlike the fixed $Y_p$ cases,
the droplet radius is comparable to the proton Debye screening length, which means that
the effect of the proton rearrangement is not pronounced in this case.
In fact, there is no enhancement of the proton number density near the surface 
in Fig.\ \ref{profbeta}, in contrast to Fig.\ \ref{proffixfull}.

\section{Kaonic pasta structures at high densities}

Next let us explore high-density nuclear matter in beta equilibrium, 
which is expected in the inner core of neutron stars.

\begin{figure*}
\begin{minipage}{0.48\textwidth}
\includegraphics[height=.25\textheight]{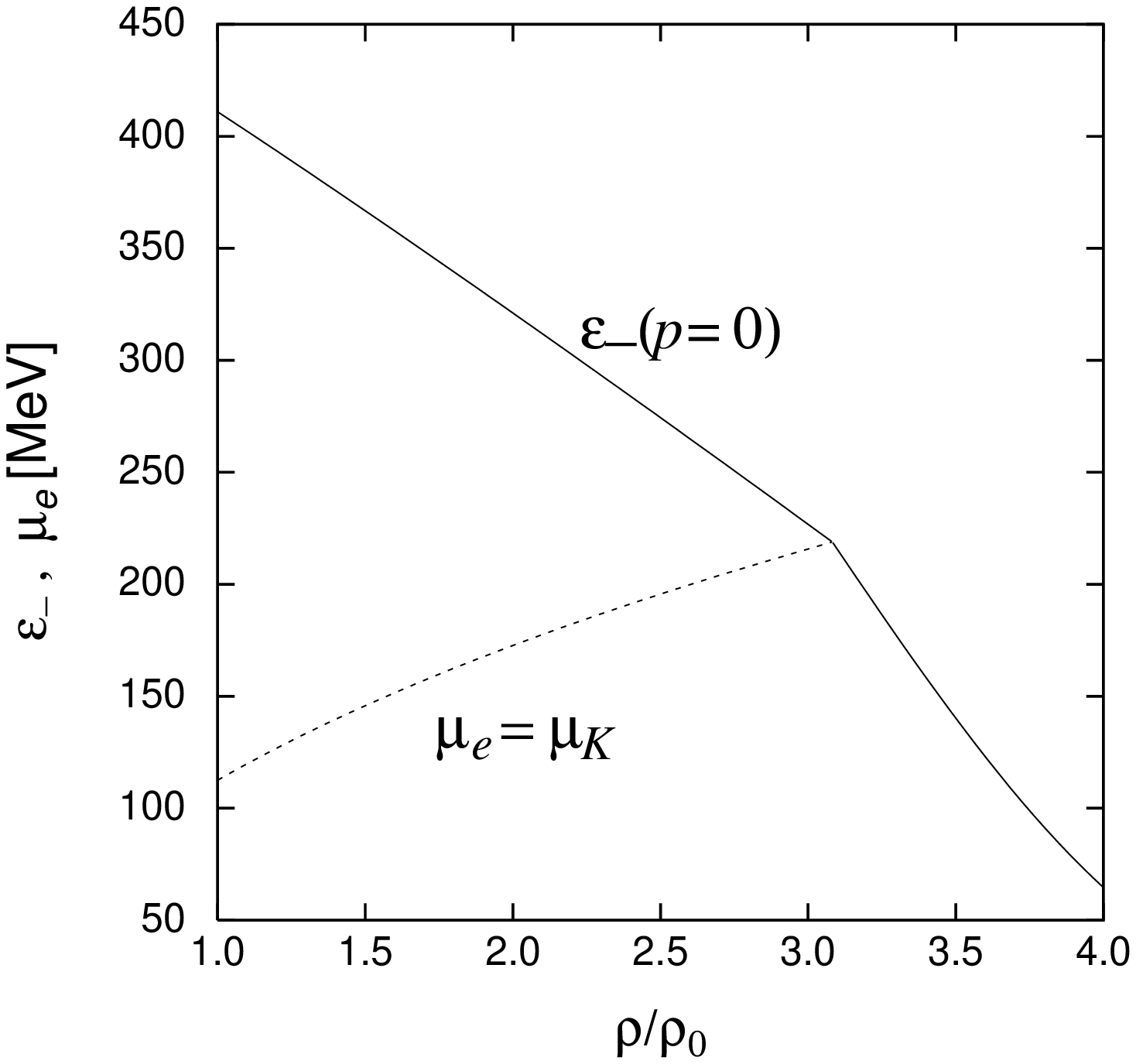}
\caption{Onset mechanism of kaon condensation within the chiral model. 
}
\label{dispr}
\end{minipage}
\hspace{\fill}
\begin{minipage}{0.48\textwidth}
\includegraphics[height=.25\textheight]{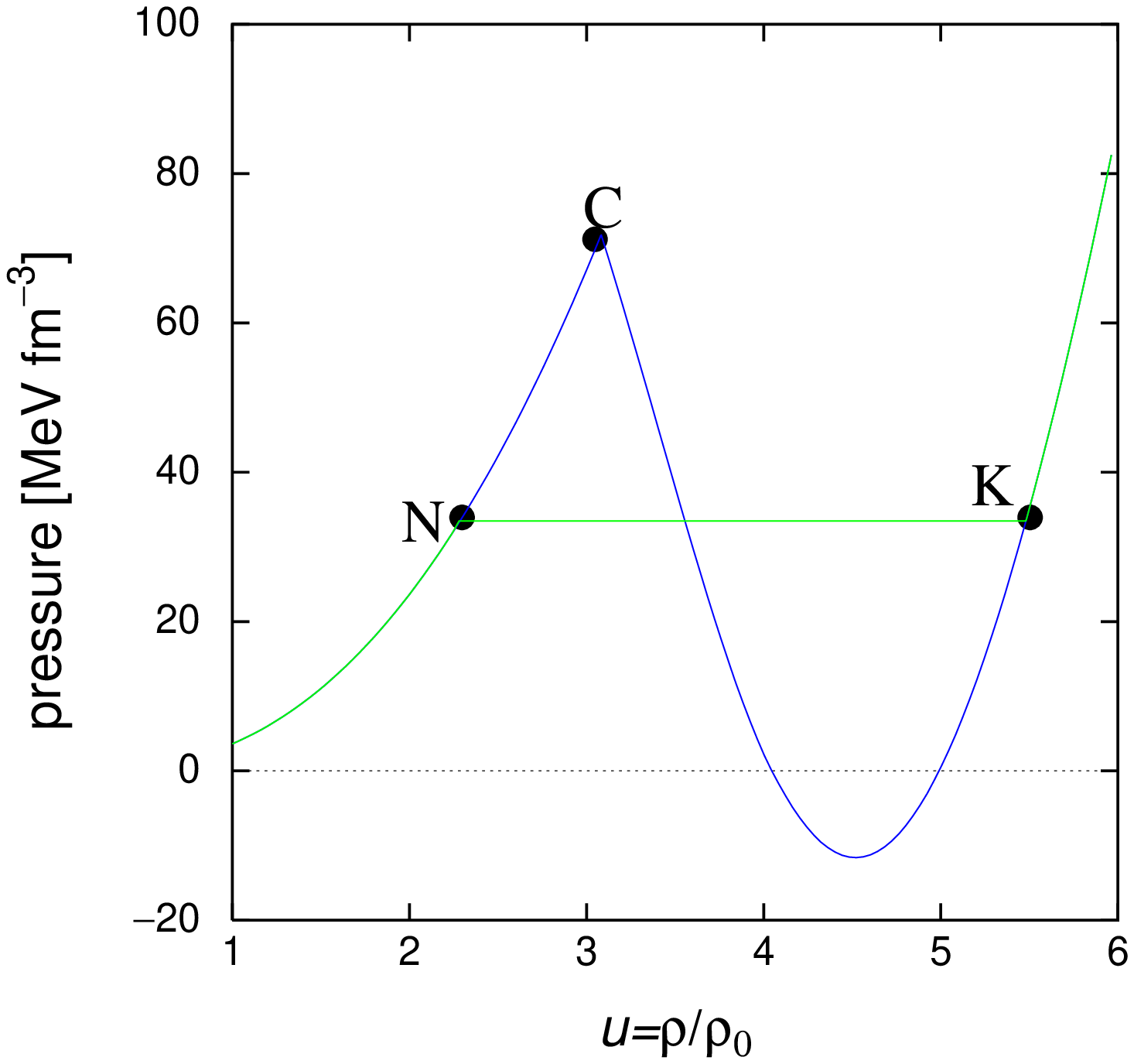}
\caption{Original EOS and the Maxwell construction given by the chiral model.
``C'' indicates the critical point between
nucleon and kaonic phases in the case of uniform matter (without mixed phase).
The mixed phase is given by the Maxwell construction between ``N'' and ``K''.
}
\label{MWCrr}
\end{minipage}
\end{figure*}

Kaons are Nambu-Goldstone bosons accompanying the spontaneous
breaking of chiral $SU(3)\times SU(3)$ symmetry 
and the lightest mesons with strangeness. 
Their effective energy is much reduced by the kaon-nucleon interaction 
in nuclear medium, which is dictated by chiral symmetry.
For low-energy kaons the $s$-wave interaction is dominant and attractive in
the $I=1$ channel, so that negatively charged kaons appear in the neutron-rich
matter once the process $n\rightarrow p+K^-$ becomes energetically allowed.
Since kaons are bosons, they cause the Bose-Einstein condensation 
at zero momentum \cite{kn86}.
The single-particle energy of kaons is given in a model-independent way (see Appendix): 
\begin{equation}
\epsilon_\pm({\bf p})=\sqrt{|{\bf p}|^2+m_K^{*2}+((\rho_n+2\rho_p)/4f^2)^2}\pm (\rho_n+2\rho_p)/4f^2,
\label{disp}
\end{equation}
where $m_K^*$ is the effective mass of kaons,
$m_K^{*2}=m_K^{2}-\Sigma_{KN}(\rho_n^{s}+\rho_p^{s})/f^2$, with the $KN$ sigma term,
$\Sigma_{KN}$, and the meson decay constant, $f\equiv f_K\sim f_\pi$.
The threshold condition then reads \cite{mut}
\begin{equation}
\mu_K=\epsilon_-({\bf p}=0)=\mu_n-\mu_p=\mu_e,
\end{equation}
which means the kaon distribution function diverges at ${\bf p}=0$ (Fig.~\ref{dispr}).

If kaon condensation occurs in nuclear matter, it has many implications on 
compact stars; one of the interesting possibilities may be the delayed
collapse of a protoneutron star to a low-mass black hole due to the large
softening of EOS \cite{YT00,bb94}, and  another one is a fast cooling mechanism
of young neutron 
stars due to the nucleon Urca process under background kaons \cite{bkpp88,t88}. 

Since many studies have shown that kaon condensation is of first order, 
we must carefully treat the phase change (Fig.~\ref{MWCrr}). 
In the following we discuss the mixed phase in the phase transition 
in a similar way to nuclear pasta.

\subsection{RMF treatment of nuclear matter with kaon condensation}

We explore high-density nuclear matter with kaon condensation by means of
RMF model as in low-density matter.
Using the same model we can discuss the non-uniform structure of nuclear matter
both at low- and high-density regime in an unified way.
To incorporate kaons into our RMF calculation,
the thermodynamic potential of Eq.~(\ref{Omega-tot}) is modified as
\begin{eqnarray}
\Omega &=& \Omega_N+\Omega_M +\Omega_e +\Omega_K,\\
\Omega_K&=&\!\!\!\int d^3r\left\{
  -{f_K^2\theta^2\over2}\left[-{m_K^*}^2+(\mu_K-V+g_{\omega K}\omega_0
  +g_{\rho K}R_0)^2 \right]+{f_K^2(\nabla\theta)^2\over2} \right\},
  \nonumber\\
\end{eqnarray}
where $m_K^*=m_K-g_{\sigma K}\sigma$, $\mu_K=\mu_e$,  
and the kaon field $K=f_K\theta/\sqrt{2}$ ($f_K$ is the kaon decay constant). 
We, hereafter, neglect a rather unimportant term $\propto \sigma^2\theta^2$.%
\footnote{We use the meson exchange model (MEM) for kaon
Lagrangian to compare our results with the earlier ones \cite{GS99,CGS00,NR00},
which is not chiral-symmetric (see Appendix). 
Note however that both can well describe the low-energy kaon-nucleon dynamics.}
The equations of motion are similar to Eqs.~(\ref{EOMsigma}) - (\ref{EOMmuNp}) 
given for the low-density case (Sec.\ 3).
We list here the modified and added ones as 
\begin{eqnarray}
\nabla^2\sigma({\bf r}) &=& m_\sigma^2\sigma({\bf r}) +{dU\over d\sigma}
-g_{\sigma N}(\rho_n^{s}({\bf r}) +\rho_p^{s}({\bf r}))
\nonumber\\&&\quad  {}+2g_{\sigma K}m_Kf_K^2 \theta^2({\bf r})
,\label{EOMsigmaK}\\
\nabla^2\omega_0({\bf r}) &=& m_\omega^2\omega_0({\bf r}) 
-g_{\omega N} (\rho_p({\bf r})+\rho_n({\bf r}))
\nonumber\\&&\quad {}-f_K^2g_{\omega K}\theta^2({\bf r})[\mu_K-V({\bf r})
+g_{\omega K}\omega_0({\bf r})+g_{\rho K}R_0({\bf r})]
,\label{EOMomegaK}\\
\nabla^2R_0({\bf r}) &=& m_\rho^2R_0({\bf r}) -g_{\rho N}
(\rho_p({\bf r})-\rho_n({\bf r}))
\nonumber\\&&\quad {}-f_K^2g_{\rho K}\theta^2({\bf r})[\mu_K-V({\bf r})
 +g_{\omega K} \omega_0({\bf r})+g_{\rho K}R_0({\bf r})]
,\label{EOMrhoK}\\
\nabla^2\theta({\bf r}) &=& \Bigl[{m_K^*({\bf r})}^2-(\mu_K-V({\bf r})
\nonumber\\&&\quad
{}+g_{\omega K}\omega_0({\bf r})+ g_{\rho K}R_0({\bf r}))^2\Bigr]
\theta({\bf r}).\label{EOMtheta}
\end{eqnarray}
The charge density now reads 
\begin{equation}
\rho_{\rm ch}({\bf r})\;=\;\rho_p({\bf r})
-\rho_e({\bf r})-\rho_K({\bf r}),
\label{eq:poissonK}
\end{equation}
with the kaon contribution,
\begin{equation}
\rho_K = f_K^2\theta^2\left[\mu_K-V+g_{\omega K}\omega_0({\bf r})
  +g_{\rho K}R_0({\bf r}) \right].
\label{rhoK}
\end{equation}
Additional parameters concerning kaons are presented in Table 2.
By a lattice QCD calculation, the kaon-nucleon sigma term
$\Sigma_{KN}$ is estimated as 290 -- 450 MeV \cite{fuk}.
Using the above values and the relation (see Appendix) 
$$
{g_{\sigma K}g_{\sigma N}\over {m_{\sigma}}^2}={\Sigma_{KN}\over 2m_K{f_K}^2},
$$
$g_{\sigma K}$ can be 0.849 -- 1.318.
On the other hand, the parameter $g_{\sigma K}$ enters the value of 
the $K^-$  optical potential $U_K$
defined by $U_K = g_{\sigma K}\sigma+g_{\omega K}\omega_0$.
There have been many works  trying to extract $U_K$ at the saturation density
from the data on the kaonic atoms \cite{fgb94,fgm99} and from
calculations \cite{wkw96,ro00,ske00,trp02,oor00},
but there is still a controversy about its depth.
We take here a somewhat deep potential, as shown in Table 2,
to compare our results with the earlier ones \cite{GS99,CG00,CGS00,NR00}.
The corresponding $g_{\sigma K}$ can be 2.209 -- 2.519, which may be rather large
compared to the lattice QCD estimation.
To understand a dependence of the results on the value of $U_K$
we further allow for its variation.

\begin{table}
\caption{
Additional parameters used in our RMF model with kaon terms. 
The kaon optical potential $U_K$ is
defined by $U_K = g_{\sigma K}\sigma+g_{\omega K}\omega_0$.
}
\begin{center}
\begin{tabular}{ccccc}
\hline
 $f_K (\approx f_\pi)$ [MeV] &
 $m_K$ [MeV]&
 $g_{\omega K}$ &
 $g_{\rho K}$ &
 $ U_K(\rho_0)$ [MeV]
\\ [1mm]
\hline\\
 93  &
 494  &
$g_{\omega N}/3$ &
$g_{\rho N}$ &
$-120$ -- $-130$ \\
\hline
\end{tabular}
\end{center}
\end{table}

\subsection{Properties of kaonic pasta structures}

\begin{figure*}
\begin{minipage}{0.48\textwidth}
  \includegraphics[width=.99\textwidth]{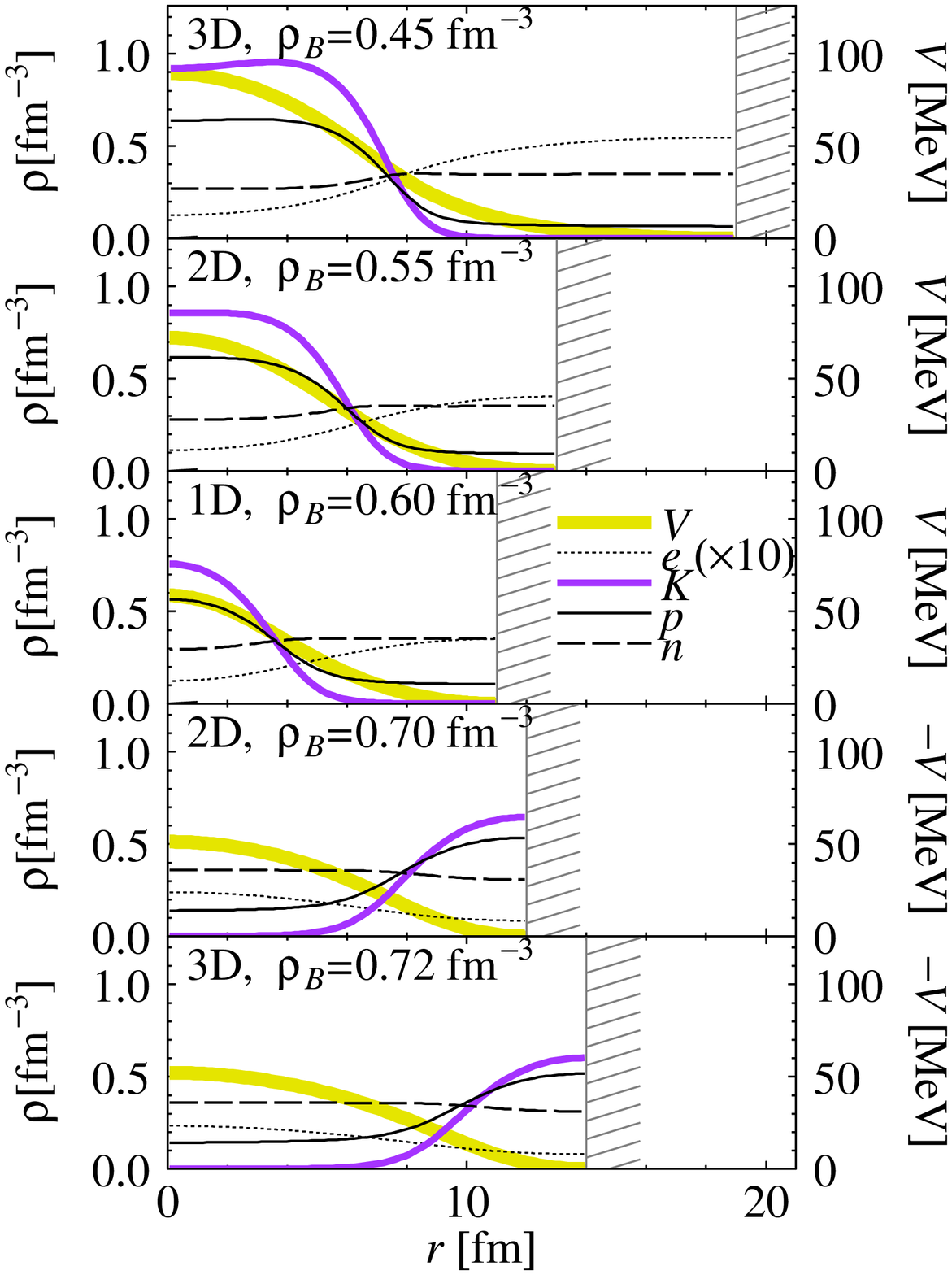}
  \caption{
   Density profiles of kaonic pasta structures.
  Here the density does not mean charge-density but number-density
  of particles.
  $U_K$, the kaon optical potential at the nuclear saturation
  density, is set to be $-130$ MeV.
  Displayed using the right axis is
  the Coulomb potential $V$.
}\label{figProfK}
\end{minipage}
\hspace{\fill}
\begin{minipage}{0.48\textwidth}
  \includegraphics[width=.96\textwidth]{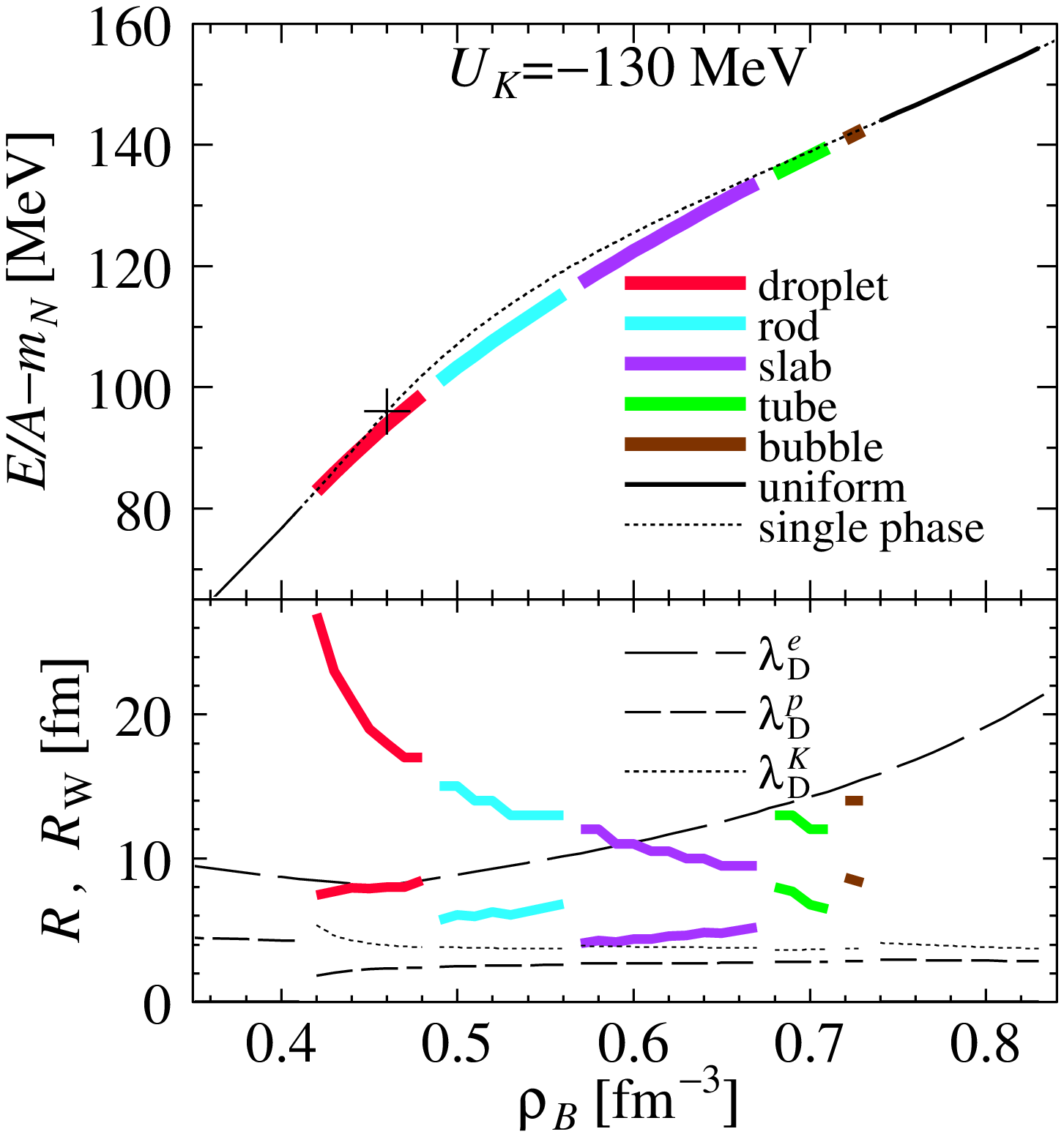}
  \caption{
  Top: binding energy per nucleon of the
  nuclear matter in beta equilibrium.
  The dotted line below the cross shows uniform normal nuclear matter
  and above the cross, uniform kaonic matter.
  Bottom: lump size $R$ (thick curves below)
  and cell size $R_{\rm W}$ (thick curves above).
  Compared are the Debye screening lengths of electron, proton and kaon.
  The Debye screening lengths are calculated using the
  explicit dependence of $\mu_a$ on $\rho_a$ (see Eq.\ (\ref{Deb})).
}\label{figKEOS}
\end{minipage}
\end{figure*}

The kaon condensation is considered to be of first order.
Therefore it may give rise to non-uniform structures with different geometries,
the structured mixed phase.
In fact, the system  exhibits a series of structure change
similar to the nuclear ``pasta'':
the kaonic droplet, rod, slab, tube, bubble (which we call ``kaonic
pasta'' structures).
Figure \ref{figProfK} displays typical density profiles and the Coulomb potential.
The neutron distribution proves to be rather flat.
The proton distribution on the other hand is strongly
correlated with the kaon distribution,
which means that the Coulomb interaction is crucial.

In the upper panel of Fig.\ \ref{figKEOS} we depict the energy per
nucleon as a function of baryon number density.
The dotted line indicates the case of single phase
(if one assumes the absence of the mixed phase).
In this case uniform matter consists of
normal nuclear matter below the critical density
and kaonic matter above the critical density.
The cross  on the dotted line ($\rho_B \simeq 0.46$~fm$^{-3}$)
shows the critical density, i.e.\ the point where kaons begin
to condensate in the case of single phase.
Pieces of solid curves, on the other hand,
indicate  energetically favored structures.
Droplets begin to appear for $\rho_B >0.41$~fm$^{-3}$
and smoothly decrease the energy of the system.
The mixed phase disappears for $\rho_B >0.74$~fm$^{-3}$.

The lower panel of Fig.\ \ref{figKEOS} shows the sizes of the lump $R$ and 
the cell $R_{\rm W}$.
We find that at the onset density the size of the cell is infinitely large
in case of the full calculation.
The corresponding steep increase of $R_{\rm W}$
with decreasing density is clearly seen in the figure.
The dashed lines and the dotted line in the lower panel of Fig.~\ref{figKEOS}
show partial contributions to the Debye screening lengths of
the electron, proton and kaon,
$\lambda_{\rm D}^{e}$, $\lambda_{\rm D}^{p}$,
and $\lambda_{\rm D}^{K}$, respectively.
We see that in most cases $\lambda_{\rm D}^{e}$ is less than the cell size
$R_{\rm W}$ but is larger than the lump size $R$.
The proton Debye length $\lambda_{\rm D}^{p}$
and the kaon Debye length $\lambda_{\rm D}^{K}$, on the other hand,
are always shorter than $R_{\rm W}$ and $R$.

\section{Charge screening and surface effects on the pasta structures}

\subsection{Charge screening effect}

\begin{figure*}
\includegraphics[width=.32\textwidth]{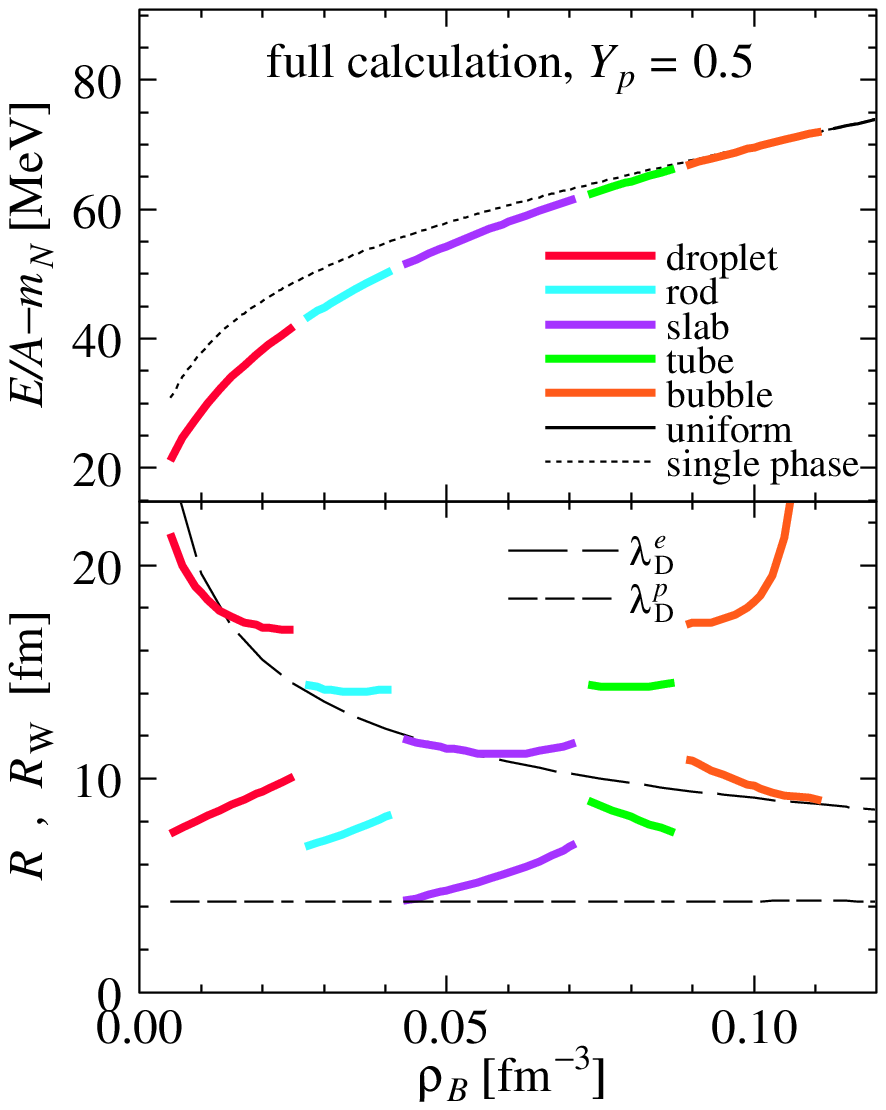}
\includegraphics[width=.32\textwidth]{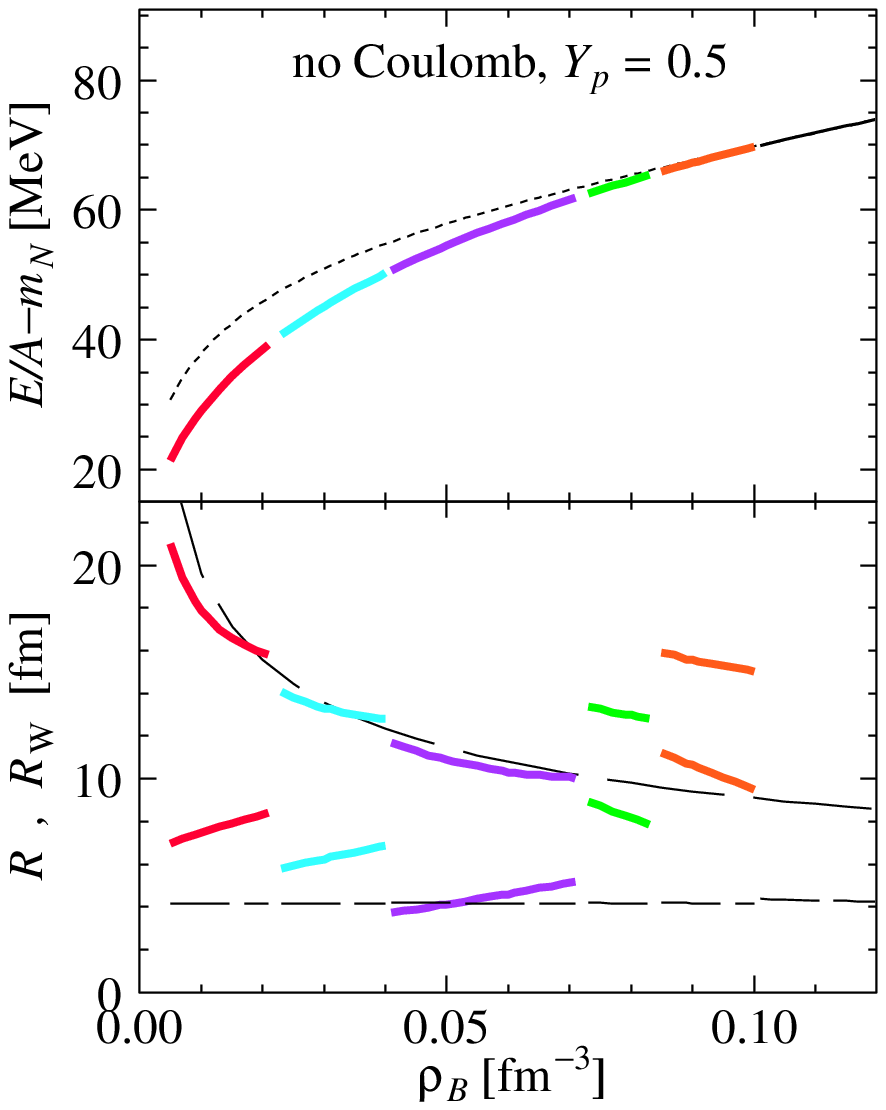}
\includegraphics[width=.32\textwidth]{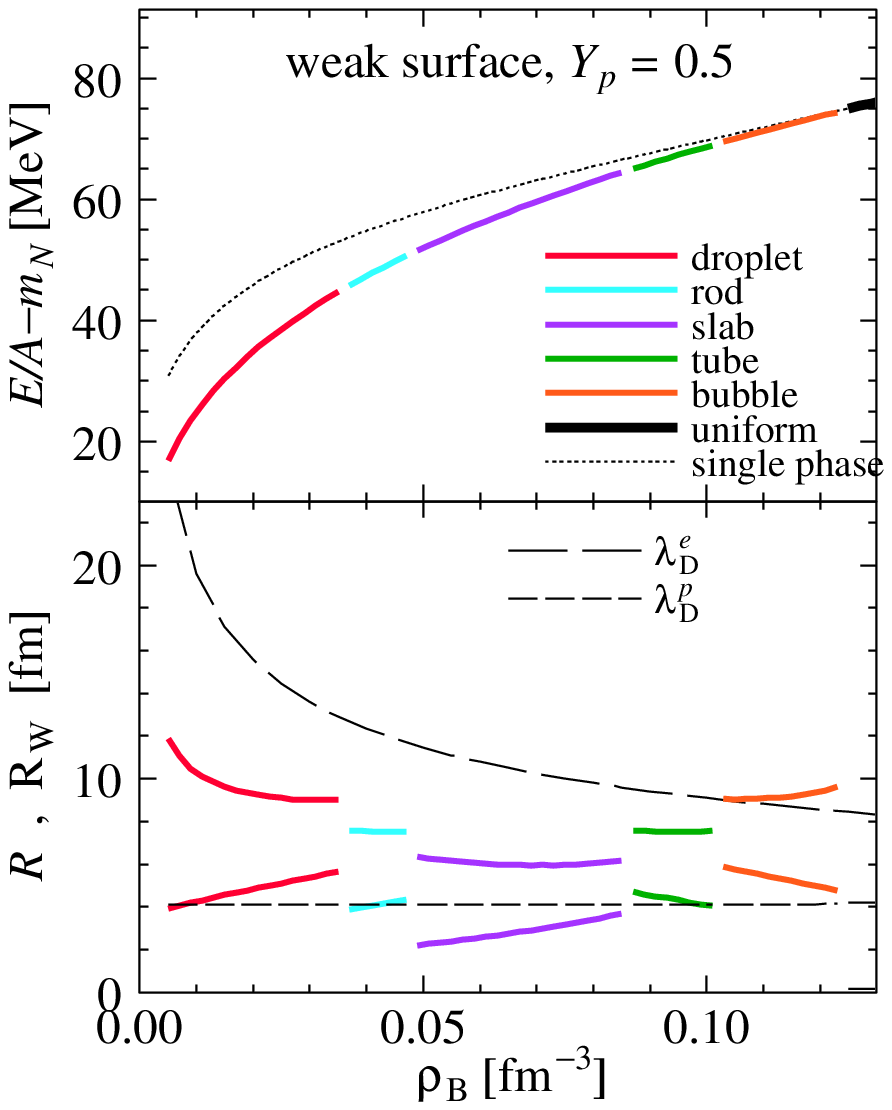}
\caption{
Top: binding energy per nucleon of nuclear matter.
Bottom: lump size $R$ and cell size $R_{\rm W}$.
The proton mixing ratio is $Y_p$=0.5 for all cases.
Different calculations are compared.
 From the left: ``full'', ``no Coulomb'' and ``weak surface'' calculations.
 In ``no Coulomb'' calculation, the electric potential is discarded 
 when determining the density profile
 and then added to evaluate the energy.
}
\label{eoscompare}
\end{figure*}

\begin{figure}
  \includegraphics[width=.32\textwidth]{Eos2WC130LK-.ps}
  \includegraphics[width=.32\textwidth]{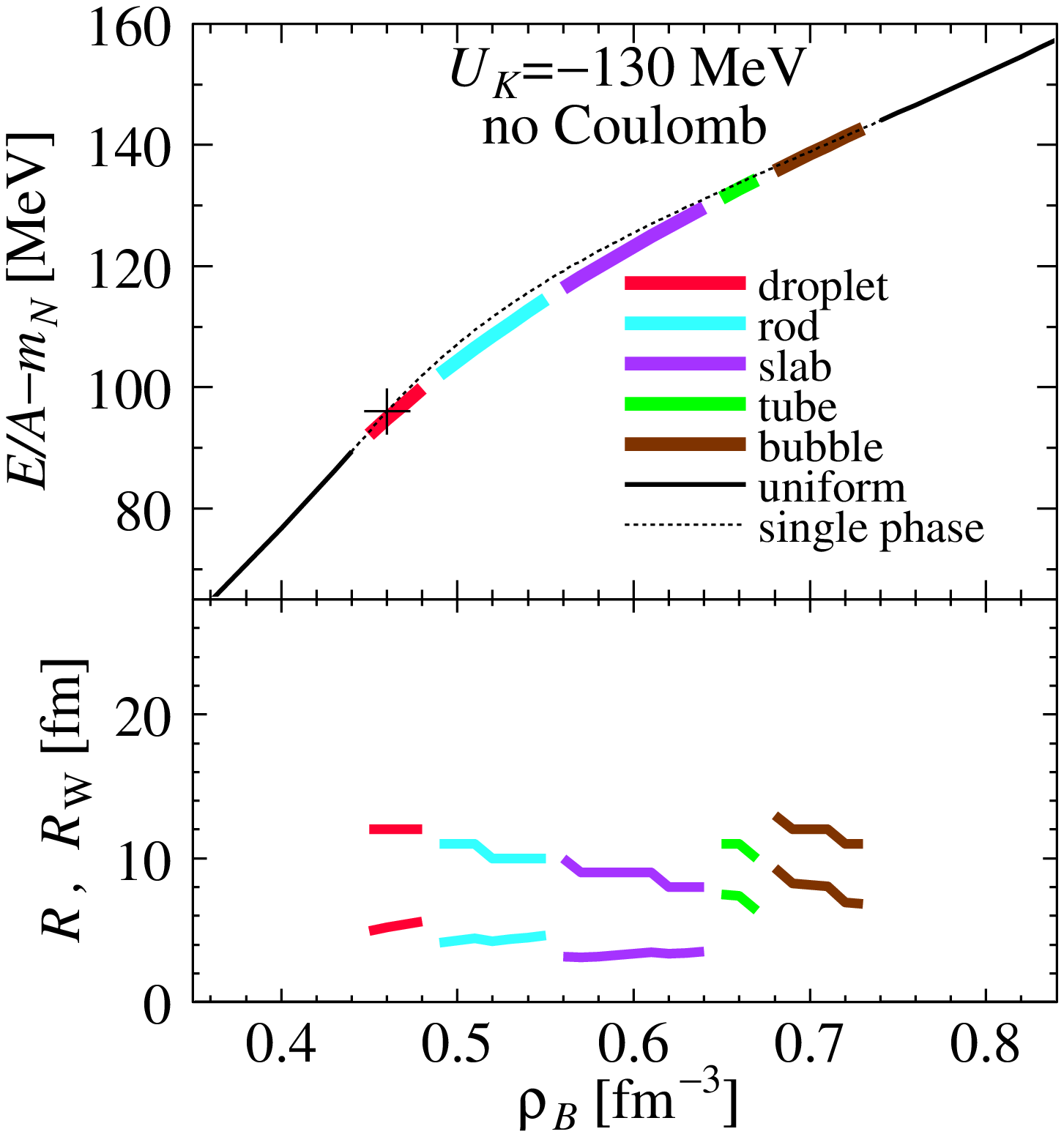}
  \includegraphics[width=.32\textwidth]{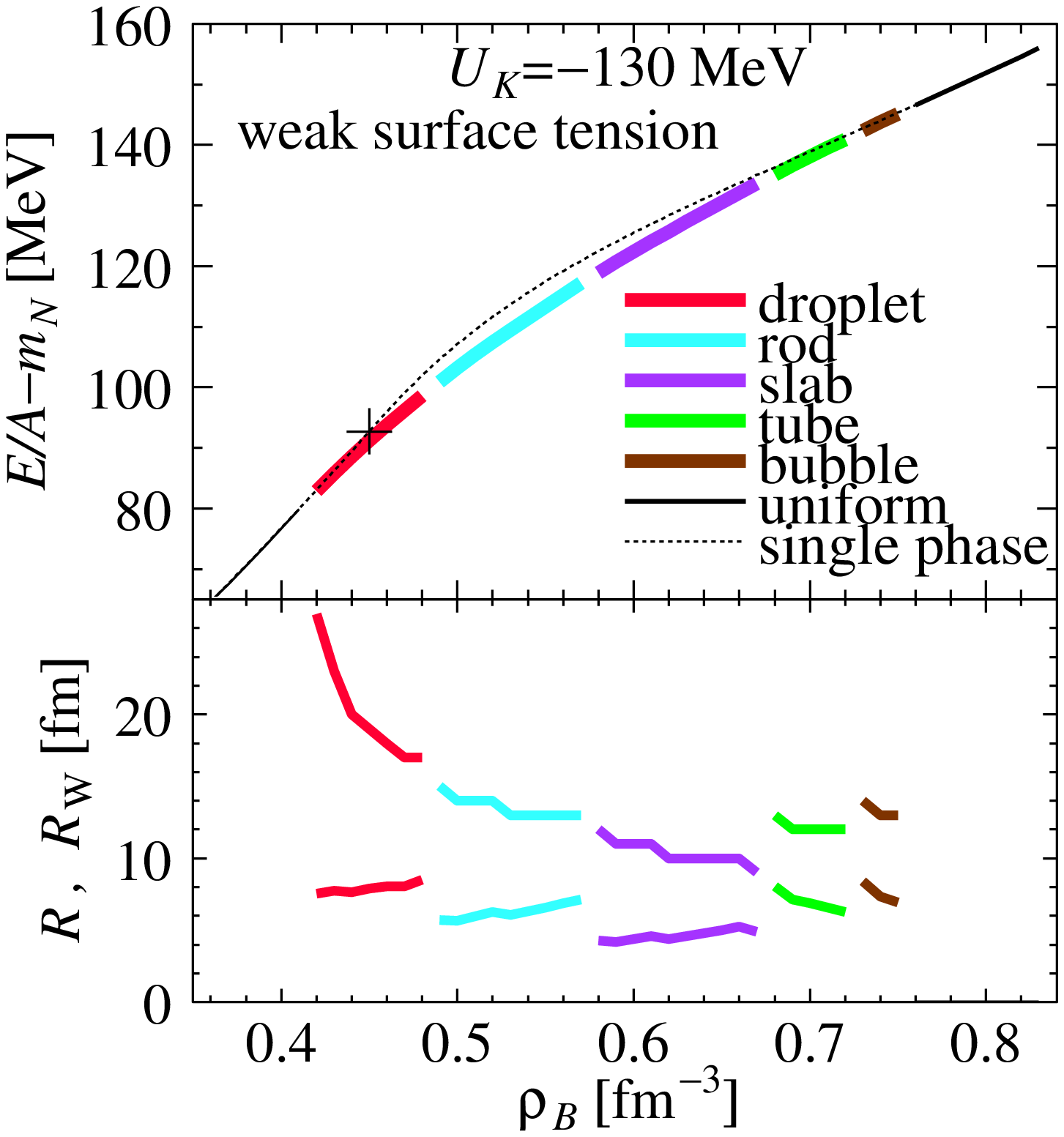}
\caption{
The same as Fig.\ \ref{eoscompare} for high-density nuclear matter
in beta equilibrium.
}\label{figKEOScompare}
\end{figure}

To demonstrate the charge screening effects we compare
results of the full calculation with
those given by a perturbative treatment of the Coulomb interaction often 
used in the literature, which we refer to as ``no Coulomb'' calculation.
In this calculation the electric potential is discarded in equations of motion
(\ref{EOMmuE}), (\ref{EOMmuNp}), (\ref{EOMomegaK})--(\ref{EOMtheta}), (\ref{rhoK}) 
which determine the density profiles.
The Coulomb energy is then added to the total energy by using the
charge density profile thus determined to find the optimal value
with respect to the cell size $R_{\rm W}$.%
\footnote{This treatment may look similar to the bulk calculation. 
However, the pressure balance condition is correctly
imposed in this calculation, while it is inconsistently used in the bulk calculation.}

First, we discuss the case of low-density symmetric nuclear matter.
In the left and the central panels of Fig.~\ref{eoscompare},
compared are different treatments of the Coulomb interaction.
The EOS (upper panels) as a whole shows
almost no dependence on the treatments of the Coulomb interaction.
However, sizes of the cell and the lump (lower panels),
especially for tube and bubbles, are different.
In the cases of the ``full calculation'', the cell radii of ``tube'' and ``bubble''
structures and that of ``slab'' structure get larger with
increase of density, while they are monotonically decreasing in
the case of ``no Coulomb'' calculation.
The other effect
is a  difference in the density range for
each pasta structure.
The ``full'' treatment of the Coulomb
interaction slightly increases the density region of the nuclear pasta.

We show the same comparison for the kaonic pasta structures 
in Fig.\ \ref{figKEOScompare}.
We see again that the density range of the mixed phase is
narrower in the case of the ``no Coulomb'' calculation than
in the full calculation, while the energy gain is almost the same.
A remarkable difference is seen in the cell size,
especially near the onset density of kaonic pasta, for $\rho_B<0.5$ fm$^{-3}$.
The cell size given by the full calculation is always larger than that
given by the ``no Coulomb'' calculation.

\begin{figure}
  \includegraphics[width=.5\textwidth]{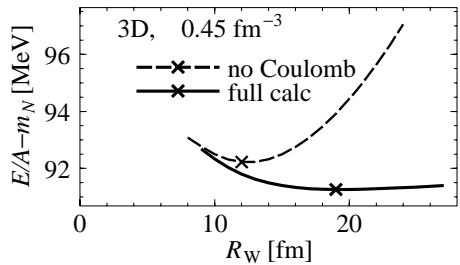}
\caption{
  The cell size $R_{\rm W}$ dependence of the energy per nucleon.
  Crosses on the curves indicate the minimum points.
}\label{figRE}
\end{figure}

To elucidate the screening effect, we depict the $R_{\rm W}$
dependence of the energy per nucleon in Fig.~\ref{figRE}.
In a general case of 3D droplet the Coulomb energy
per particle depends on the radius by its square, while the surface energy
per particle by its inverse.
Therefore the sum of the Coulomb and surface energy has a
U-shape (cf.\ ``no Coulomb'') and has a minimum at a certain radius.
If the Coulomb interaction is screened, the Coulomb contribution 
will be suppressed (cf.\ the full calculation) and the minimum
point gets larger (see Fig.\ \ref{bulk}).
Since the cell radius is approximately proportional to the droplet radius for a given
baryon density,
the above argument applies also to the cell size.

For a long time there  existed a naive view that not all the Gibbs conditions can
be satisfied in a description by the Maxwell construction
if there are two or more independent chemical components
\cite{GS99,CG00,CGS00,gle92},
because the local charge neutrality is implicitly assumed in it.
As the result of this argument,
it was suggested  that  a broad region of the 
structured mixed phase may appear in neutron stars.
However, in recent papers \cite{voskre,voskre1,dubna,emaru1,maruKaon}
we have demonstrated that
if one properly includes the Coulomb interaction,
the Maxwell construction practically satisfies the Gibbs conditions
and the range of the mixed phase will be limited.

\begin{figure}
\begin{minipage}{0.43\textwidth}
  \includegraphics[width=0.97\textwidth]{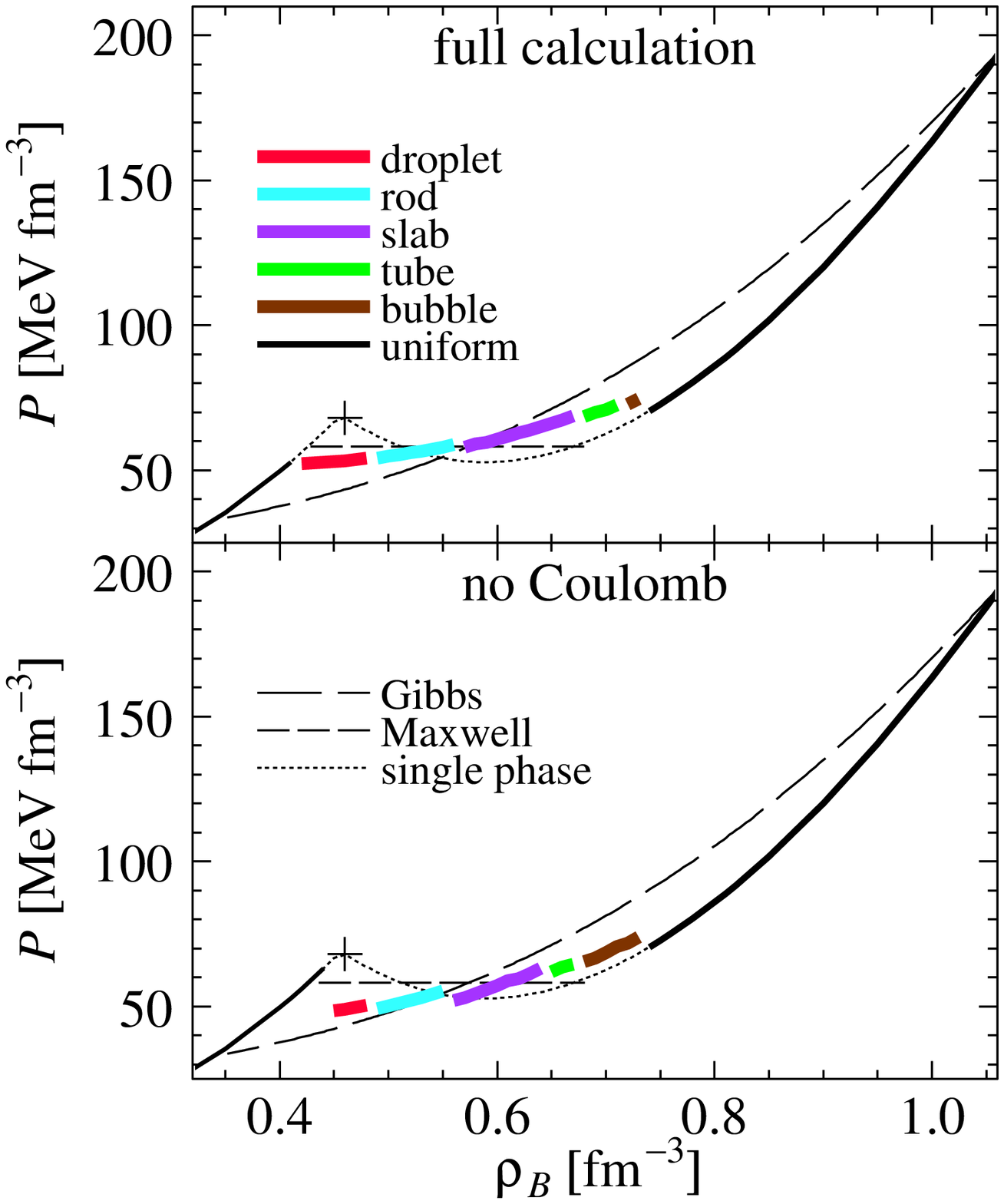}
\caption{
  Pressure versus baryon number density.
}\label{figRhoP}
\end{minipage}
\hfill
\begin{minipage}{0.55\textwidth}
\includegraphics[width=0.99\textwidth]{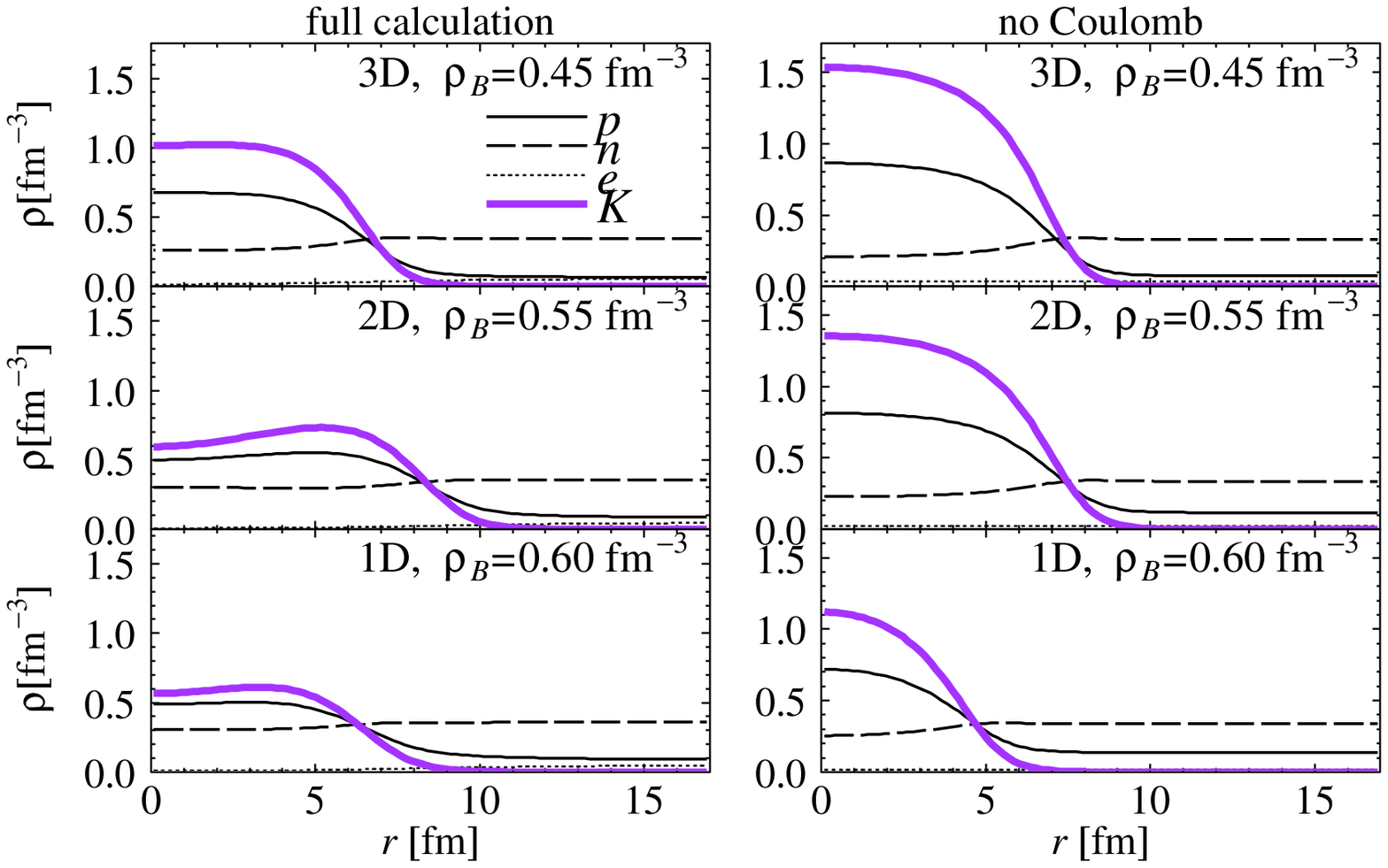}
  \caption{
  Comparison of density profiles of kaonic matter in full
  and ``no Coulomb'' calculations.
  The cell size, $R_{\rm W}=17$~fm, is not optimized
  since the optimum values would be different for different calculations.
}
\label{figKprofcompare}
\end{minipage}
\end{figure}

We present in  Fig.~\ref{figRhoP}
the pressure as a function of the baryon-number density.
We also depict the pressure when
the Gibbs conditions are applied for two semi-infinite
matters discarding the Coulomb interaction (indicated by ``Gibbs'')
and that given by the Maxwell construction (indicated by ``Maxwell'').
We see that the pieces of solid curves  lie between ``Gibbs'' and ``Maxwell''.
The full calculation case is more similar to the one given
by the Maxwell construction.
In Fig.~\ref{figKprofcompare}
compared are the density profiles
obtained by the full and ``no Coulomb'' calculations.
In case of the full calculation the difference between
the negative charge density (of kaons and electrons) and
the positive charge density of protons
is smaller,
indicating that
the system tends to have
a local charge neutrality.
These results suggest that the Maxwell construction is effectively
justified in the full calculation
owing to the charge screening effects.

\subsection{Surface effects}

\begin{figure}
  \includegraphics[width=.53\textwidth]{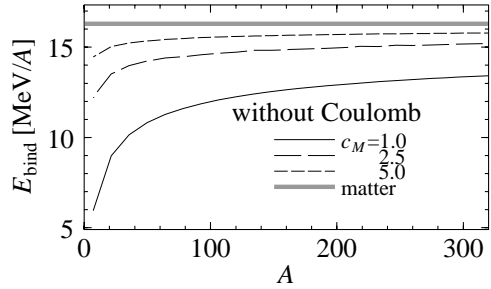}
  \caption{%
  Mass number $A$ dependence of the binding energy of finite nuclei
  $E_{\rm bind}=-(E/A-m_N)$
  without the Coulomb interaction.
  Thick gray line shows the case of
  nuclear matter. 
}\label{finiteMs}
\end{figure}

If we artificially multiply meson masses
$m_{\sigma}$, $m_{\omega}$ and $m_{\rho}$  by a factor $c_M$,
e.g.\ $c_M =1$ (realistic case), 2.5 and 5.0,
the surface tension changes.
Figure \ref{finiteMs} demonstrates
the binding energies of finite nuclei
calculated with different meson masses.
By the use of heavy meson masses, the binding energy
of finite nuclei (for finite $A$)
approaches to that of nuclear matter indicated by a thick gray line.
This shows that the surface tension is reduced
with increase of the meson masses, cf.\ \cite{MS96}.
Notice that this statement is correct only if we
fix the ratio $g_{\phi N}^2 /m_\phi^2$.

Using the above modified meson masses, we explore the effects of
surface tension in the following.
Right panels of Figs.\ \ref{eoscompare} and \ref{figKEOScompare}
are EOS and the sizes of the cell and the lump, but now for the case of
an artificially suppressed surface tension ($c_M=5.0$).
Comparing them with the left panels,
we see that there is almost no difference in the EOS.
However, there are two differences in the case of low-density nuclear matter.
First, the density range of pasta structure is slightly broader
for weaker surface tension.
Secondly, the cell size with a normal surface tension is
larger than the case of weaker one.
It means that weaker surface tension and stronger Coulomb repulsion
cause the similar effects on the cell size since the pasta structure
is realized by the balance of the both.

In the case of kaonic pasta, the meson masses have very small effects.
The $\sigma$, $\omega$ and $\rho$ mesons have
less contribution to the surface tension of
kaonic pasta but $K$-$N$ interaction is dominant.

\section{Quark pasta structures in the hadron-quark mixed phase}

Hereafter, we consider the phase transition from nuclear matter to
three-flavor quark matter in beta equilibrium at zero temperature.
Accordingly, we must introduce many chemical
potentials for the particle species, but the independent ones
in this case
are reduced to two,
the baryon-number chemical potential $\mu_\mathrm{B}$ 
and the charge chemical potential $\mu_\mathrm {Q}$
due to the beta equilibrium and the condition of total charge neutrality.
We can easily see that they are equal to the neutron and electron chemical potentials, 
$\mu_n$ and $\mu_e$, respectively. 
In the mixed phase, these chemical potentials should be spatially
constant to satisfy the Gibbs condition (GC), while the density
profiles are non-uniform. 
When we naively apply the Maxwell construction (MC) to get 
the EOS in thermodynamic equilibrium, 
we immediately notice that $\mu_\mathrm{B}$ is constant in the mixed phase, 
while $\mu_e$ is different in each phase 
because of the difference in the electron number of these phases. 
This is because the MC uses the EOS of bulk matter in each phase,
which is locally charge-neutral, uniform matter; many electrons are needed in hadron
matter to compensate for the positive charge of the protons,
while in quark matter, total charge
neutrality is almost satisfied without electrons.
We shall see how this situation is changed in the presence of the
Coulomb interaction. 
Actually we shall see that MC can be effectively
used, while it apparently violates the Gibbs conditions. 

The treatment of the mixed phase following the deconfinement transition
is somewhat different from that in previous sections due to the property
of the vacuum. 
When we consider the mixed phase, hadrons reside in the
non-perturbative vacuum, while quarks in the perturbative one. 
So the change of the vacuum should be also taken into account in this case,
which should greatly affect the properties of the interface between the hadron
and quark phases. 
Unfortunately, since dynamics of the deconfinement
transition has not been well understood yet, especially in the
finite-baryon number density case, we must have recourse to some models.
We use the MIT bag-model picture in our study, where the non-perturbative and
perturbative vacua are separated by a sharp boundary and the surface
is endowed with the surface tension. 
Note that surface tension is not needed in a consistent calculation, 
because the non-uniform density profiles naturally give its effect.

\subsection{Thermodynamic potential}

We consider the structured mixed phase (SMP) in which one
phase is embedded in the other phase with a certain geometrical form.
As in the previous studies of low-density nucleon matter and high-density kaonic matter,
we use the Wigner-Seitz cell with a size $R_\mathrm{W}$. 
A cell includes an embedded quark phase with a size $R$.

The quark phase consists of {\it u}, {\it d}, and {\it s} quarks and electrons.  
The hadron phase consists of protons, neutrons and electrons.
We incorporate the MIT bag model and assume a sharp boundary at the
hadron-quark interface.
We use the idea of density functional
 theory (DFT) and incorporate the local density approximation (LDA)
 \cite{drez,parr}.

We consider the total thermodynamic potential:
\begin{equation}
\Omega_\mathrm{total} = \Omega_\mathrm{H} +\Omega_\mathrm{Q}
 +\Omega_\tau,
\label{ometot}
\end{equation}
where $\Omega_\mathrm{H}$ and $\Omega_\mathrm{Q}$ represent the
contributions of the hadron and quark
phases, respectively.
We here introduce the surface contribution $\Omega_\tau$ 
in order to properly treat the geometrical structure.
It may be connected with the confining mechanism, but unfortunately
we have no definite idea to incorporate it.  
There have been many estimations about the surface tension in lattice QCD \cite{kaja,huan}, 
in shell-model calculations \cite{mad1,mad2,berg} 
and in model calculations based on the dual-Ginzburg Landau theory \cite{mond}.
Here, we parametrize the surface tension by $\tau$. 
Then we can write $\Omega_\tau = \tau S$, with $S$ being the area of the interface.
Moreover, there may also be a contribution from the curvature term
\cite{mad2}, and  
we regard its effect to be renormalized into $\tau$ for simplicity.
Actually, many authors have treated the strength of $\tau$ as a free parameter 
and determined how the results depend on its value \cite{gle2,pet,alf2}.
Here we also employ this approach by introducing the surface tension
parameter $\tau$ to simulate the surface effect.

We now present the expressions of
 thermodynamic potentials. 
First, the Coulomb interaction energy is expressed in terms of
particle densities, as
\begin{equation}
E_V = \frac{1}{2} \sum_{i,j} \int_{V_\mathrm{W}} d^3 r d^3 r' \frac{Q_i \rho_i({\bf r}) Q_j
\rho_j({\bf r}')}{\left| {\bf r} - {\bf r}' \right|},
\label{coulene}
\end{equation}
where $i,j=u,d,s,p,n,e$, with $Q_i$ being the particle charge ($Q=-e < 0$
for the electron) and $V_\mathrm{W}$ the volume of a Wigner-Seitz cell.
Accordingly, the Coulomb potential is defined as
\begin{equation}
V ({\bf r}) = -\sum_i \int_{V_\mathrm{W}} d^3 r' \frac{e Q_i
 \rho_i({\bf r}')}{\left| {\bf r} - {\bf r}' \right|}+V_0,
\label{vcoul}
\end{equation}
where $V_0$ is an arbitrary constant representing the gauge degree of freedom.
We here fix the gauge by stipulating the condition $V(R_\mathrm{W}) = 0$ 
 as before (see Sec.\ 2.2).

From the above considerations, we obtain
the electron contribution and the Coulomb interaction energy (in both phases) as
\begin{eqnarray}
\Omega_{\mathrm{em}} \! \! &=& \! \! \int_{V_\mathrm{W}=V_\mathrm{Q} + V_\mathrm{H}} \! d^3r \Biggl[  -\frac{1}{8\pi e^2} 
\left(
                                                               \nabla V
                                                               ({\bf r})
                                                              \right)^2
+ \epsilon_e (\rho_e ({\bf r})) - \mu_e \rho_e ({\bf r}) +  V ({\bf r}) \rho_e ({\bf r})  \Biggr] 
\nonumber \\
 &=& \Omega_\mathrm{em}^\mathrm{Q} +
 \Omega_\mathrm{em}^\mathrm{H},
\end{eqnarray}
where
$\epsilon_e (\rho_e({\bf r}))=\frac{\left( 3 \pi^2 \rho_e ({\bf r})\right)^\frac{4}{3}}{4\pi^2}$
 is the kinetic energy density of electrons.
 The contributions in the two phases, $\Omega_\mathrm{em}^\mathrm{Q}$ and
 $\Omega_\mathrm{em}^\mathrm{H}$, are obtained by simply dividing
 $\Omega_\mathrm{em}$ by the volumes $V_\mathrm{Q}$ and
 $V_\mathrm{H}$, respectively.

 Secondly, in the quark phase, the {\it u} and {\it d}
 quarks are treated as massless particles, and only the {\it s} quark is
 massive, with $m_s= 150$ MeV.
The kinetic energy density of the quark of flavor $f$ is simply expressed as \cite{tama}
\begin{equation}
\epsilon_{f \mathrm{kin}}= \frac{3}{8 \pi^2} m_f^4 \left[ x_f \eta_f \left( 2x_f^2+1 \right) 
  -\ln\left( x_f +\eta_f  \right) \right],
\end{equation}
where $m_f$ is the quark mass, $x_f = p_{\mathrm{F}f}({\bf r})/m_f$,
with the Fermi momentum $p_{\mathrm{F}f}({\bf r}) = (\pi^2 \rho_f ({\bf r})
)^\frac{1}{3}$, and $\eta_f=\sqrt{1+x^2_f}$.

For the interaction energy, we take into account the leading-order
contribution coming from the one-gluon exchange interaction:
\begin{equation}
 \epsilon_{f \mathrm{Fock}} = -\frac{\alpha_{\mathrm{c}}}{\pi^3}  m_f^4 \left\{ x_f^4 - \frac{3}{2}
 \left[ x_f \eta_f -\ln\left( x_f +\eta_f  \right)  \right]^2  \right\}.
\label{fock}
\end{equation}
Including this interaction, the quark contribution to the thermodynamic potential is expressed as
\begin{eqnarray}
& & \Omega_{\mathrm{Q}} =
 \Omega_{u}+\Omega_{d}+\Omega_{s} +\int_{V_\mathrm{Q}}
 d^3r B + \Omega_\mathrm{em}^\mathrm{Q} , \label{omeq}\\
\Omega_f  &=& \int_{V_\mathrm{Q}} \! d^3r \left[
\epsilon_f (\rho_f ({\bf r})) - \mu_f \rho_f ({\bf r}) -
                                    N_f V ({\bf r})
                                    \rho_f ({\bf r}) \right] , \hspace{10pt} N_f=\frac{Q_f}{e},
\label{omeq2}
\end{eqnarray}
where the energy density $\epsilon_f (\rho_f
({\bf r}))$ represents  $\epsilon_{f \mathrm{kin}} + \epsilon_{f
\mathrm{Fock}}$ of the $f$ quark, and $B$ is the bag constant.
The bag constant is taken as 120 MeV/fm$^3$, and the QCD fine structure constant
as $\alpha_{\mathrm{c}}=0.4$, which were also used by Heiselberg 
et al.\ \cite{pet} and in Refs.~\cite{vos,voskre,voskre1} .

Thirdly, we consider the hadron contribution.
The thermodynamic potential for the non-relativistic nucleons becomes
\begin{eqnarray}
\Omega_{\mathrm{H}} = E_\mathrm{N} - \sum_{a=p, n} \mu_a \int_{V_\mathrm{H}} d^3r \rho_a
 ({\bf r}) -  \int_{V_\mathrm{H}} d^3r \, V ({\bf r}) \rho_{p}({\bf r}) 
  +\Omega_\mathrm{em}^\mathrm{H},
\label{omeh}
\end{eqnarray}
where $E_\mathrm{N}$ is the energy of the nucleons,
\begin{eqnarray}
E_\mathrm{N} = \int_{V_\mathrm{H}} d^3r \left[ \sum_{a=p, n} \frac{3}{10m}\left( 3 \pi^2 
\right)^\frac{2}{3} \rho^\frac{5}{3}_a
 ({\bf r}) + \epsilon_\mathrm{pot}  \left( \rho_{p} ({\bf r})
                                      , \rho_{n} ({\bf r})
                                     \right) \right].
\end{eqnarray}
Here we use the effective potential $\epsilon_\mathrm{pot} (\rho_{p} ({\bf r}), \rho_{n}
({\bf r}))$ parametrized by the local densities for simplicity,
\begin{eqnarray}
\epsilon_\mathrm{pot} ({\bf r})  &=& S_0 \frac{\left( \rho_n({\bf r}) - \rho_p({\bf r})  
\right)^2}{\rho_0({\bf r})} + \left( \rho_n({\bf r}) + \rho_p({\bf r})  
\right) \epsilon_\mathrm{bind} \nonumber \\
&+&  K_0  \frac{\left( \rho_n({\bf r}) + \rho_p({\bf r})  \right)}{18} 
\left( \frac{\rho_n({\bf r}) + \rho_p}{\rho_0} - 1  \right)^2 \nonumber\\
&+&  C_\mathrm{sat}  \left( \rho_n({\bf r}) +
                             \rho_p({\bf r})  \right)  \left(
                             \frac{\rho_n({\bf r}) +
                             \rho_p({\bf r})}{\rho_0}  -  1
                              \right) ,
\label{effpot}
\end{eqnarray}
where $S_0$,  $K_0$,  $\epsilon_\mathrm{bind}$ and $C_\mathrm{sat}$ are
adjustable parameters which are chosen
to reproduce the saturation properties of nuclear
matter \cite{vos,voskre,voskre1}. 
We use $S_0 = 18$ MeV, $K_0 = 285$ MeV,
and $C_\mathrm{sat} = -14.3$ MeV at the saturation density $\rho_0$.
The value $\epsilon_\mathrm{bind}
=\tilde\epsilon_\mathrm{bind}+m_N-2\rho_0^{-1} \epsilon_p(\rho_B=\rho_0,
Y_p=1/2)$ provides the empirical binding energy
$\tilde\epsilon_\mathrm{bind}=-15.6$ MeV 

 We consider chemical equilibrium at the hadron-quark
 interface as well as in each phase, in which case we have the following:
\begin{eqnarray}
&& \mu_u+\mu_e = \mu_d,~~~\mu_d = \mu_s, \nonumber \\
&& \mu_p+\mu_e = \mu_n \equiv \mu_\mathrm{B},~~~\mu_n = \mu_u+2 \mu_d, \nonumber\\
&& \mu_p = 2 \mu_u+\mu_d.
\label{chemeq}
\end{eqnarray}

\subsection{Numerical procedure}

We get the equations of motion from the relation
$\frac{\delta \Omega_\mathrm{total}}{\delta\phi_i}=0$
($\phi_{i}=\rho_u({\bf r}), \rho_d({\bf r}), \rho_s({\bf r})$,
$\rho_p({\bf r}), \rho_n({\bf r}), \rho_e({\bf r}), V({\bf r})$).
The Poisson equation then reads
\begin{eqnarray}
\nabla^2 V ({\bf r}) = 4 \pi e^2 \rho_{\rm ch}({\bf r})
\label{poisson}
\end{eqnarray}
where the charge density $\rho_{\rm ch}^{\rm Q}({\bf r})$ is given by 
$\rho_{\rm ch}({\bf r})=\frac{2}{3}\rho_u ({\bf r})
 - \frac{1}{3}\rho_d({\bf r}) - \frac{1}{3} \rho_s- \rho_e ({\bf r})$ 
in the quark phase and 
$\rho_{\rm ch}^{\rm H}({\bf r})=\rho_{p} ({\bf r}) - \rho_e ({\bf r})$ 
in the hadron phase.
The other equations of motion give only the expressions
for the chemical potentials,
\begin{equation}
\mu_i = \frac{\delta E_\mathrm{kin+str}}{\delta \rho_i ({\bf r})} - N_i V({\bf r}),
\label{mu_i}
\end{equation}
where $\displaystyle E_\mathrm{kin+str} = \sum_{i=u,d,s,e} 
\int d^3r \epsilon_i + E_\mathrm{N}$.
Then the quark chemical potentials are expressed as
\begin{eqnarray}
\mu_{u} &=& \left( 1 + \frac{2 \alpha_{\mathrm{c}}}{3 \pi}
                     \right) \pi^\frac{2}{3}
\rho_{u}^\frac{1}{3} ({\bf r})- \frac{2}{3} V ({\bf r})  \label{eomq}\\
\mu_{d} &=& \left( 1 + \frac{2 \alpha_{\mathrm{c}}}{3 \pi}  \right) \pi^\frac{2}{3} 
\rho_{d}^\frac{1}{3}({\bf r}) + \frac{1}{3} V ({\bf r})\\
\mu_{s} &=& \epsilon_{{\mathrm{F}s}}({\bf r}) + \frac{2
 \alpha_{\mathrm{c}}}{3 \pi} \left[ p_{\mathrm{F}s}({\bf r})- 3
                              \frac{m_{s}^2}{\epsilon_{\mathrm{F}s}({\bf r})} 
\ln \left( \frac{\epsilon_{\mathrm{F}s}({\bf r})+p_{\mathrm{F}s}({\bf r})}{m_{s}} \right)  \right]
  + \frac{1}{3} V ({\bf r}), \nonumber \\
\end{eqnarray}
with $\epsilon_{\mathrm{F}s} ({\bf r})=
\sqrt{m_{s}^2+p_{\mathrm{F}s}^2({\bf r})}$,
whereas the chemical potentials of the nucleons and electrons are
\begin{eqnarray}
\mu_{n} &=& \frac{p_{\mathrm{F}n}^2}{2m} + \frac{2S_0 \left(\rho_n({\bf r})-\rho_{p}({\bf r})
\right)}{\rho_0}+ \epsilon_{\mathrm{bind}} \nonumber \\
&+& \frac{K_0}{6} \left( \frac{\rho_{n} ({\bf r}) \!+\!
                 \rho_{p} ({\bf r})} {\rho_0} - 1  \right)^2 +
\frac{K_0}{9} \left(  \frac{\rho_{n} ({\bf r})+
               \rho_{p} ({\bf r})}{\rho_0}- 1  \right) \nonumber \\
&+& 2 C_{\mathrm{sat}}  \frac{\rho_{n}({\bf r}) + \rho_{p}({\bf r})}{\rho_0} - C_{\mathrm{sat}} \\
\mu_{p} &=& \mu_{n} - \frac{p_{\mathrm{F}n}^2
 ({\bf r})}{2m}+ \frac{p_{\mathrm{F}p}^2 ({\bf r})}{2m} - \frac{4 S_0
 \left( \rho_{\mathrm{B}} - 2 \rho_{p} ({\bf r})
 \right)^2}{\rho_0} - V ({\bf r}) \\
\mu_e &=& \left( 3 \pi^2 \rho_e ({\bf r}) \right)^\frac{1}{3} + V ({\bf r}).
\label{eomh}
\end{eqnarray}
We solve these equations of motion under the GC.
The important point here is that the Coulomb potential $V({\bf r})$
is included properly in each expression.
The Coulomb potential is a
function of the charged-particle densities, and, in turn,
these densities are functions of the Coulomb potential. 
As a result, the Poisson equation becomes highly non-linear. 

 Note that we must now determine eight variables, i.e.,
six chemical potentials, $\mu_u$, $\mu_d$,
 $\mu_s$, $\mu_p$, $\mu_n$,
$\mu_e$, and  the radii $R$ and $R_\mathrm{W}$.
First, we fix $R$ and $R_\mathrm{W}$.
Here we have four conditions due to 
the beta equilibrium, expressed by (\ref{chemeq}).
Therefore, once the two chemical potentials $\mu_\mathrm{B}$ and
$\mu_e$ are given,
we can determine the other four chemical potentials,
$\mu_u$, $\mu_d$, $\mu_s$ and $\mu_p$.
Next, we determine $\mu_e$ by the global charge neutrality
condition:
\begin{equation}
\int_{V_{\rm Q}} d^3r  \rho_{\mathrm{ch}}^{\mathrm{Q}}
 + \int_{V_{\rm H}} d^3r \rho_{\rm ch}^{\mathrm{H}} = 0.
\label{globalneutral}
\end{equation}

The pressure coming from the surface tension 
\footnote{Basically the pressure is locally defined in terms of the
Gauss principal curvatures, $\kappa_1$ and $\kappa_2$, on the surface:
$
P_\tau=\tau\left(\kappa_1+\kappa_2\right).
$
For the special shapes of 
sphere, cylinder and plane, $R_i$ is simply constant to give 
$(\kappa_1,\kappa_2)=(1/R,1/R), (1/R,0)$ and $(0,0)$, respectively.}
is given by
\begin{equation}
P_{\tau}= \tau \frac{d S}{d V_{\rm Q}}.
\end{equation}
Then, we find the optimal value of $R$ ($R_\mathrm{W}$ is fixed, and
therefore $f_V$ changes with $R$)
by using one of the GC,
\begin{equation}
P^\mathrm{Q} = P^\mathrm{H} + P_\tau.
\label{pbalance}
\end{equation}
The pressure in each phase, $P^\mathrm{Q}$ or $P^\mathrm{H}$,
is given by the thermodynamic relation
$P^\mathrm{Q(H)}=-\Omega_\mathrm{Q(H)}/V_\mathrm{Q(H)}$,
where $\Omega_\mathrm{Q(H)}$ is the thermodynamic potential in each
phase and given by adding an electron and the Coulomb interaction
contributions to $\Omega_\mathrm{Q(H)}$ in Eqs.\ (\ref{omeq}) and (\ref{omeh}).
Finally, we determine $R_\mathrm{W}$ by minimizing the thermodynamic potential.
Therefore, once
$\mu_\mathrm{B}$ is given, all the other $\mu_i$ ($i=u,d,s,p,e$),
along with $R$ and $R_\mathrm{W}$, can be obtained.

Note that we keep the GC throughout the numerical procedure.
We see below how
the mixed phase would be changed by including the finite-size effects,
keeping the GC completely.
Although the MC is not strictly correct, 
our results exhibit behavior similar to that in the case of the MC,
as a result of including the finite-size effects.

In the numerical calculation, every point inside a cell is represented by a grid point
(the number of grid points is $N_\mathrm{grid} \approx 100 $).
The equations of motion are solved by using
a relaxation method for a given baryon-number chemical potential under
the constraint of global charge neutrality.

\section{Charge screening in the hadron-quark mixed phase}

\subsection{Bulk calculation}

\begin{figure}[h]
\begin{center}
\includegraphics[width=0.6\textwidth]{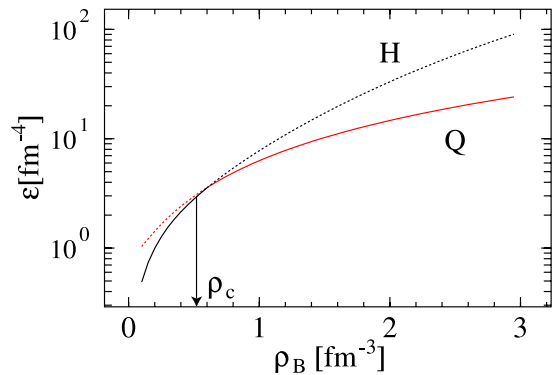}
\caption{Energy density $\epsilon$ for uniform hadron matter and quark matter. 
Uniform quark matter is energetically favorable in high-density region, 
while hadron matter in low-density region.}
\label{eosbulk}
\end{center}
\end{figure}

\begin{figure}[h]
\begin{center}
\includegraphics[width=0.6\textwidth]{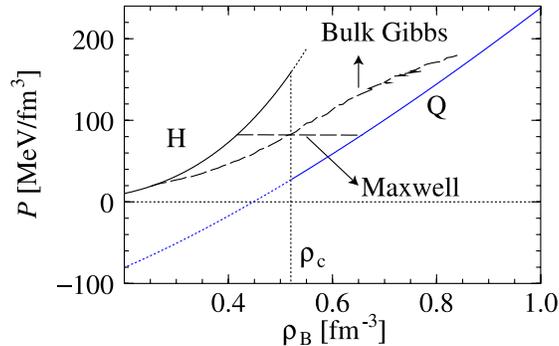}
\caption{ Pressure of uniform matter, that given by the bulk calculation with 
GC (Bulk Gibbs) and that given 
 by MC (Maxwell).}
\label{presbulk}
\end{center}
\end{figure}

First of all we depict the deconfinement transition between uniform
hadron matter and quark matter in Fig.~\ref{eosbulk} 
by using Eqs.~(\ref{omeq}), (\ref{omeh})
and (\ref{effpot}).
Then we can see that it exhibits the first-order phase transition. 
To describe the mixed phase a bulk calculation proceeds like this. 
Considering two semi-infinite matter separated by a sharp
boundary and introducing the volume fraction of quark matter $f_V$,
we apply the global charge neutrality condition,
\begin{equation}
f_V\rho_{\rm ch}^{\rm Q}+(1-f_V)\rho_{\rm ch}^{\rm H}=0.
\label{globalc}
\end{equation}
Assuming the chemical equilibrium among the particles, there are left
two independent quantities to be determined, the electron chemical
potential $\mu_e$ and the baryon-number chemical potential
$\mu_B=\mu_n$. 
Finally imposing the pressure balance condition between two phases,
\begin{equation}
P_{\rm H}=P_{\rm Q},
\label{globalp}
\end{equation} 
we can determine $\mu_e$ and $f_V$ for a given $\mu_B$ or baryon-number density. 
We show the EOS in this calculation in Fig.~(\ref{presbulk}).
Compared with the EOS given by the Maxwell construction, where the local
charge neutrality is implicitly assumed instead of Eq.~(\ref{globalc}),
we can see that the density regime of the mixed phase is considerably
widened and pressure is no more constant there. 

The above picture of the mixed phase is too simple because it
should have structures by the surface tension and the Coulomb
interaction, which are called the finite-size-effects.
Actually Heiselberg et al.\ demonstrated that 
the finite-size effects may {\it energetically} disfavor the mixed
phase and thereby its density regime is largely limited \cite{pet}. 
In their calculation, however, the non-uniform structure of 
the mixed phase is always stable due to an inconsistent inclusion 
of the Coulomb interaction. 
We shall see that the rearrangement of the charge density and the charge
screening of the Coulomb interaction together may cause a {\it mechanical} instability
of the structured mixed phase. 
As another defect they use the pressure balance condition before introducing the 
finite-size effects; they use the value of $f_V$ given by the above simple calculation.

\subsection{Mechanical instability}

Some analytic discussions have been given in Refs.\ \cite{voskre,voskre1,vos}, only by taking the
leading contribution of the Coulomb potential in the Poisson equation. 
We shall see in the next section the numerical results about
the properties of the mixed phase and EOS, but it should be useful 
to figure out some qualitative aspects by analytic discussions. 
In fact we can clearly see a mechanical instability of the structured mixed
phase due to the charge screening effect. 

If we expand the charge density in $\delta V({\bf r})=V({\bf r})-V_{\rm ref}$ 
around the reference value $V_{\rm ref}$, which is also gauge
dependent, the Poisson equation (\ref{poisson}) renders
\begin{equation}\label{Pois-lin}
\Delta \delta V^\alpha ({\bf r}) =  4\pi e^2
\rho^{{\rm ch},\alpha} (V^{\alpha}({\bf r})=V^{\alpha}_{\rm ref})
+ (\kappa^{\alpha} (V^{\alpha}({\bf r})=V^{\alpha}_{\rm ref} ))^2
\delta V^{\alpha}({\bf r})+\cdots,
\end{equation}
in each phase $\alpha=Q$ or $H$ with the Debye screening parameter,
\begin{equation}\label{debye}
(\kappa^{\alpha}(V^{\alpha}({\bf r})=V^{\alpha}_{\rm ref} ))^2 =
4\pi e^2\left[ \frac{\partial\rho^{{\rm ch},\alpha}}{\partial
V}\right]_ {V^{\alpha}({\bf r})=V^{\alpha}_{\rm ref}} =4\left.\pi
Q_i^{\alpha}Q_j^{\alpha}\frac{\partial\rho_j^{\alpha}}{\partial\mu_i^\alpha}
\right|_{V^{\alpha}({\bf r})=V^{\alpha}_{\rm ref}},
\end{equation}
where we used the gauge-invariance relation,
\begin{equation}
\frac{\partial\rho_i^\alpha}{\partial V^\alpha}=N_j^{{\rm ch}, \alpha}
\frac{\partial\rho_j^\alpha}{\partial\mu_i^\alpha}.
\end{equation}
Then we calculate contribution to
the thermodynamic potential (effective energy) of the cell
up to $O(\delta V^{\alpha}({\bf r}))^{2}$. 
The Coulomb interaction energy can be written 
by way of the Poisson equation (\ref{Pois-lin}) as
\begin{eqnarray}\label{eV}
E_V =\int_{V_{\rm Q}} d^3r\epsilon_{V}^{\rm Q} +\int_{V_{\rm H}}d^3r\epsilon_{V}^{\rm H} 
=\int_{V_{\rm Q}}\frac{(\nabla V^{\rm Q}({\bf r}))^2}{8\pi e^2}d^3r 
+ \int_{V_{\rm H}}\frac{(\nabla V^{\rm H}({\bf r}))^2}{8\pi e^2}d^3r.
\end{eqnarray}
Besides the terms given by (\ref{eV}), there
are another contributions arising from effects associated with the
inhomogeneity of the electric potential profile, through implicit
dependence of the particle densities on
$V^{\rm Q,H}({\bf r})$. 
We will call them ``correlation terms'',
$\omega_{\rm cor}^{\alpha}=\epsilon_{\rm kin+str}^{\alpha}-\mu_i^{\alpha} \rho_i^{\alpha}$.
Taking $\epsilon_{\rm kin+str}^{\alpha}$ and $\rho_{i}^{\alpha}$ as functions of
$V^{\alpha}({\bf r})$ we expand them around the reference value $V_{\rm ref}$ 
to obtain the corresponding
correlation contribution to the thermodynamic potential
$\Omega_{\rm cor} =\int_{V_{\rm Q}} d^3r\omega_{\rm cor}^{\rm Q} 
+\int_{V_{\rm H}} d^3r\omega_{\rm cor}^{\rm H}$:
\begin{eqnarray}\label{om-cor0}
&&\omega_{\rm cor}^{\alpha}= \epsilon_{\rm kin+str}^{\alpha}
(\rho_i^{\alpha}(V^{\alpha}_{\rm ref}))
-\mu_i^\alpha \rho_i^{\alpha} (V^{\alpha}_{\rm ref})
-\rho^{{\rm ch},\alpha} (V^{\alpha}_{\rm ref} )V^{\alpha}_{\rm ref}
\nonumber \\ &&+\frac{V^{\alpha}_{\rm ref}\Delta
V^{\alpha}({\bf r})}{4\pi e^2} +\frac{
(\kappa^{\alpha}(V^{\alpha}_{\rm ref }) )^2 (\delta
V^{\alpha}({\bf r}))^2 }{8\pi e^2 }+...,
\end{eqnarray}
where we also used Eqs.\ (\ref{Pois-lin}) and (\ref{debye}). 
In general $V^{\rm Q}_{\rm ref }\neq V^{\rm H}_{\rm ref }$ and they
may depend on the droplet size.  
Their proper choice should provide appropriate convergence of the above expansion 
in $\delta V ({\bf r})$. 
Taking 
\begin{equation}
V^{\rm Q}_{\rm ref }= V^{\rm H}_{\rm ref}=V_{\rm ref } = {\rm const}
\end{equation}
we find
\begin{eqnarray}\label{om-cor}
\omega_{\rm cor}^{\alpha}=\frac{ (\kappa^{\alpha}(V_{\rm ref })
)^2 ( V^{\alpha}({\bf r}) - V_{\rm ref })^2 }{8\pi e^2 } + {\rm const},
\end{eqnarray}
and one may count the potential from the corresponding constant value. 
Here we also took into account that the term
$V^{\alpha}_{\rm ref}\Delta V^{\alpha}({\bf r})/(4\pi e^2 )$
does not contribute to $\Omega$ according to the boundary
conditions in the droplet center, at the droplet boundary (at zero
surface charge), and at the boundary of the
Wigner-Seitz cell.

The Poisson equation (\ref{Pois-lin}) 
in the droplet
phase can be easily solved.%
\footnote{We only consider the quark-droplet phase here for simplicity. 
See Refs.\ \cite{voskre,voskre1,vos} for other cases.} 
For $r<R$, the charge density $\rho_{\rm ch}^{\rm Q}$ can be written as 
\begin{eqnarray}\label{qrhoch}
&&\rho_{\rm ch }^{\rm Q}\simeq \left( 1-\frac{2\alpha_c}{\pi}
\right)\nonumber \\ &&\times\left[ \frac{2\mu_B^2 (V^{\rm Q}({\bf r})-\mu_e)}
{9\pi^2 } \left( 1+O\left(\frac{(V^{\rm Q}({\bf r})-\mu_e)}{\mu_B}
\right) \right)+ \frac{\mu_B  m_s^2 }{6\pi^2 } \right] 
\end{eqnarray}
by way of Eq.~(\ref{eomq}) to find 
\begin{equation}\label{1-sol}
V^{\rm Q}(r)-\mu_e =\frac{V_{0}^{\rm Q}}{\kappa^{\rm Q} r}
\sinh(\kappa^{\rm Q} r)+ U_{0}^{\rm Q},
\end{equation}
with an arbitrary constant $V_{0}^{\rm Q}$. 
For  the Debye parameter $\kappa^{\rm Q}$ and for the constant $U_{0}^{\rm Q}$ 
we obtain:
\begin{eqnarray}\label{qscrl}
(\kappa^{\rm Q})^2 =\frac{8e^2 \mu_{B}^2}{9\pi}\left( 1-
\frac{2\alpha_c}{\pi}
\right),\,\,\,\, U_{0}^{\rm Q} \simeq -\frac{3 m_s^2}{4\mu_B}.
\end{eqnarray}
Thus, the value $U_{0}^{\rm Q}$ is rather small and the main
contribution to $V^{\rm Q}({\bf r})$ comes
from the first term in (\ref{1-sol}). 
Note that solution (\ref{1-sol}) is independent of the reference value 
$V_{\rm ref}^{\rm Q}$ in this case, cf.\ (\ref{debye}), 
since $\rho_{\rm ch}^{\rm Q}$ in (\ref{qrhoch}) is  the linear function 
of $V^{\rm Q}({\bf r})-\mu_e$ in the approximation used.

For $r>R$, expanding the charge density $\rho_{\rm ch}^{ \rm H}({\bf r})$ 
around a reference value, 
\begin{eqnarray}\label{r2}
\rho_{\rm ch }^{\rm H}
({\bf r})\simeq\rho_p (V^{\rm H}({\bf r})=V_{\rm ref}^{\rm H})
+\delta \rho_p ({\bf r})-\rho_e(V^{\rm H}({\bf r})=V_{\rm ref}^{\rm H})
-\delta\rho_e ({\bf r}), 
\end{eqnarray}
we find up to linear order
\begin{eqnarray}\label{co}
&&\delta \rho_p ({\bf r})\simeq C_0^{-1}(V^{\rm H}({\bf r})
-V_{\rm ref}^{\rm H}), \,\, \delta\rho_e ({\bf r})
=\frac{(\mu_e-V_{\rm ref}^{\rm H})^2}{\pi^2} (V^{\rm H}({\bf r})
-V_{\rm ref}^{\rm H}),
\\ &&C_0=\frac{A_{22}}{|A|},
\quad p_{{\rm F}p}=(3\pi^2\rho_p (V^{\rm H}({\bf r})=V_{\rm ref}^{\rm H}))^{1/3},
\nonumber 
\end{eqnarray}
where $A_{22}$, and $\mid A\mid$ are the corresponding matrix
element and the determinant of the matrix,
\begin{equation}
A_{ij}=\frac{\delta^2E_{\rm kin+str}}{\delta\rho_i\delta\rho_j}.
\end{equation}
Then the Poisson equation with the boundary condition
\begin{equation}
\nabla V^{\rm H}({\bf r})|_{r=R_{\rm W}}=0
\end{equation}
yields
\begin{eqnarray}\label{psi}
&&V^{\rm H}(r)-\mu_e = V_{0}^{\rm H} \frac{R}{r}
\cosh\left(\kappa^{\rm H}(r-R_{\rm W} )\right) \left(
1-\delta\right)+U_0^{\rm H}, \\ &&\delta
=\tanh\left(\kappa^{\rm H}(R_{\rm W} -r)\right) /
(\kappa^{\rm H}R_{\rm W} ), \nonumber
\end{eqnarray}
with an arbitrary constant $V_0^{\rm H}$, where the constant
$U_0^{\rm H}$ is given by
\begin{equation}
U_0^{\rm H}+\mu_e=-\frac{4\pi e^2\rho_{\rm ch}^{\rm H}(V^{\rm H}= 
V_{\rm ref}^{\rm H})} {(\kappa^{\rm H})^2}+V_{\rm ref}^{\rm H}.
\end{equation}

The charge screening in the external region is determined by the
Debye parameter
\begin{eqnarray}\label{hscrl}
(\kappa^{\rm H})^2 =\frac{4e^2 (\mu_{e}-V_{\rm ref}^{\rm H})^2}{\pi} 
+\frac{4e^2 \pi}{C_0},
\end{eqnarray}
where the second term is the contribution of the proton screening.
Taking $\rho_{B}^{\rm H}=1.5 \rho_0$,   $\mu_{e,\rm Gibbs} \simeq
170$~MeV, $\mu_{B}=\mu_n \simeq 1020$~MeV, $\alpha_c \simeq 0.4$,
we estimate typical Debye screening lengths as
$\lambda_{\rm D}^{\rm Q}\equiv 1/\kappa^{\rm Q}\simeq 3.4/m_\pi$, and
$\lambda_{\rm D}^{\rm H}\equiv 1/\kappa^{\rm H} \simeq 4.2 /m_\pi$,
whereas one would have $\lambda_{\rm D}^{\rm H}\simeq 8.5 /m_\pi$, if
the proton contribution to the screening (\ref{hscrl})  was absent ($C_0^{-1}=0$). 
With the estimate $\lambda_{\rm D}^{\rm H} \simeq 4.2 /m_\pi$ 
we get that $\kappa^{\rm H}R_{\rm W}>1$
for the droplets with the radii $R>(2f_V)^{1/3}\cdot 3.3/m_\pi$.

\begin{figure}[h]
\begin{center}
	\includegraphics[width=0.63\textwidth]{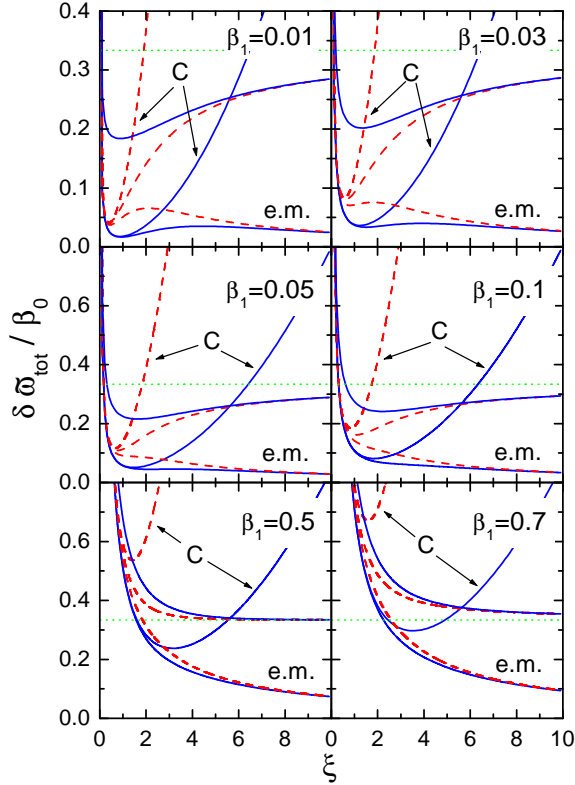}
\end{center}
\caption{Dimensionless value of the thermodynamic potential per droplet volume. 
Solid lines are given for $f_V=0.5$ and dashed lines for $f_V=1/100$. 
The ratio of the screening lengths of two phases,
 $\alpha_0=\lambda_{\rm D}^{\rm Q}/\lambda_{\rm D}^{H}$, is fixed as one. $\xi$ is a
 dimensionless radius of the droplet, $\xi\equiv R/\lambda_{\rm D}^{\rm Q}$, 
 with $\lambda_{\rm D}^{\rm Q}\simeq 5$ fm in this calculation. 
 See text for further details. }
\label{effe}
\end{figure}

In Fig.~\ref{effe} we demonstrate the radius ($R$) dependence of the total thermodynamic
potential  per droplet volume for the case of spherical droplets, 
$\delta \widetilde{\omega}_{\rm tot}/\beta_0
=(\widetilde{\epsilon}_V^{\rm Q}+ \widetilde{\epsilon}_V^{\rm H}
+ \widetilde{\omega}_{\rm cor}^{\rm Q}+ \widetilde{\omega}_{\rm cor}^{\rm H}
+ \widetilde{\epsilon}_{\rm S} )/\beta_0$, 
given by the sum of partial contributions, where tilde denotes each quantity 
scaled by the droplet volume $V=4\pi R^3/3$ and $\beta_0$ is a typical quantity 
with the dimension of the energy density \cite{vos,voskre,voskre1},
\begin{equation}
\beta_0=\frac{3(U_0^{\rm H}-U_0^{\rm Q})^2(\kappa^{\rm Q})^2}{8\pi e^2}.
\end{equation}
 Preparing some wide range for the
value of the surface tension parameter $\tau$ or 
$\beta_1= 3\kappa^{\rm Q}\tau/\beta_0$ \cite{voskre,voskre1,vos}, 
we present two cases of $f_V$, $f_V=0.01,0.5$.
The dotted green line shows the asymptotic value of 
$\delta\widetilde{\omega}_{\rm tot}/\beta_0$
in the limit $\xi\rightarrow\infty$.

The label ``C'' is given for reference to show the previous
non self-consistent case, where the Coulomb potential is not screened (see Fig.~\ref{bulk}). 
We can see that only  
in the limit of $f_V\ll 1$ and $R\ll \lambda_{\rm D}^{\rm Q}$, we are able to recover this case. 
The ``e.m.'' curve shows the partial contribution to the thermodynamic potential, 
$\widetilde{\epsilon}_{\rm e.m.}/\beta_0\equiv(\widetilde{\epsilon}_V +
\widetilde{\epsilon}_{\rm S})/\beta_0$, ignoring correlation terms. 
Comparing these curves we can see how the screening effect
changes the thermodynamic potential: we can see that the minima at the
``e.m.'' curves disappear already at $\beta_1>0.03$, 
corresponding to unphysically
small $\tau\sim$ several MeV/fm$^2$. 
However, the correlation energy
gives a sizable contribution to allow the minimum for larger value of $\tau$. 
Consequently, the minimum totally disappears between
$\beta_1=0.1$ and $\beta_1=0.5$, which may be interpreted as 
$10<\tau<50$ MeV/fm$^2$ in this calculation. 
Thus we have seen a {\it mechanical instability} of the droplet for the medium values
of $\tau$, which might be in the physically meaningful range.

\subsection{Charge screening in quark pasta structures}

\begin{figure}
\begin{minipage}[t]{0.48\textwidth}
\includegraphics[width=0.99\textwidth]{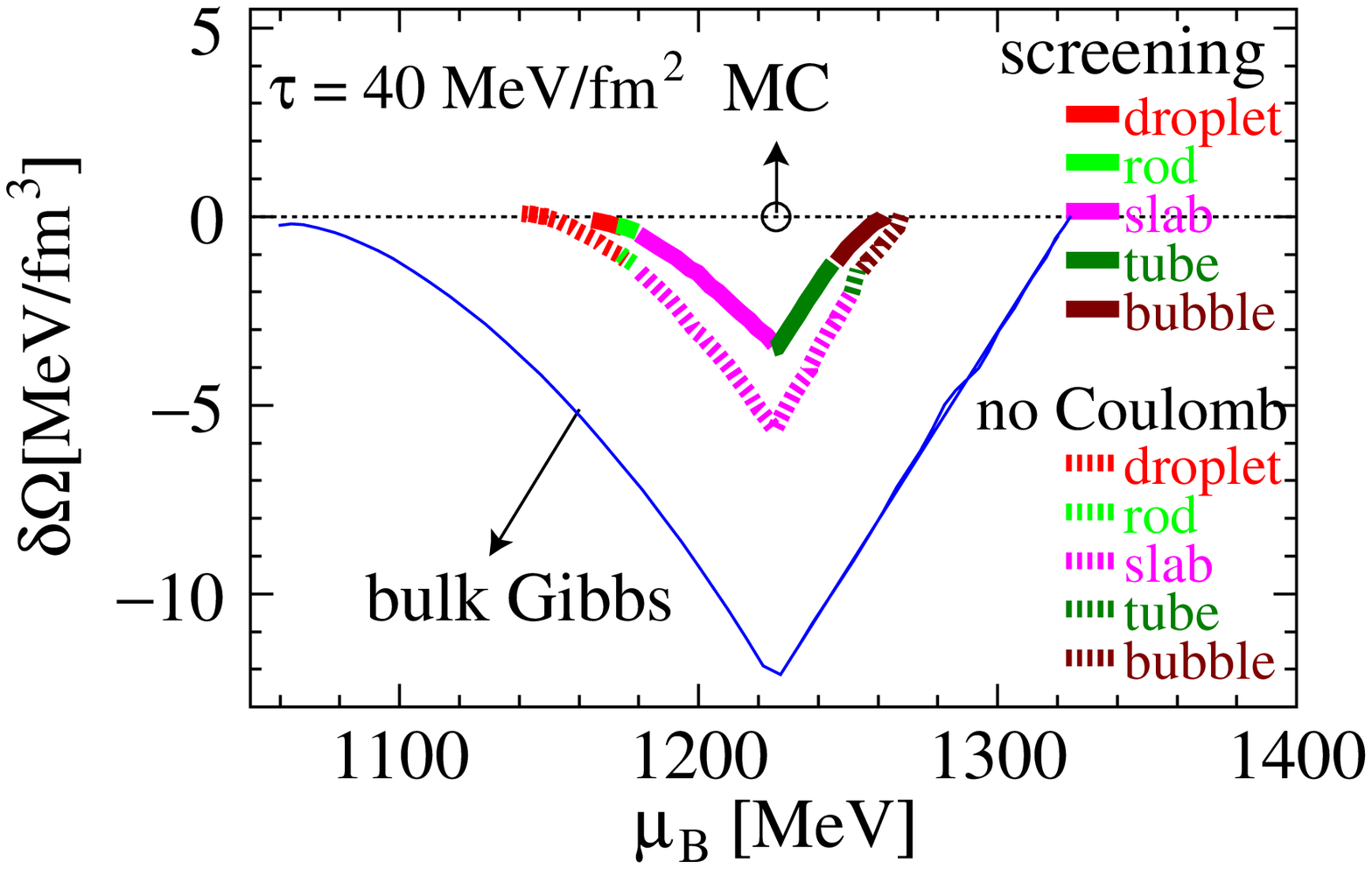}
\caption{ Difference between the thermodynamic potential densities
 as a function of baryon-number chemical potential $\mu_\mathrm{B}$ for
 $\tau=40$ MeV/fm$^2$. 
 If $\delta \omega$ is negative, the mixed phase
 is the thermodynamically favorable state. 
 The MC determines one point of the phase transition in uniform matter,
denoted as a circle in the figure.}
\label{omed40}
\end{minipage}
\hspace{8pt}
\begin{minipage}[t]{0.48\textwidth}
\includegraphics[width=0.99\textwidth]{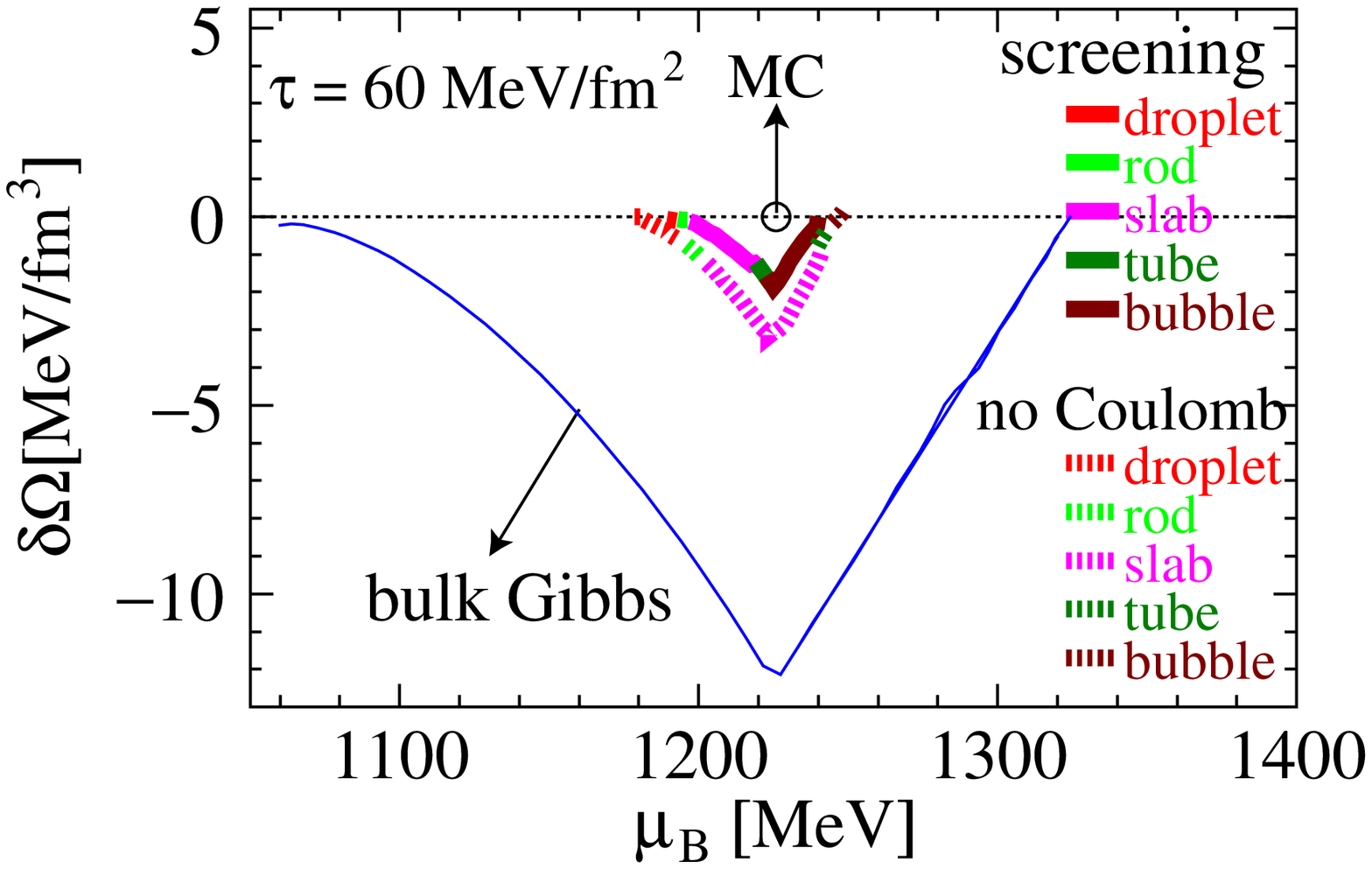}
\caption{ The same as Fig.~\ref{omed40} for $\tau=60$ MeV/fm$^{2}$. 
 The region of negative $\delta \omega$ is narrower than 
 in the $\tau=40$ MeV/fm$^{2}$ case.}
\label{omed60}
\end{minipage}
\end{figure}

We display the thermodynamic potential in Figs.\ \ref{omed40} and \ref{omed60}.
In uniform matter, the hadron phase is thermodynamically favorable for
$\mu_\mathrm{B} < 1225$ MeV, and the quark phase for $\mu_\mathrm{B} > 1225$ MeV.
Therefore we plot $\delta \omega$, the difference between the thermodynamic potential densities
of the mixed phase and each phase of uniform matter:
\begin{equation}
\delta \omega = \cases{ \omega_\mathrm{total}-\omega^\mathrm{uniform}_\mathrm{H} 
  \quad \mu_\mathrm{B} < 1225 \ {\rm MeV}, \cr
  \omega_\mathrm{total}-\omega^\mathrm{uniform}_\mathrm{Q} 
  \quad \mu_\mathrm{B} \geq 1225 \ {\rm MeV} .
}
\end{equation}
Here, $\omega_\mathrm{total}=\Omega_\mathrm{total}/V_{\rm W}$, etc.
There we also depict two results for comparison:
one is obtained from the bulk calculation,
where the finite-size effects are completely discarded (cf.\ Fig.~\ref{presbulk}),
and the  other is the thermodynamic potential obtained using a perturbative
treatment of the Coulomb interaction, which is denoted by ``no Coulomb''. 
We have employed the similar procedure as in Refs.~\cite{gle2,pet,alf2}.
In this procedure, by discarding the Coulomb potential $V({\bf r})$ to solve 
the equations of motion Eqs.\ (\ref{eomq})--(\ref{eomh}),
each density is determined to be constant in each phase for given $R$ and $R_\mathrm{W}$.
Using the chemical equilibrium relation
(\ref{chemeq}) and the charge neutrality condition
(\ref{globalneutral}), we can determine its value.
The Coulomb interaction energy (\ref{coulene}) can be separately
evaluated by using the constant densities, and
the total thermodynamic potential is obtained by adding it. 
The remaining procedure is
the same as that described in Sec.\ 6.2: we determine $R$ using the
pressure balance relation Eq.\ (\ref{pbalance}) and
the cell size ($R_\mathrm{W}$) to minimize the total thermodynamic potential.%
\footnote{
Note that they also discarded the size dependence of the particle densities
 and the surface tension for the pressure balance condition in the bulk
 calculations \cite{gle2,pet,alf2}.
}

We can see the screening effects by comparing
these ``no Coulomb'' results with the self-consistent treatment denoted by ``screening''.
The quantity $\delta \omega$ derived using the MC appears as a point denoted by a circle in
Figs.\ \ref{omed40} and \ref{omed60}, where only the two conditions 
$P^\mathrm{Q}=P^\mathrm{H}$ and
$\mu_\mathrm{B}^\mathrm{Q}=\mu_\mathrm{B}^\mathrm{H}$ are satisfied.
The mixed phase derived with the bulk Gibbs calculation appears over a
wide range of values of the $\mu_{\mathrm{B}}$.
Therefore, a narrowing the region of the mixed phase
 signals that the properties of the mixed phase have become closer
to those of the MC.
It is clearly seen that $\omega_{\mathrm{total}}$ becomes close to that
given by the MC due to the finite-size effects, the effects of the surface
tension and the Coulomb interaction.
Comparing the result of the self-consistent calculation with that in the
case of ``no Coulomb'',
we can see that the change in energy
caused by the screening effect is not so large, but still the same order of
magnitude as that caused by the surface effect.
The charge screening effect decreases the Coulomb energy. 
However, it increases the electron energy instead. 

Our parameters are the same as those of Heiselberg et al.\ \cite{pet},
and the mixed phase disappears if $\tau > 90$ MeV/fm$^2$ in the case
of ``no Coulomb''.
However, in the case of screening, it is rather difficult to estimate the 
precise value of
$\tau$ at which the mixed phase disappears, because a mechanical instability
would appear for values of $\tau$ that are not too large \cite{vos,voskre,voskre1}.
Nevertheless we could infer that the mixed phase would disappear for
$\tau = 70$ -- $80$ MeV/fm$^2$ by referring to the difference between
the cases of 40 MeV/fm$^2$ (Fig.\ \ref{omed40}) and 60 MeV/fm$^2$ (Fig.\ \ref{omed60}).

If the surface tension becomes stronger, the relative importance of the
screening effect becomes smaller, and the effect of the surface tension
becomes more dominant, as seen in Figs.\ \ref{omed40} and \ref{omed60}.
Although the charge screening does not have so large effect on the bulk properties of matter,
we shall see that its effect is significant for the charged particles in
the mixed phase, and this brings about a significant change of the properties of the mixed phase.

\begin{figure}
\begin{minipage}[t]{0.48\textwidth}
\includegraphics[width=0.95\textwidth]{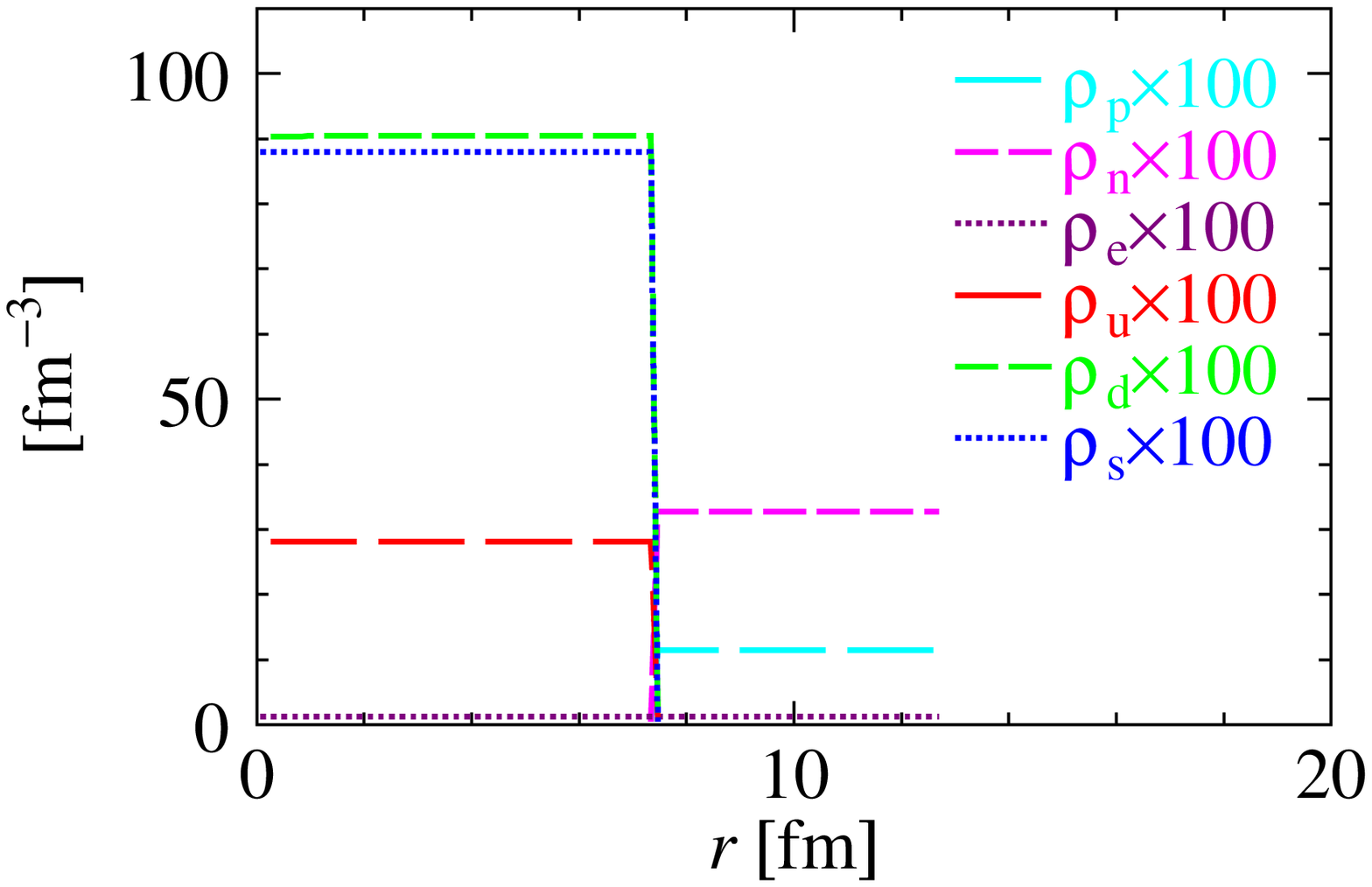}
\caption{ Density profiles in the droplet phase obtained with ``no Coulomb''
for $\mu_\mathrm{B}=1189$ MeV and $\tau=60$ MeV/fm$^{2}$.
It is seen that they are uniform in each phase.
Here, $R=7.2$ fm and $R_\mathrm{W}=12.8$ fm.}
\label{densprof-no}
\end{minipage}
\hspace{8pt}
\begin{minipage}[t]{0.48\textwidth}
\includegraphics[width=0.95\textwidth]{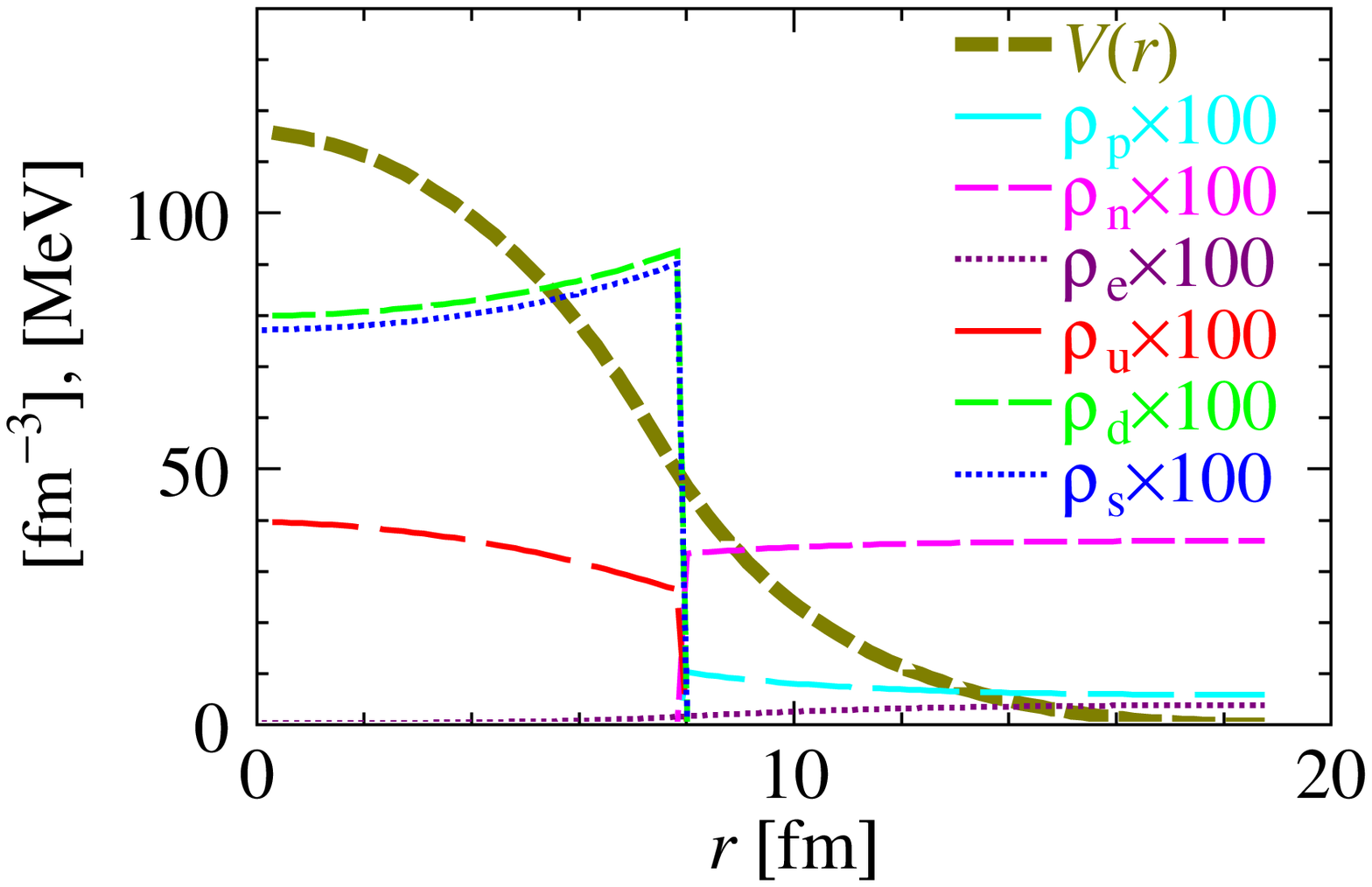}
\caption{Density profiles and the Coulomb potential derived using the self-consistent
 calculation for the same parameter set as in Fig.\ \ref{densprof-no}. 
 Here $R=7.7$ fm and $R_\mathrm{W}=18.9$ fm.}
\label{densprof-sc}
\end{minipage}
\end{figure}

The charge screening effect induces
a rearrangement of the charged particles.%
\footnote{In this context, it would be interesting to refer the paper by
Heiselberg \cite{hei}, who studied the screening effect on a quark
droplet (strangelet) in the vacuum, and also suggested the importance of
the rearrangement of the charged particle densities.}
We can see this screening effect by comparing
Fig.\ \ref{densprof-no} with Fig.\ \ref{densprof-sc}.
The quark
phase is negatively charged, and the hadron phase is positively charged.
The negatively charged particles in the quark phase, such as {\it d},
{\it s} and {\it e}, and the positively charged particle in the hadron
phase, {\it p}, are attracted toward the boundary.
In contrast, the positively charged particle in the quark phase, {\it u},
 and the negatively charged particle in the hadron phase, {\it e}, are repelled
from the boundary.

\begin{figure}
\includegraphics[width=0.50\textwidth]{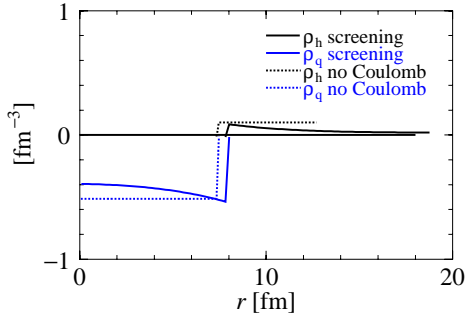}
\caption{ Local charge densities in the case of ``no Coulomb'' and the
 case of the self-consistent calculation with the screening effect.
 For ``no Coulomb'', the charge
 density of each phase is constant over the region. 
 The absolute value of the charge density is
 larger than that derived from the self-consistent calculation in each phase.
 In the hadron phase, the charge density almost
 vanishes near the cell boundary, $r=R_\mathrm{W}$.}
\label{chdensprof}
\end{figure}
The charge screening effect also reduces the net charge in each phase.
In Fig.\ \ref{chdensprof}, we display the local charge densities
of the two cases considered in Figs.\ \ref{densprof-no} and \ref{densprof-sc}.
The change of the number of charged particles due to the screening is as follows:
in the quark phase, the numbers of $d$ and $s$ quarks and
electrons decrease,
while the number of $u$ quarks increases.
In the hadron phase, contrastingly,
the number of protons should decrease and the number of electrons
should increase.
Consequently, the local charge decreases in both phases.
In Fig.\ \ref{chdensprof}, we can see that the core region of the droplet
tends to be charge neutral, and near the boundary of the Wigner-Seitz cell
it is almost charge neutral.

\begin{figure}
\begin{minipage}[t]{0.48\textwidth}
\includegraphics[width=0.99\textwidth]{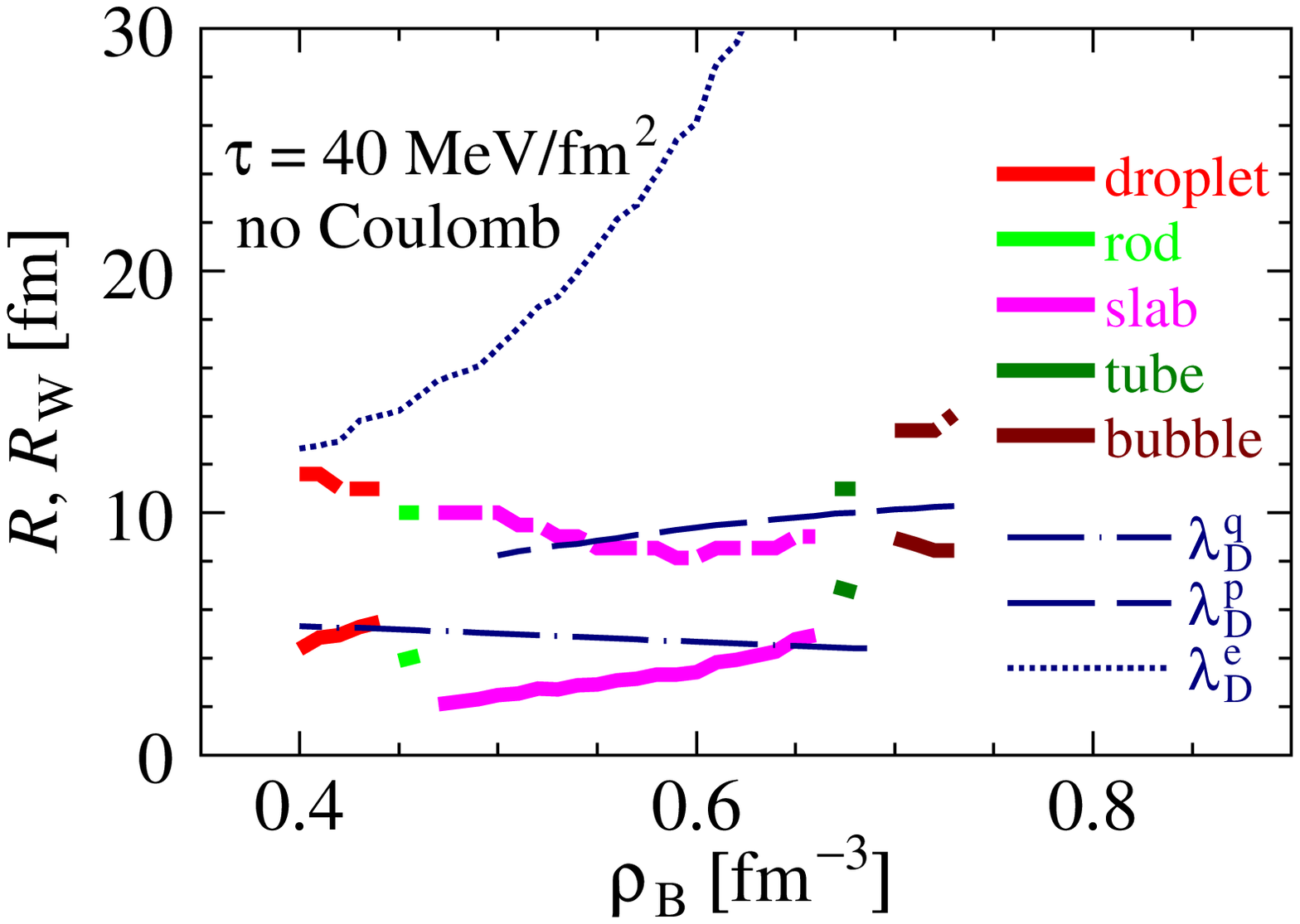}
\caption{ Lump and cell radii derived in the ``no Coulomb'' calculation. 
 The Debye screening length is also depicted for comparison. 
 $R$ is  represented by the thick, solid line and $R_\mathrm{W}$ by
 the thick, dashed line.
We can see that the size of the structure becomes less than the
 Debye screening length.}
\label{rho-cell40no}
\end{minipage}
\hspace{8pt}
\begin{minipage}[t]{0.48\textwidth}
\includegraphics[width=0.99\textwidth]{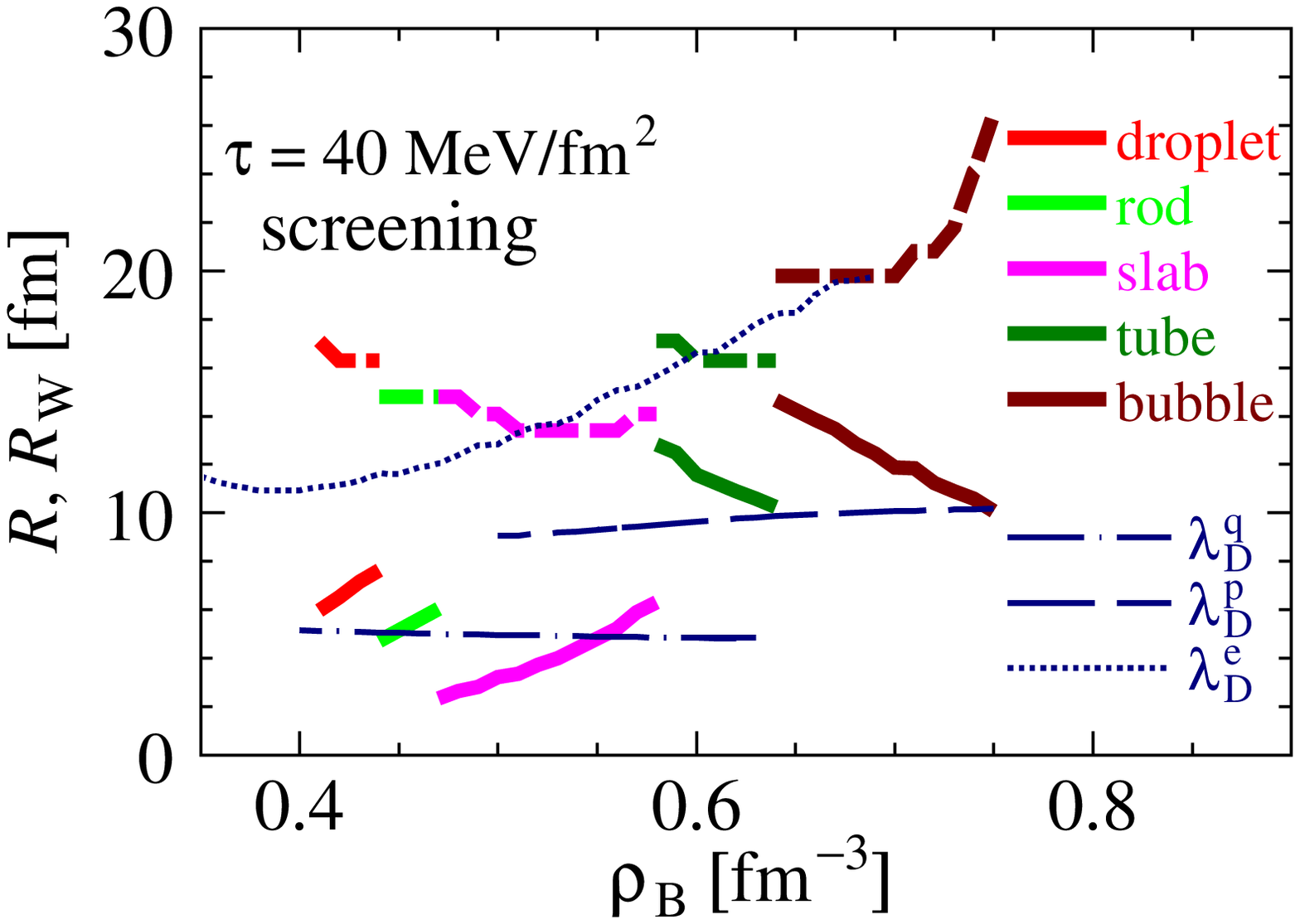}
\caption{ The same quantities as in Fig.\ \ref{rho-cell40no}
 derived using the self-consistent calculation with the screening effect. 
 The size of the structure here is larger than
that obtained with ``no Coulomb'', and exceeds the
 Debye screening length.}
\label{rho-cell40sc}
\end{minipage}
\end{figure}

In Figs.~\ \ref{rho-cell40no} and \ref{rho-cell40sc} we present radii of 
the lump and cell as functions of density.
The $R$ dependence of the total thermodynamic potential comes from the
contributions of the surface tension and the Coulomb interaction.
The optimal radius representing the minimum of the thermodynamic potential
is then determined by the balance between these two contributions.
If the Coulomb interaction energy is suppressed, 
the minimum of the thermodynamic potential is shifted
in the direction of larger radius.
As a result, the size of the embedded phase $R$ and the cell size $R_\mathrm{W}$ become large.
In the previous section we have seen that the minimum disappears for
a large value of the surface tension parameter, and hence the structure becomes
mechanically unstable.
We cannot show this directly in our framework, because such unstable solutions
are automatically excluded during the numerical procedure,
although we can see this tendency
in Figs.\ \ref{rho-cell40no} and \ref{rho-cell40sc}:
$R$ and $R_\mathrm{W}$ become larger through the screening effect.

\begin{figure}
\begin{center}
\includegraphics[width=0.3\textwidth]{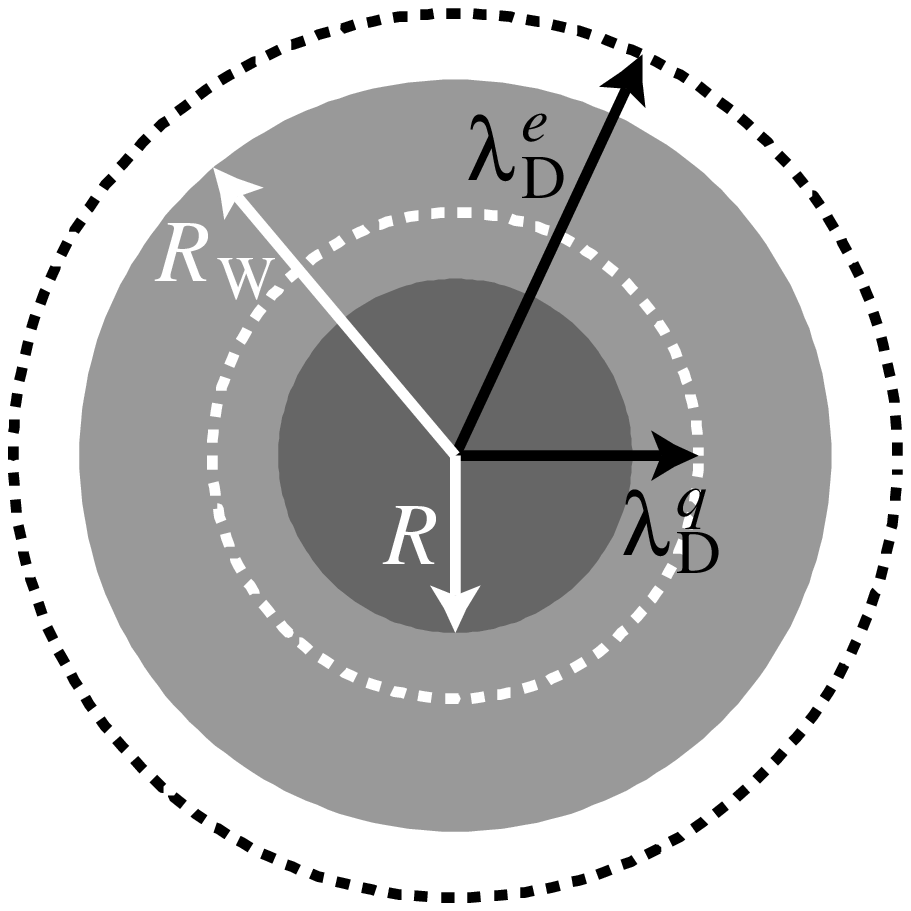}
\hspace{10pt}
\includegraphics[width=0.3\textwidth]{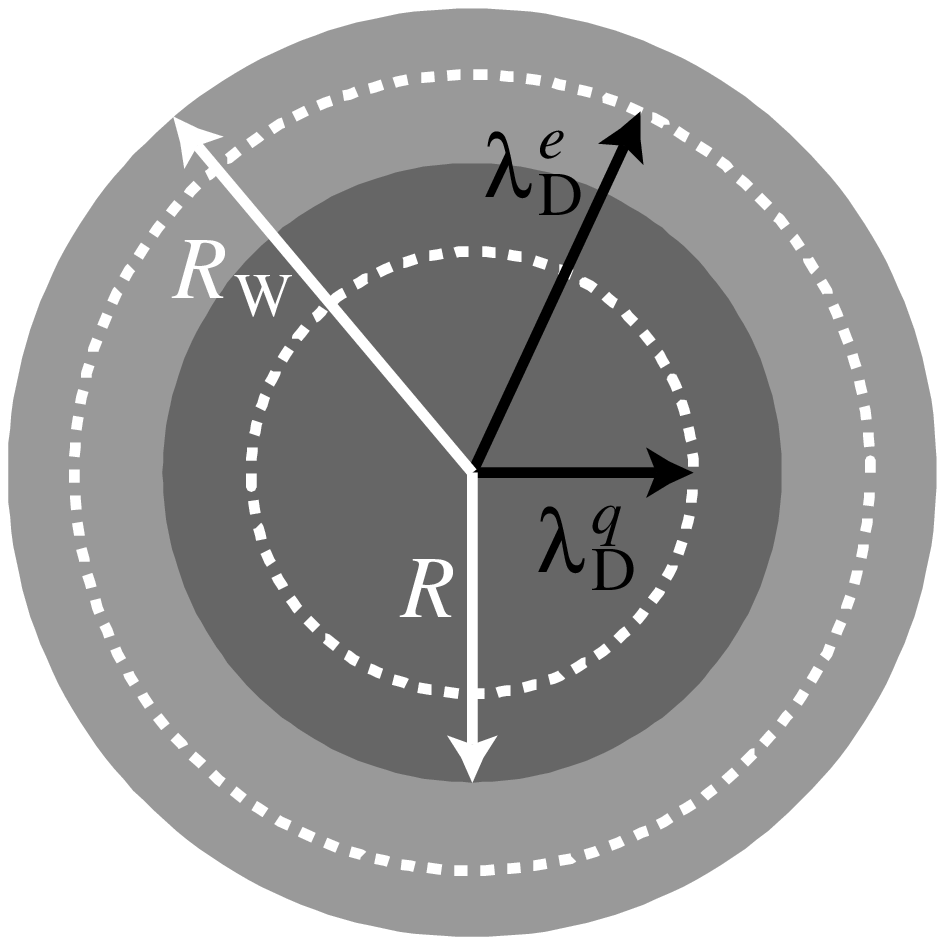}
\caption{Schematic graphs of the droplet size and the Debye screening length. 
 The right figure corresponds to the case of the self-consistent calculation with the
 screening effect, and left figure corresponds to the case of ``no Coulomb''.}
\label{lambdadrop}
\end{center}
\end{figure}

We also see the relation between the size of the geometrical structure and the Debye
screening length.
The Debye screening length appears in the {\it linearized} Poisson
equation and is given by
\begin{equation}
\left(\lambda^{q}_{\rm D}\right)^{-2}\!\!=\! 4 \pi \sum_f Q_f \! \left( \frac{\partial \langle
                                     \rho_f^\mathrm{ch}
                                     \rangle}{\partial \mu_f} \right), \hspace{5pt}
\left(\lambda^{p}_{\rm D}\right)^{-2}\!\!=\! 4 \pi Q_p \! \left( \frac{\partial \langle
                                     \rho_p^\mathrm{ch}
                                     \rangle}{\partial \mu_p} \right), \hspace{5pt}
\left(\lambda^{e}_{\rm D}\right)^{-2}\!\!=\! 4 \pi Q_e \! \left( \frac{\partial \langle
                                     \rho_e^\mathrm{ch}
                                     \rangle}{\partial \mu_e} \right),
\label{lambda}
\end{equation}
where $\langle \rho_f^\mathrm{ch} \rangle$ is
the average density in the quark phase, $\langle \rho_p^\mathrm{ch}
\rangle$ is the average proton number density in the hadron phase,
and $\langle \rho_e^\mathrm{ch} \rangle$
is the average electron number density inside the cell.
The Debye screening length gives a rough measure of the screening effect:
at a distance larger than the Debye screening length,
the Coulomb interaction is effectively suppressed.

In Fig.\ \ref{rho-cell40no}, we present sizes of the geometrical structure
in the case of ``no Coulomb''.
If we ignore the screening effect, the size of the embedded phase is
comparable to or smaller than the corresponding quark Debye screening
length, $\lambda_{\rm D}^q$, for a droplet, rod and slab, or the proton Debye screening
length, $\lambda_{\rm D}^p$, for a tube
and bubble (Fig.\ \ref{lambdadrop}).
This may imply that the Debye
screening is not so important.
Actually, many authors have ignored the screening
effect for the reason elucidated by this argument \cite{pet,alf2}.
In Fig.\ \ref{rho-cell40sc}, however,
we see that the size of the embedded phase
can be larger than $\lambda_{\rm D}^q$ (Fig.\ \ref{lambdadrop})
in the self-consistent calculation.
We can also see a similar situation concerning $R_\mathrm{W}$ and $\lambda_{\rm D}^e$.
This means that the screening has important
effects in this mixed phase.
We cannot expect such an effect
without solving the Poisson equation,
because of the non-linearity.\footnote{
Within the linear approximation for the charge screening, Iida ans Sato have studied
the screening effect on quark droplets in a different context \cite{iida}.
In their case, the electron number density
is much larger than in our case, because they considered two-flavor quark matter. 
Accordingly, the electron screening effect may be dominant, in contrast to 
the situation in our case.
}

\begin{figure}
\begin{minipage}[t]{0.48\textwidth}
\includegraphics[width=0.99\textwidth]{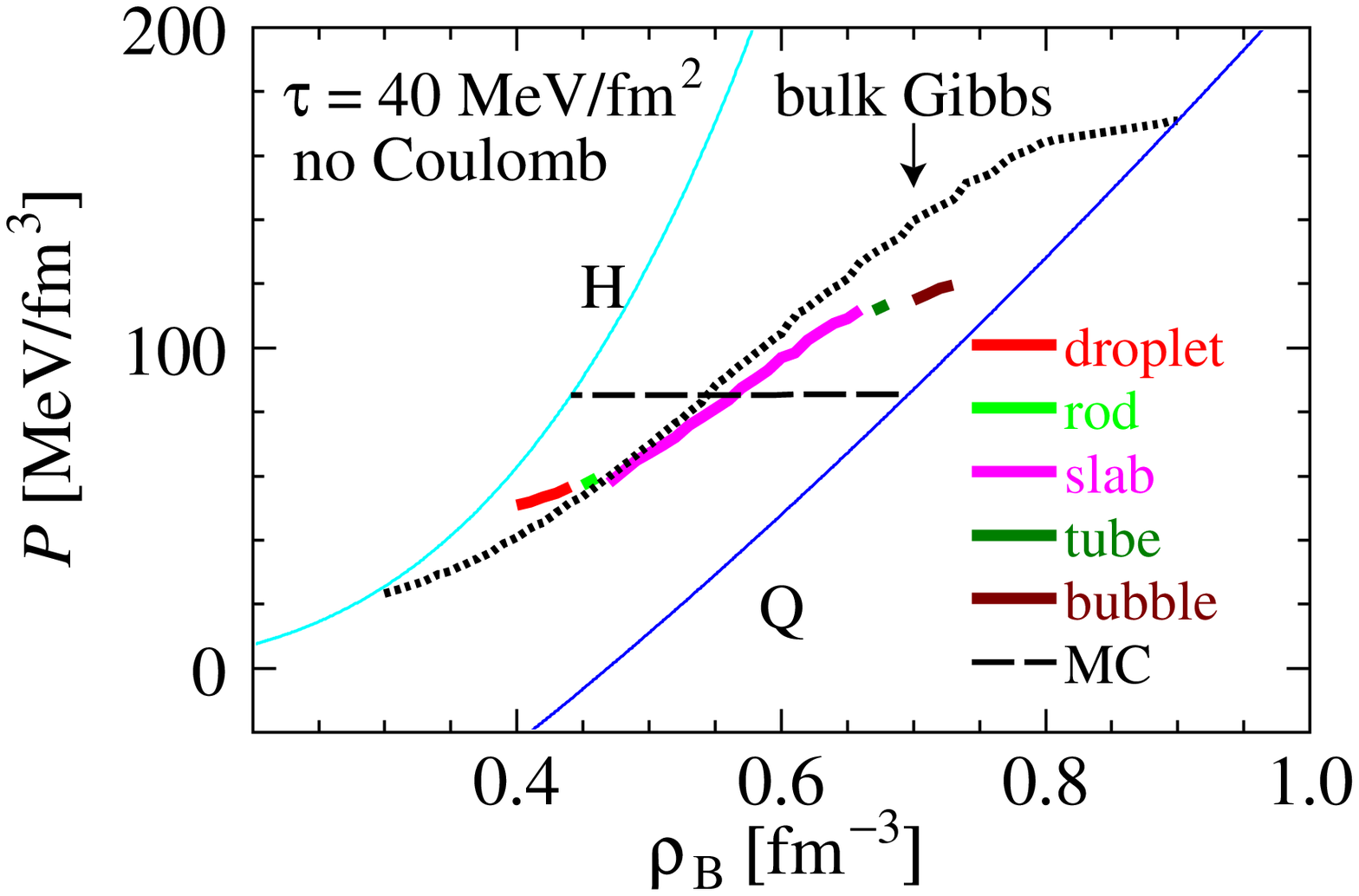}
\caption{ Pressure as a function of the baryon-number density obtained from the
 ``no Coulomb'' calculation for $\tau=40$ MeV/fm$^2$. 
 The results found with the bulk Gibbs calculation and the MC are 
 also presented for comparison.}
\label{pres40no}
\end{minipage}
\hspace{8pt}
\begin{minipage}[t]{0.48\textwidth}
\includegraphics[width=0.99\textwidth]{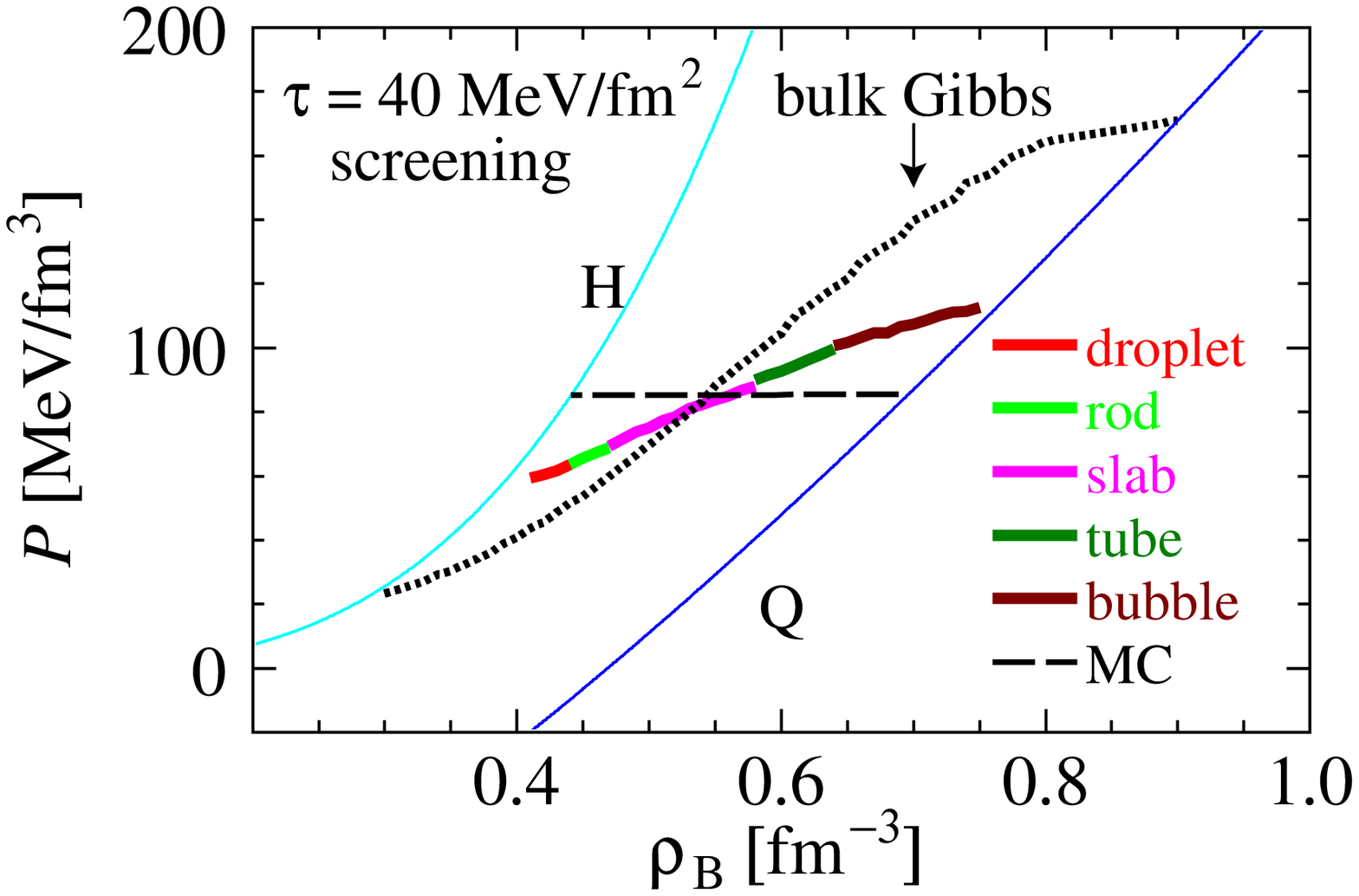}
\caption{ The same quantities as in Fig.\ \ref{pres40no}
 computed using the self-consistent calculation with the
 screening effect.}
\label{pres40sc}
\end{minipage}
\end{figure}

We plot the EOS
in Figs.\ \ref{pres40no} and \ref{pres40sc}. 
It is seen that the pressure of the mixed phase becomes similar to that found using the MC.
Because a region of local charge neutrality appears
 due to the screening effect (see Fig.\ \ref{chdensprof}), its
 properties correspond to those derived using the  MC.

\section{Summary and concluding remarks}

We have investigated three forms of non-uniform hadron matter 
at different density ranges, which correspond to 
the mixed phases in the first-order phase transitions. 
As has been discussed by many authors, such non-uniform structures appear
as ``pasta'' structures, i.e.\ droplet, rod, slab, tube and bubble.
By means of mean-field approach under the Wigner-Seitz cell approximation,
we have calculated the ground-state density profiles of matter. 
The size and the dimensionality of the cell
were chosen to their optimal values for a given baryon density.
Thus we could determine optimal structure of matter along the
change of density.

First we have studied the pasta structures in low-density nucleon matter.
In this study we have used the relativistic mean-fields (RMFs) as an
effective interaction.
The parameter set we have used gives the incompressibility of matter $K=240$ MeV 
and the symmetry energy coefficient of 32.5 MeV 
at the saturation density of symmetric nuclear matter, 
and can well reproduce the bulk properties of finite systems (nuclei).
In nucleon matter with a fixed proton mixing ratio,
which corresponds to supernova matter in the collapsing stage,
we have observed the nuclear pasta: with increase of density the favorable
structure changes from nuclear droplet, rod, slab, tube, and to bubble. 
Near the saturation density (normal nuclear density), 
nuclear pasta dissolved to uniform matter.
By the appearance of such non-uniform structure, the equation of state
of the system becomes significantly soft (the energy gets lower up to 15 MeV/$A$).
In the case of the beta equilibrium (neutron star matter),
we did not observe the full pasta structures but only proton droplets
appeared in the neutron sea.

Our second target was the kaon condensation at around three times 
the normal nuclear density,
which is relevant to the inner core of neutron stars.
Using the same RMF model, we have added the kaonic degree of freedom
and could naturally describe both normal nucleon matter and kaonic matter.
The surface tension between nucleon matter and kaonic matter is
automatically included in the framework.
With this model we observed the appearance of pasta structures:
kaonic droplet, kaonic rod, etc.

An important advantage of our framework is the consistent 
treatment of the Coulomb interaction and charged particle distribution: 
we have included the Coulomb potential in the thermodynamic potential 
in a gauge-independent way and solved the Poisson equation consistently with 
the charged particle densities. 
If we take uniform charge densities as in the bulk calculations, 
we overestimate the potential energy due to the neglect
of the charge screening effect. 
In fact, we have seen the importance of
the charge screening for the properties of the mixed phases; the region
of the mixed phase is largely modified, 
and it may cause a mechanical instability of the geometrical structure in some cases.
As an important consequence of the charge screening effect, the EOS becomes 
more or less similar to that given by the Maxwell construction 
which is irrelevant for multiple particle composition. 
Actually the electron chemical potential should be different in the absence of 
the Coulomb potential because of the different 
electron number in two phases. 
However, we have emphasized that charge chemical potential is only defined after 
the gauge fixing of the Coulomb potential, so that number densities are not solely 
given by the chemical potential but by the linear combination of the chemical 
potential and the Coulomb potential. 
The different strength of the Coulomb potential in two phases can give 
the different particle densities, 
which can resolve an apparent contradiction in the Maxwell construction.  

So far, we have explored only nuclear matter in the ground state (zero temperature).
However, supernova matter, for example, should be as hot as several MeV.
It is naturally expected that the pasta structure will be modified and
dissolved at around the critical temperature \cite{Gen04}.
Our model is also applicable for not-very-high temperature 
by modifying the baryon momentum distribution.
The structural phase diagram in density-temperature plane
should be useful for the discussion of supernova explosion.

We have used the meson exchange model (MEM) to describe the
kaon-nucleon interaction. 
If we use the chiral model instead, it gives a
nonlinear Lagrangian including the kaon self-interaction. 
The thermodynamic inconsistency of the chiral
model has been shown within the mean-field
approximation \cite{PREPL00,YT00,kub}: 
a bulk calculation of the mixed phase consisting of two semi-infinite matter
have concluded that we can not construct the mixed phase satisfying the
Gibbs conditions. 
It may be caused by the non-linearity of the kaon field. 
So, it would be worth studying the chiral model with our
framework to see whether the chiral model is really ill defined in a
thermodynamic sense.

We have seen that our framework based on RMF can well reproduce the bulk
properties of finite nuclei as well as nuclear matter. 
It should be interesting to describe kaonic nuclei \cite{ay02,d04,k99,i04} as a 
direct application of our framework. 
Then we can get a consistent description of kaonic
nuclei by taking into account the Coulomb interaction as well as
the $K$-$N$ interaction \cite{maru052}.

Then our third subject was the mixed phase of hadron-quark phase transition.
We have used a rather simple model for this study.
Nowadays there have been many studies about the deconfinement transition 
and its implications on compact stars \cite{glebook,web,gle2,gle4,pet} and
relativistic heavy-ion collisions \cite{mul,rhic21,rhic22,rhic23}. 
Let us consider some implications of these results for the structure of
hybrid stars.
Glendenning and Pei \cite{gle2} considered some geometrical structures
in the mixed phase and conjectured that ``crystalline
structures'' (quark pastas) would appear 
in the core region of a hybrid star by using the bulk Gibbs calculation.
His results predict that the mixed phase should appear for several kilometers.
However, we can say that the region of SMP
should be narrow in the $\mu_\mathrm{B}$ space, and the EOS is more similar
to that obtained with the MC,
due to the finite-size effects.

We have considered the deconfinement transition from non-strange nuclear
matter for simplicity, while realistically we must take into account hyperons. 
Many people believe that hyperons will mix in nuclear matter 
at several times of the saturation density \cite{tak,panda,baldo,vidana,nishizaki,sch,mei}. 
In such a case we must consider the mixed phase by taking into account hyperons as well. 
One of the interesting consequences may be related to the properties of hybrid stars:
actually it has been shown that the admixture of hyperons considerably
softens EOS, while the deconfinement transition stiffens
EOS again at high-density region \cite{sch,mei}. 
If the stiffening works well, 
it becomes rather easy to construct hybrid stars.
For quark matter we should consider the possibilities of color
superconductivity \cite{alf1,alf3} and ferromagnetism
\cite{tat1,tat2,tat}.
In particular, magnetic properties of quark matter might be related to
the strong magnetic field observed in compact stars \cite{mag1,mag2,mag3}. 

Our framework may be applicable for other subjects. 
One of the interesting subjects may be color superconductivity. 
There have been many works about color superconductivity, 
where various first-order phase transitions have been suggested between 
different types of superconducting phases \cite{neu,shov,redd}. 
Color neutrality as well as charge neutrality 
should be considered in that case. 
Accordingly 3rd and 8th gluons
should play an important role for global color neutrality.

\section*{Acknowledgments}

The authors thank M.~Alford, M.~Baldo, D.~Blaschke, A.~Bonasera, K.~Iida, A.~Iwamoto, 
N.~Iwamoto, E.~Kolomeitsev, S.~Kubis, U.~Lombardo, Tomoyuki Maruyama, T.~Muto,  
S.~Nishizaki, K.~Oyamatsu, 
C.~J.~Pethick, S.~Reddy, H.-J.~Schulze, T.~Takatsuka, R.~Tamagaki,  T.~Tanigawa,  
S.~Tsuruta, D.~N.~Voskresensky, G.~Watanabe, F.~Weber and M.~Yasuhira 
for their interests and discussions.
This work is partially supported by the Grant-in-Aid for the 21st Century COE
``Center for the Diversity and Universality in Physics'' from
the Ministry of Education, Culture, Sports, Science and
Technology of Japan. 
It is also partially supported by the Japanese
Grant-in-Aid for Scientific
Research Fund of the Ministry of Education, Culture, Sports, Science and
Technology (13640282, 16540246).

\appendix

\section*{APPENDIX: Kaon interactions within the chiral model}

We have used the meson exchange model (MEM) to describe kaon condensation.
When we use the chiral model instead, the thermodynamic potential is given as follows:
\begin{eqnarray}
\Omega&=&\Omega_B+\Omega_K+\Omega_M+\Omega_e,\nonumber\\
\Omega_B  &=&
  \sum_{a=p,n}
  \int  d^3r
  \left[
  \int_0^{k_{{\rm F},a}}
  { d^3k \over 4\pi^3}
  \sqrt{{m_N^*}^2+k^2}-\rho_a\nu_a
  \right],\nonumber\\
\Omega_K\!\!&=&\!\!\int\!\! d^3r\!\! \left[ 
-f^2(\cos\theta\!-\!1)\left(m_K^{*2}\!-\!2(\mu-V) X_0\right)\!-\!\frac{1}{2}(\mu\!-\!V)^2f^2\sin^2\theta
+\frac{f^2}{2}(\nabla\theta)^2\right],\nonumber\\
\Omega_M\!\!&=&\!\!\int\!\! d^3r\!\!\left[\frac{1}{2}(\nabla\sigma)^2
+\frac{1}{2}m_\sigma^2\sigma^2
\!+\!U(\sigma)
\!-\!\frac{1}{2}(\nabla\omega_0)^2\!-\!\frac{1}{2}m_\omega^2\omega_0^2
\!-\!\frac{1}{2}(\nabla R_0)^2\!-\!\frac{1}{2}m_\rho^2 R_0^2\right],\nonumber\\
\Omega_e&=&\int d^3r\left[-\frac{1}{8\pi e^2}(\nabla V )^2
-\frac{(V-\mu)^4}{12\pi^2}\right],
\label{chiralomega}
\end{eqnarray}
where the kaon field is defined as 
\begin{equation}
K=\frac{f}{\sqrt{2}}{\rm sin}\theta
\end{equation}
and
\begin{eqnarray}
X_0&=&g_{\omega K}\omega_0+g_{\rho K}R_0,\nonumber\\
\mu_p+V&=&\nu_p+g_{\omega N}\omega_0+g_{\rho N}R_0,\nonumber\\
\mu_n&=&\nu_n+g_{\omega N}\omega_0-g_{\rho N}R_0,\nonumber\\
m_K^{*2}&=&m_K^2-2g_{\sigma K}m_K\sigma,\nonumber\\
m^*_N&=&m_N-g_{\sigma N}\sigma.
\end{eqnarray}
Then the equations of motion can be easily written down,
\begin{eqnarray}
-\nabla^2\sigma+m_\sigma^2\sigma&=&-\frac{dU}{d\sigma}+g_{\sigma B}(\rho_n^s+\rho_p^s)
-2g_{\sigma K}m_Kf^2(\cos\theta-1),\nonumber\\
-\nabla^2\omega_0+m_\omega^2\omega_0&=&g_{\omega N}(\rho_n+\rho_p)
+2f^2g_{\omega K}(\cos\theta-1)(\mu-V),\nonumber\\
-\nabla^2 R_0+m_\rho^2 R_0&=&g_{\rho N}(\rho_p-\rho_n)
+2f^2g_{\rho K}(\cos\theta-1)(\mu-V),\nonumber\\
\nabla^2 V&=&4\pi e^2\rho^{ch},\nonumber\\
\nabla^2\theta&=&\sin\theta\left[m_K^{*2}-2(\mu-V)X_0-(\mu-V)^2\cos\theta\right],
\label{chiraleom}
\end{eqnarray}
where
\begin{eqnarray}
\rho^{ch}&=&\left[ \rho_p -\rho_K -\frac{(\mu-V)^3}{3\pi^2} \right],\nonumber\\
\rho^K&=& (\mu-V)f^2\sin^2\theta+2f^2(\cos\theta-1)X_0.\nonumber
\end{eqnarray}
We can see how the thermodynamic potential (\ref{chiralomega}) or the
equations of motion (\ref{chiraleom}) can recover the previous formulae
used in the studies of kaon condensation in uniform matter \cite{PREPL00,YT00}.
The $K$-$N$ interaction terms are easily extracted from Eq.~(\ref{chiraleom}).
Discarding  $V$, the kaon source terms for the mean-fields $\sigma,
\omega,\rho$ and the nonlinear potential for $\sigma$, $U(\sigma)$, 
we have 
\begin{eqnarray}
\sigma&=&\frac{g_{\sigma N}(\rho_n^s+\rho_p^s)}{m_\sigma^2},\nonumber\\
\omega_0&=&\frac{g_{\omega N}(\rho_n+\rho_p)}{m_\omega^2},\nonumber\\
R_0&=&\frac{g_{\rho N}(\rho_p-\rho_n)}{m_\rho^2},
\end{eqnarray}
and
\begin{eqnarray}
X_0&=&\frac{g_{\omega N}g_{\omega K}}{m_\omega^2}(\rho_n+\rho_p)
+\frac{g_{\rho N}g_{\omega K}}{m_\rho^2}(\rho_p-\rho_n),\nonumber\\
m_K^{*2}&=&m_K^2-2m_K\frac{g_{\sigma N}g_{\sigma K}}{m_\sigma^2}(\rho_n^s+\rho_p^s),
\end{eqnarray}
for soft kaons. 
If we impose the following relations among the coupling constants:
\begin{eqnarray}
\frac{g_{\sigma N}g_{\sigma K}}{m_\sigma^2}&=&\frac{\Sigma_{KN}}{2m_Kf^2},\nonumber\\
\frac{g_{\omega N}g_{\omega K}}{m_\omega^2}&=&\frac{3}{8f^2},\nonumber\\
\frac{g_{\rho N}g_{\rho K}}{m_\rho^2}&=&\frac{1}{8f^2},
\end{eqnarray}
we can recover the $K$-$N$ interaction terms dictated by chiral symmetry,
\begin{eqnarray}
X_0&=&\frac{1}{4f^2}(\rho_n+2\rho_p),\nonumber\\
m_K^{*2}&=&m_K^2-\frac{\Sigma_{KN}}{f^2}(\rho_n^s+\rho_p^s).
\end{eqnarray}
Thus we can derive the previous formulae for kaon
condensation within the chiral model \cite{PREPL00,YT00}

To get MEM, we linearize
Eqs.~(\ref{chiralomega}) and (\ref{chiraleom}) with respect to $\theta$
and further add the non-linear terms, $X_0^2\theta^2$ and
$\sigma^2\theta^2$, in $\Omega_K$.


\end{document}